%% file: 0_main.tex
\documentclass[sigconf, nonacm, pdfa]{acmart}

\newcommand\vldbdoi{10.14778/3819518.3819561}
\newcommand\vldbpages{2426 - 2438}
\newcommand\vldbvolume{19}
\newcommand\vldbissue{9}
\newcommand\vldbyear{2026}
\newcommand\vldbauthors{\authors}
\newcommand\vldbtitle{\shorttitle} 
\newcommand\vldbavailabilityurl{https://github.com/YannianNiu/STEM2_VLDB2026}
\newcommand\vldbpagestyle{empty} 

\usepackage{xcolor}
\usepackage{xspace}
\usepackage{subfigure}
\usepackage{subcaption}
\usepackage{paralist}
\usepackage{balance}
\usepackage[normalem]{ulem}
\usepackage{textcase}
\usepackage{hyperref}
\usepackage{enumitem}
\usepackage{mdframed}
\usepackage[a-2b]{pdfx}

\usepackage{booktabs} % 必选，用于绘制专业粗细线条
\usepackage{pifont}   % 必选，提供更好看的对号和叉号
\usepackage{array}    % 优化列表格宽度

\newcommand{\cmark}{\ding{51}}                      % 普通对号
\newcommand{\xmark}{\textcolor{red}{\ding{55}}}     % 红色叉号，一劳永逸

\newcommand{\sysname}{STEM$^2$\xspace}
\newcommand{\wang}[1]{\textcolor{red}{#1}}

\begin{document}

\title{\sysname: A Fast and Space-efficient Data Structure for Exact Multi-Set Membership Queries}

%%
%% The "author" command and its associated commands are used to define the authors and their affiliations.

\author{Yannian Niu}
\affiliation{
  \institution{University of Connecticut}
  \city{}
  \state{}
  \country{}}
\email{yannian.niu@uconn.edu}

\author{Song Han}
\affiliation{
  \institution{University of Connecticut}
  \city{}
  \state{}
  \country{}}
\email{song.han@uconn.edu}

\author{Minmei Wang}
\authornote{Corresponding author.}
\affiliation{
  \institution{University of Connecticut}
  \city{}
  \state{}
  \country{}}
\email{minmei.wang@uconn.edu}

\begin{abstract}
Multi-set membership queries are ubiquitous in networking and database systems. Current solutions force a difficult compromise: hash tables guarantee correctness but suffer from high memory footprints, while filter-based approaches optimize space at the cost of probabilistic errors. In this paper, we propose \sysname, a fast and space-efficient data structure that achieves 100\% query accuracy and can support dynamic key updates for multi-set membership queries. \sysname utilizes a balanced binary tree architecture where each non-leaf node incorporates a novel Exact Binary Set Separator (XBSS) to partition keys into two disjoint groups. A key innovation of our design is a minimized hashing scheme that requires only two hash computations per key lookup, significantly reducing computational overhead. Additionally, \sysname separates the control plane and the data plane: the control plane handles construction and dynamic updates, while the data plane is dedicated to serving efficient membership queries. Extensive experiments show that \sysname achieves over 120 million operations per second (Mops) in lookup throughput, outperforming the state-of-the-art Coloring Embedder by 20\% and the Ludo hashing by up to $21.6\times$, while maintaining compact memory cost and exact correctness.

\end{abstract}

\maketitle

\pagestyle{\vldbpagestyle}
\begingroup\small\noindent\raggedright\textbf{PVLDB Reference Format:}\\
\vldbauthors. \vldbtitle. PVLDB, \vldbvolume(\vldbissue): \vldbpages, \vldbyear.\\
\href{https://doi.org/\vldbdoi}{doi:\vldbdoi}
\endgroup
\begingroup
\renewcommand\thefootnote{}\footnote{\noindent
This work is licensed under the Creative Commons BY-NC-ND 4.0 International License. Visit \url{https://creativecommons.org/licenses/by-nc-nd/4.0/} to view a copy of this license. For any use beyond those covered by this license, obtain permission by emailing \href{mailto:info@vldb.org}{info@vldb.org}. Copyright is held by the owner/author(s). Publication rights licensed to the VLDB Endowment. \\
\raggedright Proceedings of the VLDB Endowment, Vol. \vldbvolume, No. \vldbissue\ %
ISSN 2150-8097. \\
\href{https://doi.org/\vldbdoi}{doi:\vldbdoi} \\
}\addtocounter{footnote}{-1}\endgroup

\ifdefempty{\vldbavailabilityurl}{}{
\vspace{.3cm}
\begingroup\small\noindent\raggedright\textbf{PVLDB Artifact Availability:}\\
The source code, data, and/or other artifacts have been made available at \url{\vldbavailabilityurl}.
\endgroup
}

\input{1_introd}
\input{2_related}
\input{3_motivation}

\input{4_XBSS}
\input{4_design_revision}

\input{5_eval}
\input{6_conclusion}

\begin{acks}
 This work was supported by the National Science Foundation under Grant IUCRC-1916756 and by industry funding from the Center for Hardware Embedded Systems Security and Trust (CHEST). The work of Yannian Niu and Minmei Wang was partially supported by the National Science Foundation under Grant CNS-2426030. The work of Song Han was partially supported by the National Science Foundation under Grant CNS-2008463. 
 % This work was supported by the [...] Research Fund of [...] (Number [...]). Additional funding was provided by [...] and [...]. We also thank [...] for contributing [...].
\end{acks}

\balance
\bibliographystyle{ACM-Reference-Format}
\bibliography{reference}

\clearpage
\setcounter{page}{1}
% \input{7_response}
% \appendix
% \input{appendixA}

\end{document}

%% file: 1_introd.tex
\section{Introduction}
\label{sec:intro}

Given a universal key set $U = \{x_1, x_2, ..., x_i, ..., x_n\}$, which is partitioned into $m$ disjoint subsets such that $S_1 \cup S_2 \cup ...\cup S_{m-1} \cup S_m = U$ and $S_i \cap S_j = \emptyset$ for all $i \neq j$, the multi-set membership queries (MS-MQ) is defined as follows: Given a key $x \in U$, determines the set ID $i \in \{1,2,..., m\}$ such that $x \in S_i$.

MS-MQ is a fundamental primitive in a wide range of network and database applications, including network traffic management and routing \cite{yang2014guarantee, yu2009buffalo, shi2019re}, distributed data storage/caching \cite{fan2000summary, zhang2015mega}, security and privacy-preserving systems \cite{zhang2023linear, nevo2021simple}, and a variety of data mining tasks such as customer purchase analysis \cite{chen2021multiset,lo2016understanding}. Below, we present two representative case studies.

%For example, computer networking, data science, distributed file system and information retrieval of data center. We lists some real-world case studies in the following:

%and other related fields, such as network packet classification, intrusion detection system

\noindent \textbf{Case 1: Layer-2 packet forwarding}. Packet forwarding~\cite{shi2019re} is a fundamental network function that directs incoming packets to their corresponding output ports. Specifically, layer-2 switches perform this operation by querying a MAC table. This querying process can be viewed as a MS-MQ problem, where the MAC address of a packet serves as the key and the associated port number represents the set ID. A typical MAC table may contain tens of thousands of entries spanning multiple ports~\cite{coloringembedder}. Efficiently determining the correct port for a given MAC address is therefore essential for high-speed packet processing.

%Sets with identical IDs contain multiple elements (MAC address). In practice, tens of millions packets should be processed to support line-rate forwarding, which pose great challenge to switches with very limited hardware resource (memory and computation resource).

%one MAC table typically contains ten thousands of entries(MAC address-port pairs), which poses  to support line-rate forwarding. 

%\noindent \textbf{Case 2: Distributed cache retrieval.} Large language models require key-value caching to reduce inference complexity and significantly improve long-context generation efficiency. When the model is large and the context is long, the KV cache will consume a significant amount of memory. Distributed KV-cache technology stores cache across multiple GPUs/nodes/devices and fetch them when it is needed. Currently, multiple studies are proposed to alleviate this challenge by providing error-allowed query algorithm, which means the query accuracy is not 100\%. However, caching on the wrong device and correct errors usually means more resource consumption and time delay, which seriously affect the token generation rate and user experience.

\noindent \textbf{Case 2: Entity tag query.} Entity tag query is a common operation in many practical systems, where each entity is assigned exactly one tag (or identifier) from a finite set. For example, customers may be classified into different VIP levels (Bronze, Silver, Gold, Platinum), or IoT devices may be grouped into priority tiers for resource allocation. The tag query can be formulated as a MS-MQ problem, where the entity identifier (e.g., customer ID or device ID) serves as the key, and the corresponding tag (e.g., VIP level or priority tier) represents the set ID.

%\noindent \textbf{Data science.} In modern commercial application, data analysis has important applications in areas such as user profiling and advertisements targeting. For example, a credit card transaction database contains billions of entries and each entry is a tuple of transaction id, card number, paid amounts and other necessary information. We want to quickly map the consuming record corresponding to the range of consuming amounts, which is a typical multiple set query problem.

%\noindent \textbf{Privacy protection}. In many commercial applications, how to analyze users' business backgrounds without compromising their privacy has become an important topic. For example, a credit card transaction database contains billions of entries including user name, card number, paid amounts and name of the productions paid for. 

A desired MS-MQ solution should meet the following properties:

%\subsection{prior arts and their limitation, motivation}

%Above three use cases bring the following requirements of this multi-set query solution is as follows:

\begin{itemize}[nosep, leftmargin=*]
    \item \textbf{High lookup throughput.} Efficiently query the set ID for a given key is essential. High lookup throughput directly impacts the scalability and responsiveness of the applications, enabling real-time query processing even under large-scale workloads.
    %This metric is particularly important for network applications. The throughput of multi-set query algorithm have to match the packet transmission/data transfer to avoid becoming a bottleneck in network/database services.
    \item \textbf{Correctness of query results.} Ensuring 100\% accuracy in query results is critical. Although some applications may tolerate a small error rate, even minor inaccuracies can cause significant resource inefficiencies, such as wasted computing power, I/O capacity, and network bandwidth. For instance, an incorrect forwarding decision may send a packet to the wrong port, leading to misdelivery and subsequent retransmissions. % Although most current set query algorithms introduce very small error rates, the query errors will result in additional effort to design a error tolerance mechanism and penalties in terms of time and resources. 
    
    \item \textbf{Low memory cost.} Devices such as switches, routers, and IoT nodes that perform MS-MQ often operate under strict resource constraints. Designing a compact data structure is essential to enable efficient operation within limited memory budgets.
    
    %\item Compact memory consumption. For tables and dataset which contains hundreds of millions of entries, compact memory consumption is critical since the limited memory resources of switches/edge node/GPUs.
    \item \textbf{Dynamic update support.} Since set memberships can change over time, the proposed data structure and algorithm should efficiently handle dynamic key updates, including insertions, deletions, and migrations.
    %\item Supporting dynamic update (insertion and deletion). High speed update is required to handle scenarios of network state update and database update. For solution which don't support dynamic update, rebuild the whole data structure is required once there is a update of sets, which cause a long latency.
\end{itemize}

Existing solutions for MS-MQ problem can be divided into two categories: hash tables and filter-based methods. Hash tables are conventional data structures for key-value storage and can be adapted for MS-MQ by storing the set ID as the value associated with each key. However, traditional hash tables incur high memory cost due to the requirement of explicit key storage. Recently, several efficient key-value stores have been proposed, including partial key Cuckoo hashing~\cite{fan2014cuckoo, lim2011silt}, SetSep~\cite{fan2013cycles, zhou2015scaling}, Bloomier/Othello~\cite{chazelle2004bloomier, charles2008bloomier, othello}, and Ludo hashing~\cite{ludo}. These key-value stores are memory-efficient and can support high lookup throughput. Specifically, Ludo hashing achieves the lowest space cost for dynamic key-value lookups among existing solutions~\cite{ludo}. However, traditional key-value lookup tables are designed for general-purpose applications, where both keys and values may vary arbitrarily. In contrast, MS-MQ involves value drawn from a fixed and finite set (e.g., set identifiers). Moreover, the size distribution of sets in such applications is often highly skewed. For instance, network traffic flows in large-scale systems typically follow Zipfian or Zipfian-like distributions~\cite{sen2002analyzing}. These unique characteristics make general-purpose hash tables less efficient in both memory usage and lookup performance. Some research work focuses on filter-based methods that trade off memory cost, lookup throughput, and accuracy, allowing small error rates~\cite{shiftingfilter, bhbf, luo2021mcfsyn, coloringembedder}. However, these approaches do not guarantee 100\% query accuracy. Consequently, there is a need for an efficient structure and algorithm for MS-MQ that ensure full accuracy while maintaining high performance, low memory cost, and support for dynamic updates.

In this paper, we present a holistic design for MS-MQ, called \textbf{S}et \textbf{T}ree for \textbf{E}xact \textbf{M}ulti-set \textbf{M}embership Queries (\sysname). \sysname organizes keys from multiple sets in a binary tree. At each internal node, a binary set separator recursively partitions the sets into two disjoint groups until each leaf corresponds to a single set. To support this design, we develop an efficient Exact Binary Set Separator ($\mathtt{XBSS}$), which serves as the core building block of \sysname.

To reconcile lookup efficiency with update agility, \sysname presents a decoupled architecture with separate control and data planes. The control plane design, $\mathtt{STEM^2_C}$, is responsible for structure construction and incremental updates, whereas the data plane design, $\mathtt{STEM^2_D}$, is optimized for high-speed lookups. This decoupling is also reflected in the separator design. Specifically, $\mathtt{XBSS_C}$, the separator used in $\mathtt{STEM^2_C}$, combines a Counting Bloom Filter~\cite{fan2000summary} with an Othello hashing structure~\cite{othello}. It is then transformed into the data-plane separator $\mathtt{XBSS_D}$ by replacing the Counting Bloom Filter with a standard Bloom Filter~\cite{broder2004network} while retaining the Othello structure. Another key novelty of our design is that we apply less hashing~\cite{kirsch2006less} to both $\mathtt{XBSS}$ and the overall \sysname structure to reduce per-operation computational overhead. In particular, lookups in both $\mathtt{XBSS_D}$ and $\mathtt{STEM^2_D}$ require only two hash computations, thereby improving query performance.

This paper makes the following contributions. 

\begin{itemize}[nosep, leftmargin=*]
    %\item{\wang{We propose a holistic design for exact for multi-set membership queries }}
    \item {We propose \sysname, a holistic system for exact MS-MQ that adopts a decoupled control-plane/data-plane architecture to jointly achieve 100\% query accuracy, high lookup throughput, compact memory usage, and dynamic update support.}

    \item We design $\mathtt{XBSS}$, a novel Exact Binary Set Separator that serves as the core building block of \sysname. $\mathtt{XBSS}$ adopts a decoupled control-plane/data-plane design and incorporates less-hashing technique to support compact, high-throughput exact lookups and efficient updates.

   % \item We design an efficient exact binary set separator called $\mathtt{XBSS}$ to only use two hash computations for key lookups and then design \sysname based on $\mathtt{XBSS}$ for exact multi-set membership queries \textcolor{red}{with high throughput, low memory cost and supporting dynamic update.}

   \item We propose two splitting strategies for \sysname, namely fully greedy splitting (FGS) and balanced-then-greedy splitting (BGS), and comprehensively analyze their trade-offs across different scenarios.
   
    \item We implement \sysname and run comprehensive experiments to show that \sysname provides the highest lookup throughput, with the throughput exceeding 120 million queries per second, is memory-efficient, and can support dynamic key updates. Notably, \sysname guarantees 100\% correctness in query results.
\end{itemize}

The remainder of this paper is organized as follows. Section~\ref{sec:related} reviews related work for multi-set membership queries and binary set separator designs. In Section~\ref{sec:motivation}, we present the background and motivation for our study. Section~\ref{sec:XBSS} introduces our presented binary set separator $\mathtt{XBSS}$. Section~\ref{sec:sysdesign} details the design of our proposed system, \sysname. The evaluation results of \sysname are presented in Section~\ref{sec:eval}, and we conclude the paper in Section~\ref{sec:conclu}.

%\textcolor{red}{define of multi-set query}

%\textcolor{red}{Two real world cases}

%\textcolor{red}{prior arts and their limitation, motivation}

%\textcolor{red}{our work and contribution}

% Packet processing. A router classifies packets it receives before forwarding at the network layer. After parsing the packet, it quickly finds the corresponding forwarding port and forwards the packet out. In this case, the data packet is the element to query, and the correct port needs to be returned from the set of possible ports.

% State machines monitoring [3, 25]. When monitoring the connection states of TCP flows for attack detection, a specified TCP flag may indicate the potential problem. For example, when a large number of TCP flows on the server are in the semi-connected state of SYN-RECEIVED, it may indicate potential SYN attack. Here, the TCP flow is the element and the states represented by flags (such as ACK, RST, SYN, and FIN) form the set.

%the location of user data is located by querying the user IDs; Multi-tenant security groups. Access Control List/IDS: Severs need to operate different rules for different packet. The 5-tuple is key and rule groups are set ID; Protocol-Based Traffic Classification: Difference service is offered for different protocols, such as HTTPS, FTP and VoIP-RTP.

%% file: 2_related.tex
\section{Related Work}
\label{sec:related}

%In this section, we review existing designs for multi-set membership queries. As our proposed data structure builds upon techniques for binary set queries, we also discuss the underlying designs for binary set queries.

\subsection{Multi-set Membership Queries}

%Multiple set query, as a topic closely related to data science and computer science, have been wildly and extensively researched.
Conventional MS-MQ solutions map each key to its associated set ID using hash tables, yet explicit key storage often incurs high memory overhead. To optimize performance, the 3D Hash Table \cite{flachs20223d} employs a hierarchical structure that maintains a single key for multiple values, substantially increasing lookup throughput. Similarly, Cuckoo hashing \cite{pagh2004cuckoo} provides $O(1)$ lookups, with space-efficient variants \cite{fan2014cuckoo, lim2011silt, wang2019vacuum} utilizing compact fingerprints instead of full keys to further minimize memory footprints.
%Conventional approaches to solving the multi-set membership query problem typically rely on hash tables, where each key is associated with its corresponding set ID. However, most existing hash table implementations must store the keys explicitly, resulting in significant memory cost. \textcolor{blue}{To address the performance degradation caused by duplicate build keys in main-memory hash joins, Flachs et al.~\cite{flachs20223d} proposed the 3D Hash Join, which leverages a hierarchical hash table to cluster collisions by distinct values and applies deferred unnesting to achieve significant performance gains.} Cuckoo hashing~\cite{pagh2004cuckoo}, a well-known key-value mapping structure, achieves $O(1)$ lookup time in the worst case and amortized $O(1)$ update time. To reduce memory cost, partial-key Cuckoo hashing and its variants store compact key digests instead of full keys~\cite{fan2014cuckoo, lim2011silt, wang2019vacuum}. 
More recent work proposes key-value lookup structures that avoid storing full keys altogether, such as SetSep~\cite{fan2013cycles, zhou2015scaling}, Bloomier/Othello hashing~\cite{chazelle2004bloomier, charles2008bloomier, othello}, and Ludo hashing~\cite{ludo}. The Bloomier filter was originally designed for static lookup tables and therefore does not support dynamic updates.  While Othello hashing~\cite{othello}, an extension of Bloomier filters, enables runtime updates. Among these methods, Ludo hashing achieves the lowest memory cost for dynamic key-value lookups and provides high lookup throughput~\cite{ludo}. Another representative work, Coloring Embedder~\cite{coloringembedder}, adopts a Bloomier-like design. Its key idea is to embed each key into a high-dimensional space to minimize hashing collisions, followed by dimensionality reduction to encode set IDs efficiently.

Another line of research explores filter-based data structures for MS-MQ~\cite{yu2009buffalo, shiftingfilter, bhbf, dai2016noisy, luo2021mcfsyn, fan2000summary, ShiftingBF,liu2018id, sun2019magic, yang2017difference}. These methods typically extend Bloom filters~\cite{broder2004network} or Cuckoo filters~\cite{fan2014cuckoo}. For example, Buffalo~\cite{yu2009buffalo} builds approximate membership query structures, such as Bloom filters~\cite{broder2004network} or Cuckoo filters~\cite{fan2014cuckoo}, for each set individually, and then checks whether a key belongs to a given set. The Shifting Bloom Filter (ShBF)~\cite{ShiftingBF} stores set IDs using offset bits within the filter. However, ShBF does not support dynamic sets or key deletions and suffers from performance degradation as the number of represented sets increases. The Shifting Filter (SF)~\cite{shiftingfilter} introduces a modified Cuckoo filter design that supports efficient set queries and key deletions with modest memory usage. The $B_h$ Sequence-based Bloom Filter ($\mathtt{B_hBF}$)~\cite{bhbf}, a variant of the Bloom filter, encodes set IDs as a $B_h$ sequence to compactly represent multi-set memberships and reduce auxiliary storage overhead while supporting insertions and membership queries. The Marked Cuckoo Filter (MCF)~\cite{luo2021mcfsyn} extends the Cuckoo filter by attaching identifiers to slots for storing set IDs, enabling MS-MQ with improved space efficiency. While these filter-based approaches are memory-efficient, they cannot guarantee 100\% query accuracy.

\subsection{Binary Set Queries}

The binary set queries problem determines whether a key $x \in {U = P \cup N}$ belongs to the positive set $P$ or the negative set $N$. A straightforward solution is to use a hash table that stores the binary set information as the value for each key. Approximate membership query structures, such as Bloom filters~\cite{broder2004network}, Cuckoo filters~\cite{fan2014cuckoo}, and their variants~\cite{breslow2018morton, liu2020stable, bonomi2006improved, chen2017dynamic, mitzenmacher2020adaptive, xie2017d}, address this problem by encoding keys from one set into the filter. However, this approach introduces false positives and therefore cannot guarantee 100\% query accuracy. More recent designs propose memory-efficient structures that achieve exact query results without false positives~\cite{tinycr, crlite}. The filter cascade~\cite{crlite} constructs a multi-layer Bloom filter system for binary set queries, resolving false positives produced by earlier layers. TinyCR~\cite{tinycr} introduces a structure called DASS, which adopts a two-layer design: the first layer is a Cuckoo filter, the second layer is an Othello \cite{othello}. Both filter cascade and DASS are memory-efficient, especially when two sets differ significantly in size.

%% file: 3_motivation.tex
\section{Preliminaries and Motivation}
\label{sec:motivation}
%In this section, we introduce the motivation of the design of \sysname in detail.
This section first introduces several fundamental data structures that serve as building blocks for both prior state-of-the-art solutions and our proposed design for the MS-MQ problem. We then present a tree-based framework for solving the MS-MQ problem. %These include Bloom filters (BFs) ~\cite{broder2004network} and counting Bloom filters (CBFs)~\cite{fan2000summary} for approximate membership queries, Othello~\cite{othello} for key-value lookup, filter cascade \cite{crlite} and DASS \cite{tinycr} for exact set membership queries.}

%Then, we elaborate tree-based framework and potential related solutions for MS-MQ problem. Finally, we conduct spme preliminary evaluation, analysis and reveal drawbacks of the potential tree-based solutions and existing the state-of-the-art solutions for MS-MQ problem to introduce our design.

%\textcolor{red}{In this section, we first briefly introduce some fundamental data structures, including approximate membership query data structures such as Bloom filter~\cite{broder2004network}, counting bloom filter~\cite{fan2000summary}, key-value store and lookup structure such as Othello~\cite{othello}, and binary set separator such as filter cascade and DASS, all of them could be used as basic building blocks for existing start-of-the-art solutions and our design for MS-MQ problem. Then, we elaborate tree-based framework and potential related solutions for MS-MQ problem. Finally, we conduct spme preliminary evaluation, analysis and reveal drawbacks of the potential tree-based solutions and existing the state-of-the-art solutions for MS-MQ problem to introduce our design.}

\subsection{Introduction of Basic Data Structures}
\label{sec:intro_base}

\subsubsection{Bloom filters and counting Bloom filters.}
\label{sec:bf}
Bloom filters \cite{broder2004network} are the most well-known data structures for approximate membership queries. A Bloom filter represents a set of $n$ keys by an array of $m$ bits. Each key $x$ is mapped to $k$ positions in the array by $k$ independent hash functions $h_1, h_2, \ldots, h_l$, and the bits at positions $h_i(x)$ are set to 1 for all $0 \leq i \leq k-1$. To test whether a key $x$ belongs to the set, the Bloom filter checks the value in the $h_i(x)$-th bit. If all $k$ bits are 1, the Bloom filter returns true. Otherwise, it returns false. A Bloom filter yields false positives. The false positive rate is $\epsilon \approx \left(1-e^{-kn/m}\right)^k$. For a target false positive rate $\epsilon$, the optimal number of hash functions $k = \frac{m}{n}\ln{2}$, which requires an optimal bit array size of $m \approx 1.4427\, n \log_2 \frac{1}{\epsilon}$. One limitation of Bloom filter is that it cannot support key deletions. Counting Bloom filters~\cite{fan2000summary} address this limitation and support deletions by replacing each bit in the array with a counter that records how many times the corresponding position has been set.

\begin{figure}[t!]
\begin{center}
\includegraphics[width=0.85\linewidth]{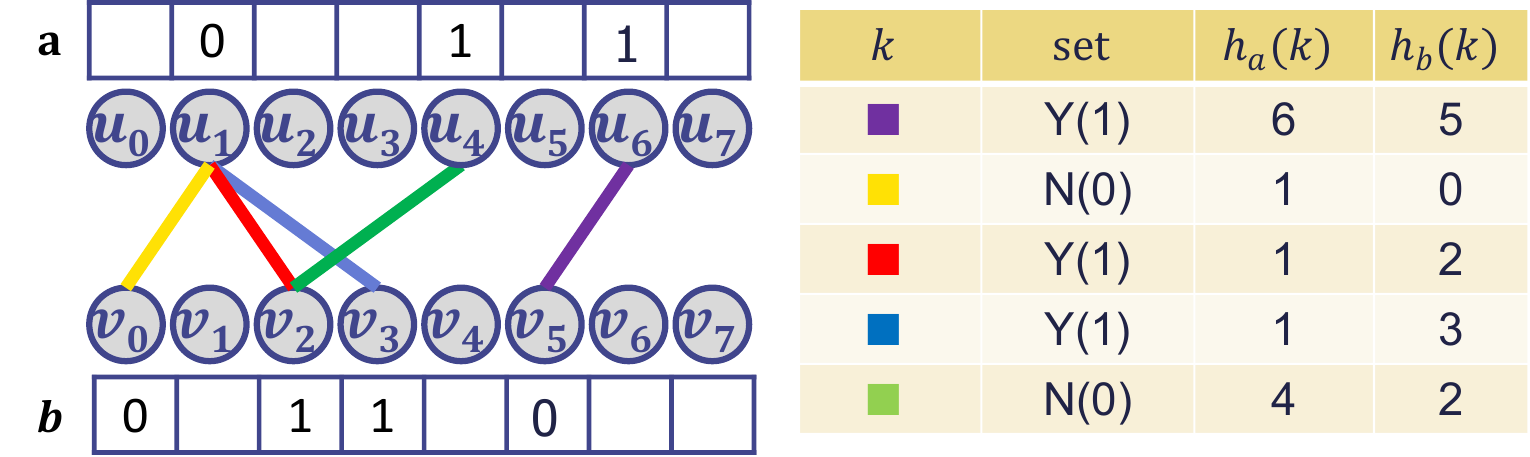}
\vspace{-2ex}
\caption{An Example of 1-bit Othello Hashing.}
\label{fig:1bitOthello}
\end{center}
\end{figure}

\subsubsection{Othello.}
\label{sec: Othello}
Othello hashing~\cite{othello} is a key-value lookup structure which can represent a set of key-value pairs. In 1-bit Othello hashing, each value is either 0 or 1, therefore 1-bit Othello is an efficient candidate for membership query problem. Given a set $S$ with $n$ keys, a 1-bit Othello hashing structure is defined as a seven-tuple $\mathtt{<n_a, n_b, a, b, h_a, h_b, G>}$, where:

\begin{itemize}[nosep, leftmargin=*]
\item $n_a$ and $n_b$ are the sizes of the Othello arrays, with $n_a \approx 1.33n$ and $n_b \approx n$.

\item $a$ and $b$ are arrays of $n_a$ and $n_b$ bits, respectively.

\item $h_a$ and $h_b$ are uniform random hash functions mapping keys to integers in $\{0, 1, \ldots, n_a-1\}$ and $\{0, 1, \ldots, n_b-1\}$, respectively.

\item $G$ is an acyclic bipartite graph used to determine the bit values in $a$ and $b$ during Othello construction.
\end{itemize}

During construction, Othello hashing must form an acyclic bipartite graph when inserting all keys. Once the graph is acyclic, the bit values in $a$ and $b$ are determined accordingly. 
%If two or more keys generate edges that create a cycle (i.e., share the same vertices), the graph becomes cyclic, and a new pair of hash functions must be selected until an acyclic graph is obtained. 
If two or more keys generate edges that create a cycle, the graph becomes cyclic. In this case, the structure must be reconstructed by selecting a new pair of hash functions until an acyclic graph is obtained. Figure~\ref{fig:1bitOthello} illustrates an example of a 1-bit Othello hashing. The Othello structure is built using five keys, where each key is associated with a set ID of either 0 or 1. To query the value associated with a key $x$, the lookup result is computed as $\tau(x)=a[h_a(x)] \oplus b[h_b(x)]$.

\begin{figure*}[t!]
    \centering
    \subfigure[Lookup thro. vs. \# of keys]{
        \includegraphics[width=0.23\textwidth]{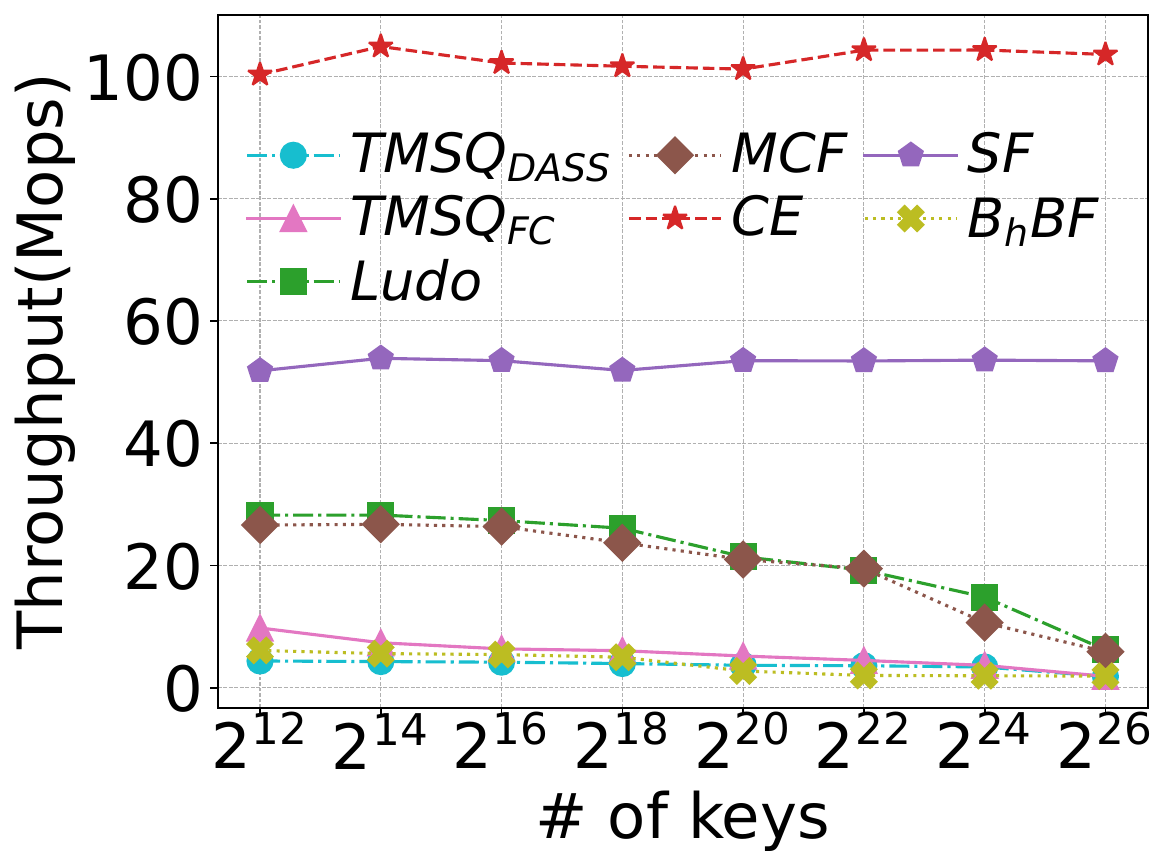}
        \label{fig:lkp4keys}
    }
    %\hfill
    \subfigure[Memory cost vs. \# of keys]{
        \includegraphics[width=0.23\textwidth]{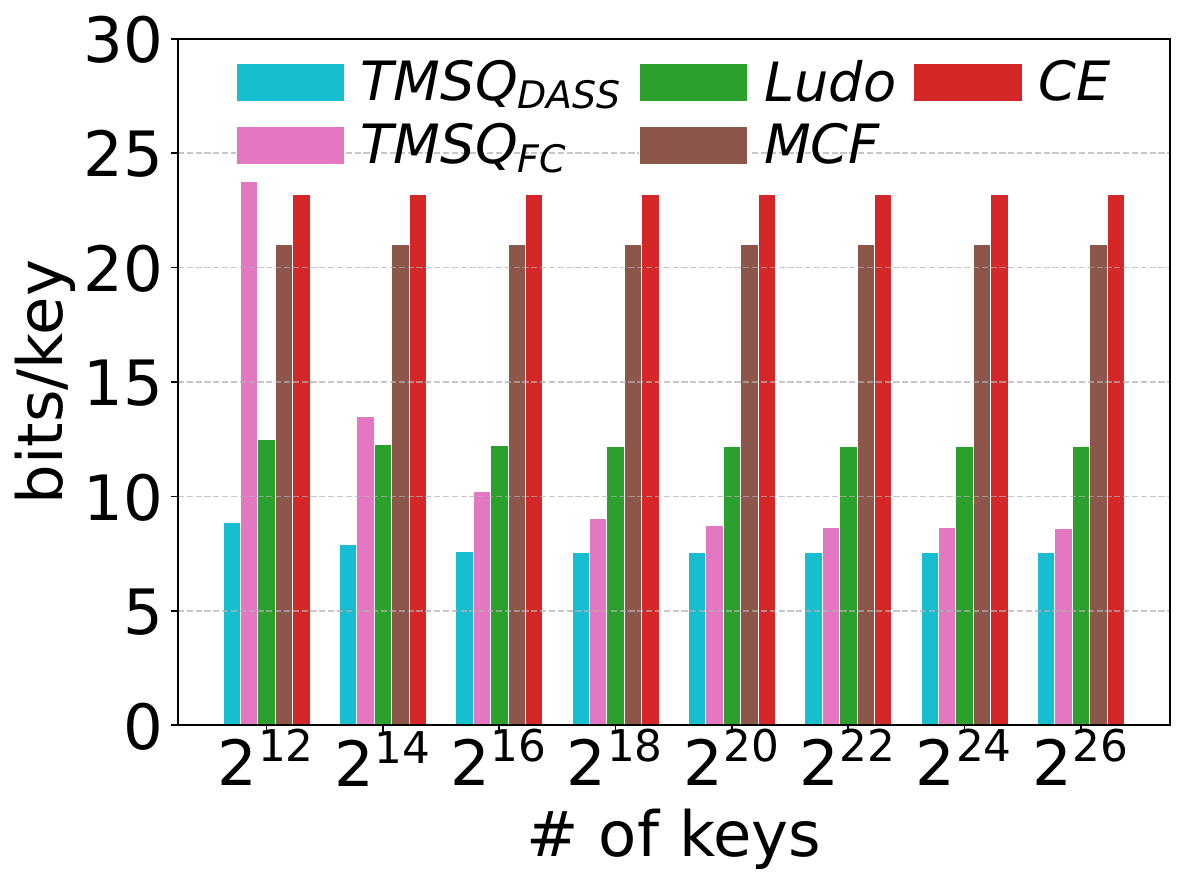}
        \label{fig:lkm4keys}
    }
    \subfigure[Lookup thro. vs. \# of sets]{
        \includegraphics[width=0.23\textwidth]{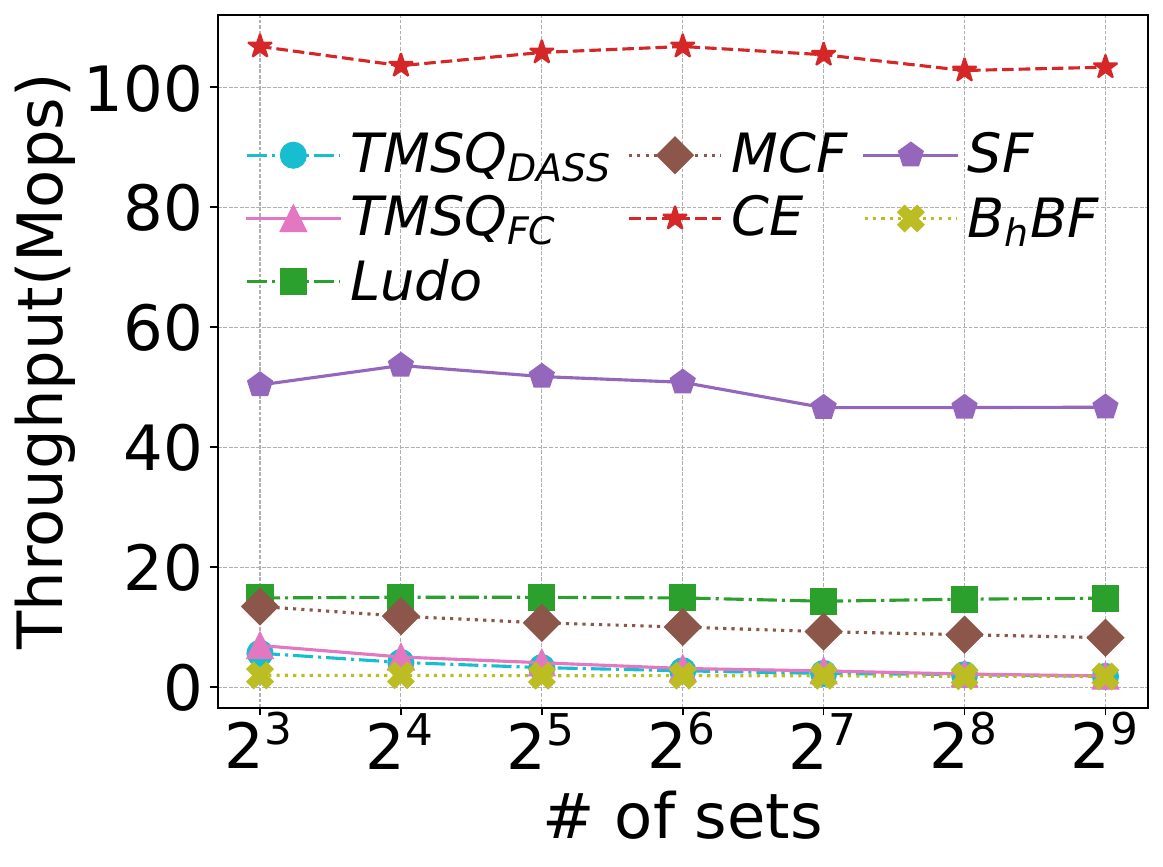}
        \label{fig:lkp4sets}
    }
    \subfigure[Memory cost vs. \# of sets]{
        \includegraphics[width=0.23\textwidth]{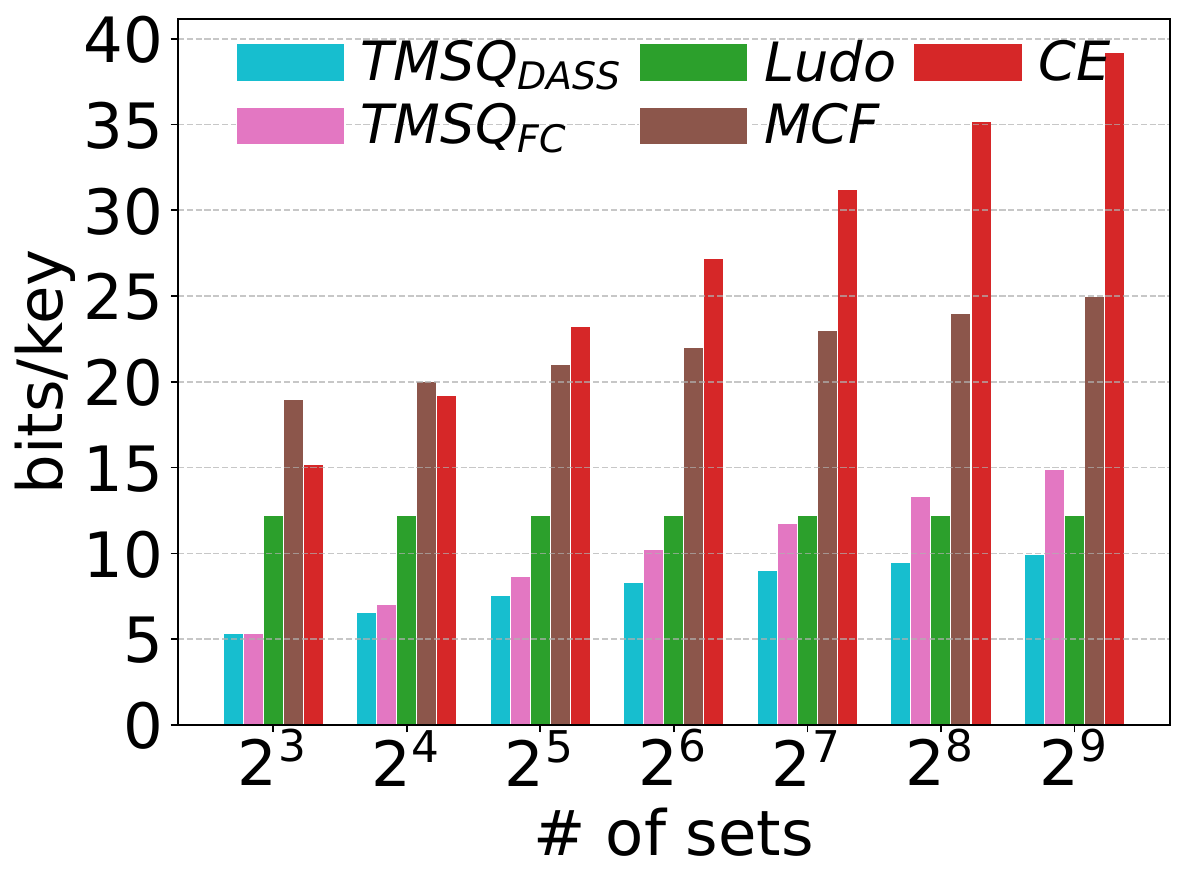}
        \label{fig:lkm4sets}
    }
    \vspace{-3.5ex}
    \caption{Comparison of Lookup Throughput and Memory Cost with Varied Number of Keys and Sets.}
    \label{fig:MotivationEvaluation}
\end{figure*}

\subsubsection{Binary Set Separators for exact set membership queries}
Approximate membership query data structures, built on a finite set $P$, can answer whether a key belongs to the set. However, they may yield false positives. Given the finite set $P$ and the corresponding negative set $N$, a binary set separator determines whether a key $x \in U = \{P \cup N\}$ belongs to $P$ or $N$. Representative examples include Filter Cascade~\cite{crlite} and DASS~\cite{tinycr}, which provide 100\% query accuracy and are more memory-efficient than general-purpose hash tables for key-value storage and lookup.

%Given a finite universe of keys $U$ that can be partitioned into two disjoint subsets $P$ and $N$, where $U = \{P \cup N\}$, a binary set separator determines whether a key $k \in U$ belongs to $P$ or $N$. The key distinction between separators and conventional membership query structures lies in the definition of the negative set. In membership query structures, only a finite positive set $P$ is specified, while the negative set is typically considered to be the infinite complement of $P$. In contrast, separators assume that both $P$ and $N$ are finite and explicitly defined. Representative examples of such separators include filter cascade~\cite{crlite} and $\mathtt{DASS}$~\cite{tinycr}. The primary advantage of these approaches is that they guarantee 100\% query accuracy, while remaining more memory-efficient than general-purpose hash tables for key–value storage and lookup.

\noindent \textbf{Filter cascade.} The filter cascade introduced in CRLite~\cite{crlite} is a multi-layer Bloom filter structure designed for binary set membership queries.
%as illustrated in Figure~\ref{fig:fc}. 
The first-layer filter, $\mathtt{BF_1}$, encodes set $P$. Due to the false positives inherent to Bloom filters, some keys from set $N$ may be incorrectly classified as members of $P$. These false positives are collected to construct the second-layer filter, $\mathtt{BF_2}$. Similarly, $\mathtt{BF_2}$ may still yield false positives, which are then used to build the third layer, and so on. This cascading process continues until the false-positive set becomes empty, ensuring 100\% query accuracy.

In this data structure, the odd-numbered Bloom filters represent whitelists (encoding keys from $P$), while the even-numbered filters represent blacklists (encoding keys from $N$). To determine whether a key $x$ belongs to $P$ or $N$, the query starts at $\mathtt{BF_1}$ and proceeds layer by layer until the first filter $\mathtt{BF_i}$ is found such that $x \notin \mathtt{BF_i}$. Because Bloom filters never yield false negatives, if $i$ is odd, then $x$ must belong to $N$; if $i$ is even, then  $x$ must belong to $P$. If no such $\mathtt{BF_i}$ is found (i.e.,  $x$ is contained in all filters), the result depends on the total number of layers $l$: if $l$ is odd, $x \in P$; otherwise, $x \in N$.

\begin{figure}[t!]
\begin{center}
\includegraphics[width=0.85\linewidth]{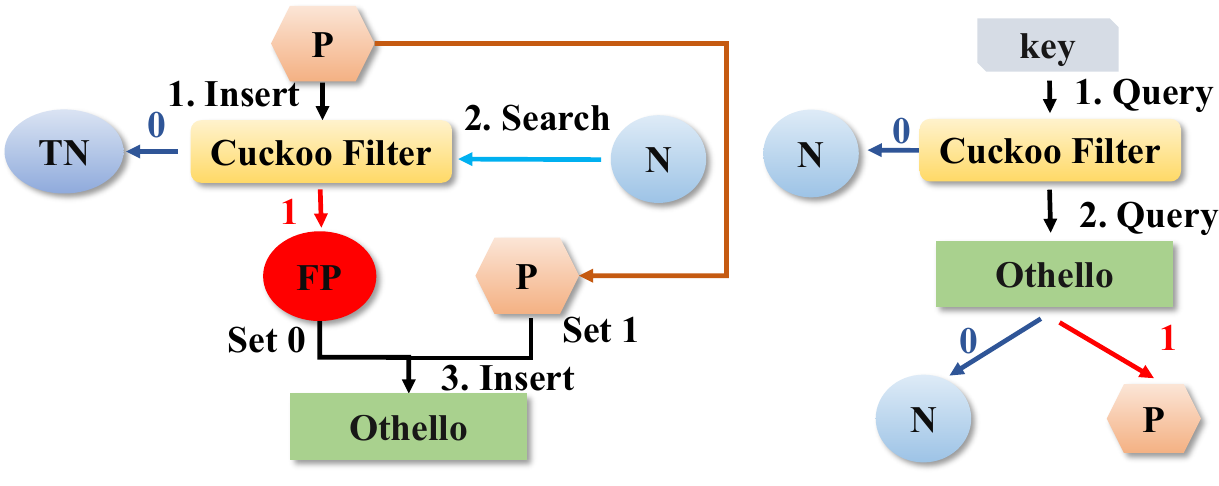}
\vspace{-2ex}
\caption{Insertion and Query Process of DASS.}
\label{fig:dass}
\end{center}
\end{figure}

\noindent \textbf{DASS.} TinyCR~\cite{tinycr} introduces a compact data structure called DASS for binary set membership queries. DASS adopts a two-layer architecture, as illustrated in Figure~\ref{fig:dass}. The first layer is a filter implemented using a Cuckoo filter~\cite{fan2014cuckoo}, while the second layer is an Othello structure~\cite{othello}. Specifically, all keys from set $P$ are first inserted into the Cuckoo filter. Then, keys from set $N$ are tested against the filter. Most keys in $N$ will be correctly identified as negatives, forming the true negative set (TN). However, a small subset of $N$ may be incorrectly identified as positives due to the inherent false positive rate of the filter, forming the false positive set (FP). An Othello is then constructed to distinguish between keys from $P$ (labeled as 1) and keys from FP (labeled as 0). During a query for a key $x$, if the Cuckoo filter returns a negative result, then $x \in N$, and the query terminates. If the Cuckoo filter returns a positive result, Othello is used to complete the lookup. If Othello returns 1, $x$ is a true positive and thus belongs to $P$; otherwise, $x$ is a false positive and belongs to $N$.

Both filter cascade and DASS achieve low memory cost when $|N| \gg |P|$. Compared with filter cascade, DASS additionally supports efficient key updates without requiring reconstruction of the entire data structure.

\subsection{Tree-based Framework for Multi-Set Query}
\label{sec:tb}
Given a binary set separator which can separate keys from two disjoint sets, it is natural to extend it to support MS-MQ by organizing such separators into a hierarchical tree structure, an approach also discussed in TinyCR \cite{tinycr}. %Figure ~\ref{fig:msq} shows \wang{an example of} this tree-based design.
Given $n$ sets $\mathtt\{S_1, S_2, \ldots, S_n\}$, we construct a binary decision tree in which each node employs a binary set separator to partition the keys into two disjoint subsets. The tree growth continues until each leaf node becomes pure, meaning that all keys associated with that leaf belong to a single set $S_f$. To determine the set ID for a given key $x$, a binary set query is performed at each node along the path from the root to a leaf. At each step, the query result decides the traversal direction, left or right, until a leaf node is reached.

%Given a binary set separator which can separate keys from two disjoint sets, it is natural to extend it to support multi-set membership query by incorporating the separator into a tree-based structure, as introduced in TinyCR \cite{tinycr}. Figure ~\ref{fig:msq} shows this tree-based design. Specifically, given multiple sets $\mathtt\{S_1, S_2, \ldots, S_8\}$, we construct a binary decision tree in which each node employs a binary set separator to partition the keys into two disjoint subsets. The tree growth continues until each leaf node becomes pure, meaning that all keys associated with that leaf belong to a single set $S_f$. To determine the set ID for a given key $k$, a binary set query is performed at each node along the path from the root to a leaf. At each step, the query result decides the traversal direction, left or right, until a leaf node is reached.

Despite this intuitive design, existing tree-based solutions for MS-MQ suffer from suboptimal lookup throughput and memory overhead~\cite{li2021building}. Although incorporating DASS into the tree-based framework for MS-MQ on skewed datasets reduces memory cost~\cite{tinycr}, it still suffers from low lookup throughput, as shown in Figure~\ref{fig:MotivationEvaluation}.

\noindent\textbf{Preliminary evaluation}. We incorporate DASS and filter cascade into the tree-based framework for MS-MQ, denoted as $\mathtt{TMSQ_{DASS}}$ and $\mathtt{TMSQ_{FC}}$, respectively, and evaluate their lookup throughput and memory cost. To exploit the property that both DASS and filter cascade achieve lower memory cost when $|N| \gg |P|$, we adopt a greedy strategy to determine the splitting point at each tree node. Specifically, to construct a binary set separator at a node for a given collection of sets $\{S_1, S_2, \ldots, S_n\}$, we first sort by their sizes and select the median set as the splitting point. This approach maximizes the size ratio between the left and right child nodes, resulting in a balanced tree with minimal depth. We generate synthetic datasets with Zipfian-distributed set sizes to evaluate their performance against existing state-of-the-art methods, including Ludo Hashing~\cite{ludo}, Shifting Filter (SF)~\cite{shiftingfilter}, $B_h$ Sequence-based Bloom Filter ($\mathtt{B_hBF}$)~\cite{bhbf}, Marked Cuckoo Filter (MCF)~\cite{luo2021mcfsyn}, and Coloring Embedder (CE)~\cite{coloringembedder} (see Figure~\ref{fig:MotivationEvaluation}). Specifically, we fix the number of sets to $2^5$ in Figure~\ref{fig:lkp4keys} and Figure~\ref{fig:lkm4keys}, and fix the total number of keys to $2^{24}$ in Figure~\ref{fig:lkp4sets} and Figure~\ref{fig:lkm4sets}. The results show that both $\mathtt{TMSQ_{FC}}$ and $\mathtt{TMSQ_{DASS}}$ achieve high memory efficiency in most scenarios, with $\mathtt{TMSQ_{DASS}}$ even outperforming other baselines. However, both approaches exhibit lower lookup throughput compared to other methods.

%\textcolor{red}{However, this tree-based solution is still facing challenges when deploy it to industrial application, in both throughput and memory throughput. Firstly, tree structure is considered as an memory-intensive approach by previous works \cite{li2021building}, although the data structure at each node is memory compact algorithm such as bloom filter. This kind of misconception typically stems from the fact that a key may be inserted into multiple binary separators to support query. Besides, all binary set separators at tree nodes must be stored in the memory. The second challenge is, query a random key from $n$ sets needs to traverses at least $\lceil \log_2(n) \rceil$ binary separator, which causes high latency. Our work challenges such perception by proposing an efficient tree-based approach for MS-MQ problems, which is one of the contributions of this paper.}

\subsection{Analysis of Limitations and Design Insights}
\label{sec:insights}
The low lookup throughput in the tree-based framework is mainly due to the traversal of multiple binary set separators, each of which requires several hash computations and memory accesses for a single key lookup. Moreover, $\mathtt{TMSQ_{FC}}$ does not support key deletions because of the inherent limitations of Bloom filters, and key insertions can incur significant overhead, often necessitating reconstruction of the filter cascade. Despite these limitations, the tree-based structure remains a promising approach for achieving 100\% query accuracy with low memory usage. The key challenge, therefore, is to design a tree-based structure that delivers high lookup throughput while supporting efficient updates, including insertions, deletions, and key migrations.

\noindent \textbf{New design insights.} The design of MS-MQ data structures requires high lookup throughput, 100\% query accuracy, low memory cost, and support for dynamic updates. The experimental results in Figure~\ref{fig:MotivationEvaluation} show that the tree-based structures $\mathtt{TMSQ_{FC}}$ and $\mathtt{TMSQ_{DASS}}$ achieve low memory cost, but both suffer from low lookup throughput. To understand this limitation, we analyze the design of Bloom filters and Cuckoo filters. Bloom filters have simpler memory access patterns than Cuckoo filters and their variants, but their lookup efficiency is often limited by the computational overhead of multiple hash functions. This observation suggests that, if Bloom filters can be effectively adapted to construct binary set separators while minimizing their hashing overhead within a tree-based framework, the resulting structure could support highly efficient MS-MQ.

Furthermore, practical MS-MQ deployments often operate over a heterogeneous memory hierarchy. While the complete and dynamically evolving collection of sets resides in large-capacity memory, the MS-MQ structure itself must fit within a much smaller fast-memory footprint. This separation may arise across devices—for example, when a centralized server maintains the full set collection while a compact query structure is offloaded to resource-constrained devices such as routers or IoT nodes. It can also arise within a single device, where structure construction and updates are handled in larger but slower memory, while lookup queries are served from smaller and faster memory. This resource asymmetry is particularly well aligned with MS-MQ workloads, where query processing must be fast and memory-efficient, but construction and updates can be handled separately. It therefore motivates a decoupled design with two distinct planes: a control plane for constructing the data structure and processing incremental updates, and a data plane optimized exclusively for high-speed queries.

Guided by these dual-plane insights, the following sections first detail the \textbf{Exact Binary Set Separator ($\mathtt{XBSS}$)} as a dual-plane building block, followed by the comprehensive design for MS-MQ.

%% file: 4_XBSS.tex
\section{Design of the Lightweight XBSS}
\label{sec:XBSS}
In this section, we present $\mathtt{XBSS}$, a lightweight binary set separator designed for exact set membership queries, which serves as the core building block of our MS-MQ design.

%\subsection{Design of $\mathtt{XBSS}$}
Adhering to the architectural principles in Section~\ref{sec:insights}, $\mathtt{XBSS}$ adopts a decoupled design to reconcile query efficiency with update agility, as shown in Figure~\ref{fig:XBSS}. Structure construction and update maintenance are handled in the control plane ($\mathtt{XBSS_C}$), while the data plane ($\mathtt{XBSS_D}$), is optimized for high-throughput lookups and memory compactness. This separation isolates maintenance overhead from the performance-critical query path and enables seamless transformation into a query-optimized representation.

%\wang{Following the architectural principles outlined in Section~\ref{sec:insights}, we design $\mathtt{XBSS}$ with a decoupled functional partition. The data plane data structure $\mathtt{XBSS_D}$ is optimized for high-performance lookups and compact memory usage, whereas the control plane structure $\mathtt{XBSS_C}$ is for construction and dynamic updates. Moreover, the design should support efficient transformation of the control-plane representation into its query-optimized data-plane form.} \wang{Figure~\ref{fig:XBSS} presents the design of $\mathtt{XBSS}$.}

\subsection{Design of $\mathtt{XBSS_C}$}
\label{sec:xbss_c}

\begin{figure}[t!]
\begin{center}
\includegraphics[width=0.9\linewidth]{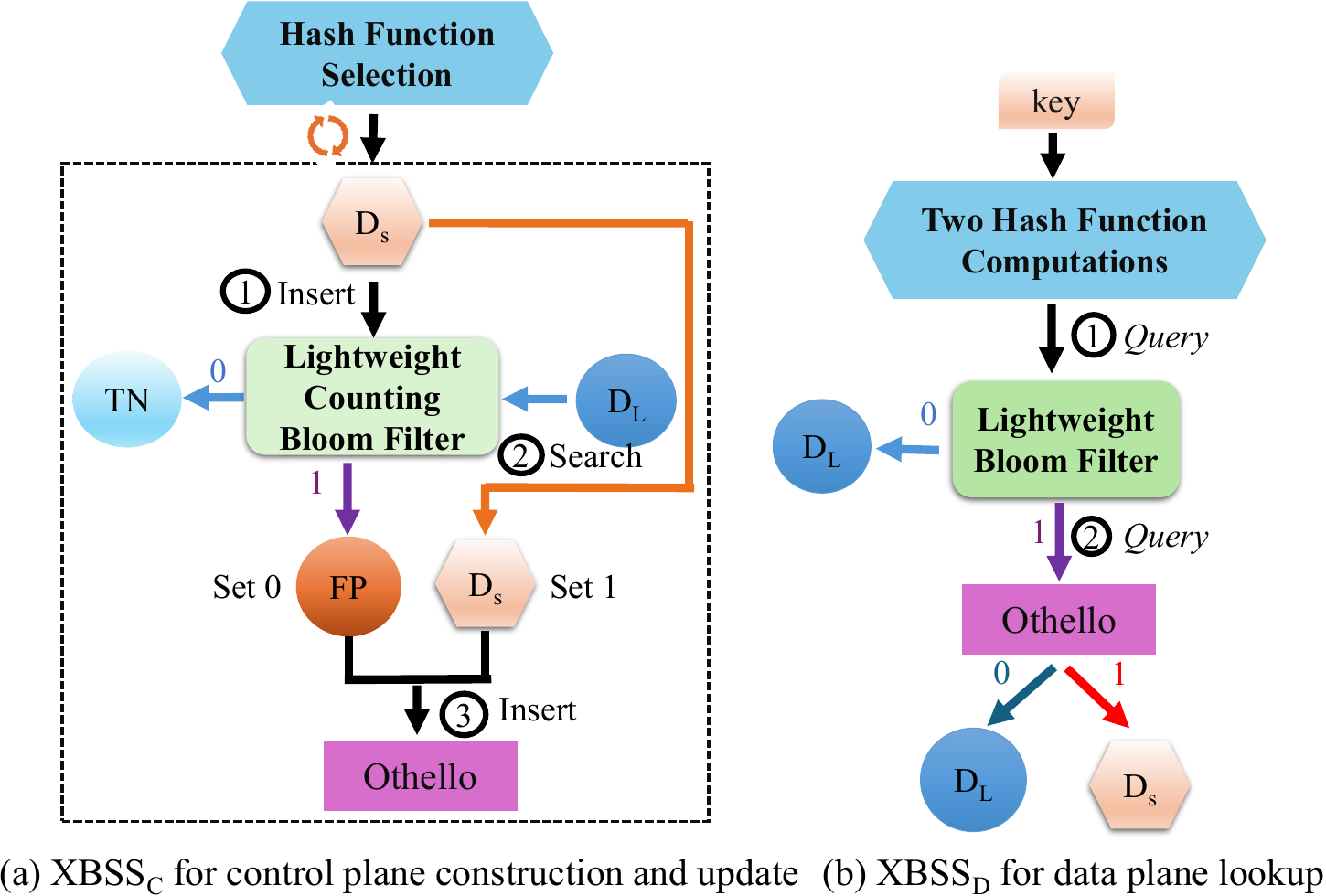}
\vspace{-2ex}
\caption{Design of $\mathtt{XBSS}$ on Control Plane and Data Plane.}
\label{fig:XBSS}
\end{center}
\end{figure}

$\mathtt{XBSS_C}$ adopts a two-layer structure inspired by DASS \cite{tinycr}, as shown in Figure~\ref{fig:XBSS}(a). The first layer is a counting Bloom filter (CBF)~\cite{fan2000summary} which supports dynamic updates, including key insertion and deletion. The second layer is an Othello. However, directly incorporating the CBF into the design leads to substantial computational overhead, as it requires multiple hash computations for each operation. Inspired by the findings that two independent hash functions $h_1(x)$ and $h_2(x)$ can simulate multiple hash functions by Equation~\ref{equ:hashing}:
\begin{equation}
        g_i(x) = h_1(x) + ih_2(x) \qquad(i=0,1,2,...) 
        \label{equ:hashing}   
\end{equation}

\noindent without any loss in the asymptotic false positive probability \cite{kirsch2006less}, this \textbf{less hashing technique} provides a promising solution to reduce the computational overhead. Accordingly, we adopt this technique in $\mathtt{XBSS_C}$, reducing the total number of hash computations required for key operations (e.g., insertion, flipping, and deletion) to two, thereby enabling a lightweight counting Bloom filter. Thus, $\mathtt{XBSS_C}$ selects two base hash functions to derive the hash functions required by the CBF and Othello. Since Othello may trigger a reconstruction process when the selected hash functions fail to produce an acyclic graph, the two base hash functions may need to be reselected. $\mathtt{XBSS_C}$ supports the following operations:

%For $\mathtt{XBSS}$, we can use two hash functions to simulate multiple hash functions needed for Bloom filters and Othello. \sout{Thus, key lookup on $\mathtt{XBSS}$ only requires two hash computations.} It is worth noting that while the hash functions used within a single data structure, such as a Bloom filter, should remain independent, different data structures, such as a bloom filter and an Othello, can share the same hash functions. Since \sysname consists of multiple $\mathtt{XBSS}$, we can use just two hash functions to simulate all hash functions used in \sysname. %The shared hash functions rule can also be applied here, since different $\mathtt{XBSS}$ can share same hash functions. With the integration of the less hashing technique, $\mathtt{XBSS}$ and \sysname are guaranteed to require only two hash computations for each key lookup.

\noindent \textbf{Construction ($\mathtt{XBSS_C.build}$).} Given two disjoint groups of keys, $\mathtt{D_L}$ and $\mathtt{D_S}$, the control plane constructs $\mathtt{XBSS_C}$ to distinguish between them. A $\mathtt{CBF}$ is first built using the keys from $\mathtt{D_S}$. Then, the keys from $\mathtt{D_L}$ are queried against this filter, where most are expected to yield negative results. Keys that return positive results constitute the false positive set $\mathtt{FP}$ (set 0). An Othello structure is then built to distinguish between $\mathtt{FP}$ (set 0) and the true positive set $\mathtt{D_S}$ (set 1).

\noindent \textbf{New key insertion ($\mathtt{XBSS_C.insert(x)}$).} For a new key $x$ inserted into set $\mathtt{D_S}$, the key should be added to both the $\mathtt{CBF}$ and the Othello $O$. If $\mathtt{CBF.query}(x) == 1$, then $x$ only needs to be inserted into $O$ with $O.\mathtt{query}(x) == 1$. Otherwise, $x$ is first inserted into the $\mathtt{CBF}$.  After the insertion, keys from $\mathtt{D_L}$ are rechecked against the updated $\mathtt{CBF}$ to identify any new false positives. These newly identified keys $x'$ are inserted into $O$ with $O.\mathtt{query}(x') == 0$, while the newly inserted key $x$ from $\mathtt{D_S}$ is inserted into $O$ with $O.\mathtt{query}(x) == 1$.  For a new key $x$ inserted into set $\mathtt{D_L}$, if $\mathtt{CBF.query}(x) == 0$, no further action is required. Otherwise, $x$ should be inserted into $O$ with $O.\mathtt{query}(x) == 0$.

\noindent \textbf{Key flipping ($\mathtt{XBSS_C.flip(x)}$).} When a key $x$ is moved from $\mathtt{D_S}$ to $\mathtt{D_L}$, $\mathtt{CBF}$ first removes $x$ since $x$ has already been inserted into $\mathtt{CBF}$ and $O$, and then checks whether $x$ would be recognized as a false positive key after removal. If $x$ is a false positive, then Othello $O$ flips the value of $x$ by making $O.\mathtt{query}(x)==0$. Otherwise, $x$ is deleted from $O$. When a key $x$ is moved from $\mathtt{D_L}$ to $\mathtt{D_S}$, $\mathtt{CBF}$ first checks whether $x$ is tested positive. If so, Othello $O$ flips the value of $x$ by making $O.\mathtt{query}(x)==1$. Otherwise, $x$ is inserted into $\mathtt{CBF}$ and $\mathtt{D_L}$ are tested against $\mathtt{CBF}$ to get new false positives. Then $x$ is inserted into $O$ with $O.\mathtt{query}(x)==1$ and new false positives $x'$ is inserted into $O$ with $O.\mathtt{query}(x')==0$.

\noindent \textbf{Key deletion ($\mathtt{XBSS_C.delete(x)}$).} When a key $x$ is removed from $\mathtt{D_S}$, then both $\mathtt{CBF}$ and $O$ delete $x$. When a key $x$ is removed from $\mathtt{D_L}$, $\mathtt{CBF}$ first checks whether $x$ is a false positive. If so, $O$ deletes $x$. Otherwise, no further operation is required.

It is worth noting that insertions or deletions in $\mathtt{CBF}$ may alter the set of false positives when testing keys from $\mathtt{D_L}$. Therefore, these keys need to be re-evaluated. To reduce the overhead of re-checking all keys in $\mathtt{D_L}$, we maintain an index table associated with the $\mathtt{CBF}$. The index table records, for each slot in the $\mathtt{CBF}$, the indices of the $\mathtt{D_L}$ keys that hash to that slot. When the value of a slot changes due to an insertion, flipping, or deletion, only the keys linked to that slot need to be re-checked, and the false positive set can then be updated accordingly.

%\textcolor{red}{\noindent \textbf{Install($\mathtt{XBSS_C.install}$}). The control plane call this operation to generate data plane structure, $\mathtt{XBSS_D}$, by converting the $\mathtt{CBF}$ into a standard Bloom filter ($\mathtt{BF}$), setting all nonzero counters to 1 and duplicating the Othello structure. The resulting $\mathtt{XBSS_D}$ is then deployed to the data plane of \sysname ($\mathtt{STEM^2_D}$) for key lookup.}

\subsection{Design of $\mathtt{XBSS_D}$}
\label{sec:xbss_d}

After $\mathtt{XBSS_C}$ has been successfully constructed using the selected two base hash functions, $\mathtt{XBSS_D}$ is obtained by converting the $\mathtt{CBF}$ into a standard Bloom filter ($\mathtt{BF}$), setting all nonzero counters to 1, and copying the Othello structure, as shown in Figure~\ref{fig:XBSS}(b). The resulting $\mathtt{XBSS_D}$ can then be deployed in the data plane to support efficient exact binary set membership queries. The main supported operation of $\mathtt{XBSS_D}$ is key lookup.

\noindent \textbf{Key lookup ($\mathtt{XBSS_D.lookup(x)}$).} For a given key $x$, only two hash computations are required for the query. if $\mathtt{BF.query}(x)==0$, then $x \in D_L$. Otherwise, the key is checked in the Othello $O$. If $O.\mathtt{query}(x)==0$, then $x \in D_L$. Otherwise, $x \in D_S$.

\subsection{Memory Analysis of $\mathtt{XBSS}$}
\label{sec:mem}

This section analyzes the memory cost of $\mathtt{XBSS}$, focusing on the trade-offs under different false positive rate configurations of the filter. We examine $\mathtt{XBSS_D}$, which is deployed in the data plane and may operate on resource-constrained devices. $\mathtt{XBSS_D}$ shares the same structural properties as $\mathtt{XBSS_C}$, except that the $\mathtt{CBF}$ in $\mathtt{XBSS_C}$ requires approximately $c$ times more memory than the $\mathtt{BF}$ used in $\mathtt{XBSS_D}$, where $c$ denotes the number of bits per counter in the $\mathtt{CBF}$.

$\mathtt{XBSS_D}$ consists of a Bloom filter $\mathtt{BF}$ and an Othello $O$. A key design trade-off exists between the sizes of $\mathtt{BF}$ and $O$, determined by the false positive rate of $\mathtt{BF}$. Specifically, reducing the false positive rate of $\mathtt{BF}$ requires more memory but results in fewer false positives, thereby reducing the size of $O$ needed to distinguish between false positives and true positives. Suppose $\mathtt{XBSS_D}$ is built using keys from two groups $\mathtt{D_S}$ and $\mathtt{D_L}$ with size $\mathtt{n_s}$ and $\mathtt{n_l}$ respectively ($\mathtt{n_l} \geq \mathtt{n_s}$), and keys from $\mathtt{D_S}$ are inserted into the $\mathtt{BF}$. If the false positive rate of $\mathtt{BF}$ is set to $\epsilon$, then $\mathtt{BF}$ requires at least $1.4427\mathtt{n_s} \cdot \mathtt{log_2(\frac{1}{\epsilon})}$ bits \cite{broder2004network}. The expected number of false positives is $\epsilon \cdot \mathtt{n_l}$. The Othello $O$ then requires $2.33 \cdot (\epsilon \cdot \mathtt{n_l} + \mathtt{n_s})$ bits \cite{othello}. Thus the total memory cost of $\mathtt{XBSS}$ is $1.4427 \cdot \mathtt{n_s} \cdot log_2(\frac {1}{\epsilon}) + 2.33 \cdot \mathtt{n_l} \cdot \epsilon + 2.33 \cdot \mathtt{n_s}$.

Let $r=\mathtt{n_l} / \mathtt{n_s}$, then the total memory cost is $M = \mathtt{n_s} \cdot (1.4427 \cdot log_2(\frac {1}{\epsilon}) + 2.33 \cdot r \cdot \epsilon + 2.33)$. Since $\mathtt{n_s}$ and $r$ are constant for the given sets, we can minimize $M$ with respect to $\epsilon$. The minimum memory cost is achieved when $\epsilon = \frac{0.8925}{r}$, yielding $M_{min} = (2.08\ln r+4.65)n_s$. Accordingly, the amortized memory cost per key is $M_a=(2.08\ln r +4.65)/(r+1)$. By taking the derivative of $M_a$ with respect to $r$ and setting $\frac{\partial M_a}{\partial r} = 0$, we find that $M_a$ reaches its maximum at $r = 0.892$. Since $r \geq 1$ in practical settings, increasing $r$ monotonically reduces the amortized memory cost per key. 

%Compared to Othello, which requires memory $M_o = 2.33(n_l + n_s) = (2.33r + 2.33)n_s$, $\mathtt{XBSS}$ significantly reduces the total memory cost to $\Theta(n_s \log r)$.

%\textcolor{red}{\noindent \textbf{Parameter setting of \sysname.} All parameters in \sysname are automatically derived from the sizes of the key sets, with the objective of minimizing data-plane memory overhead. As a result, users are not required to manually compute or tune any parameters. This design choice is motivated by the fact that parameter configuration directly determines critical structural properties, including the number of hash functions in Bloom filters, the bitmap size, and the bitmap length in Othello. Since \sysname employs less hashing techniques, adjusting parameters to vary the number of hash functions does not yield meaningful performance benefits. Therefore, configuring parameters to minimize memory consumption represents the most principled and effective choice.}

%% file: 4_design_revision.tex
\section{Multi-set Membership Query Design}
\label{sec:sysdesign}
\begin{figure}[t!]
\begin{center}
\includegraphics[width=0.9\linewidth]{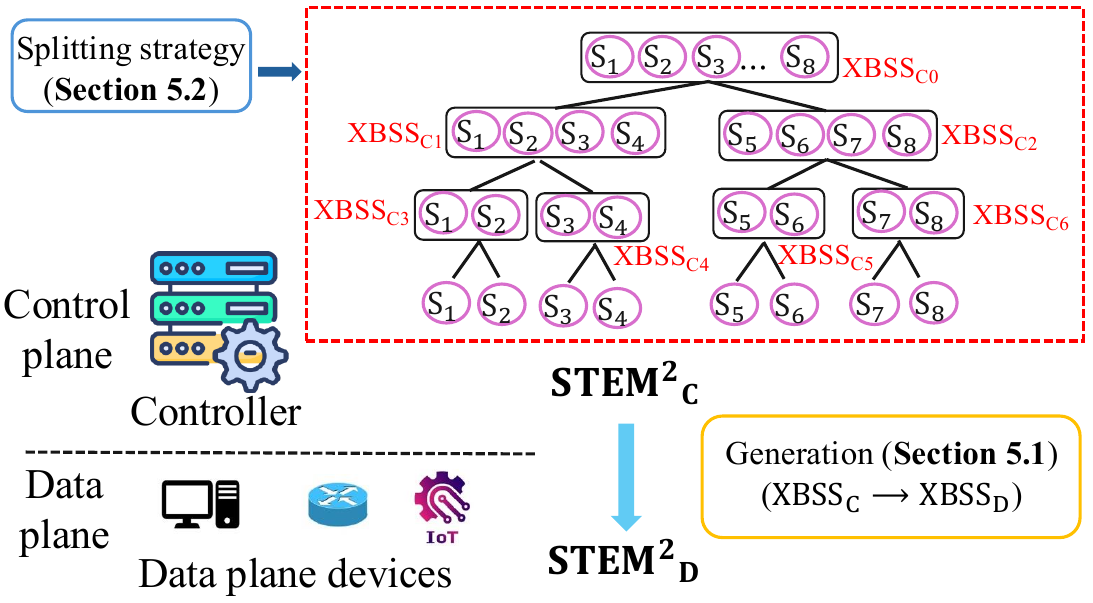}
\vspace{-2ex}
\caption{System Overview of \sysname for Supporting  MS-MQ.}
\label{fig:sys}
\end{center}
\end{figure}

This section presents the holistic design of \sysname. As shown in Figure~\ref{fig:sys}, \sysname adopts a decoupled architecture composed of a control plane and a data plane. Following the principles established in Section~\ref{sec:insights}, maintenance-intensive operations, including structure construction and incremental updates, are assigned to the control plane $\mathtt{STEM^2_C}$. In contrast, the data plane $\mathtt{STEM^2_D}$ remains a lightweight, read-only structure for high-throughput, low-latency lookups on data-plane devices. We then present the splitting strategies used to construct $\mathtt{STEM^2_C}$ in Section~\ref{sec:splitting}. %and operations of \sysname in Section~\ref{sec:stem_operation}. %and the memory layout used to improve lookup throughput in $\mathtt{STEM^2_D}$ in Section~\ref{sec:memmamagement}.

\iffalse
\begin{figure}[t!]
\begin{center}
\includegraphics[width=0.85\linewidth]{figures/MSQ_crop.pdf}
\caption{\wang{Tree-based Framework for Multi-Set Queries.}}
\label{fig:msq}
\end{center}
\end{figure}
\fi

\subsection{Design of $\mathtt{STEM^2_C}$ and $\mathtt{STEM^2_D}$}
\label{sec:stem}

We adopt the tree-based framework to construct $\mathtt{STEM^2_C}$. At each node of the tree, we incorporate our proposed $\mathtt{XBSS_C}$, introduced in Section~\ref{sec:XBSS}, as the binary set separator. The corresponding data-plane structure, $\mathtt{STEM^2_D}$, is obtained by replacing each $\mathtt{XBSS_C}$ in $\mathtt{STEM^2_C}$ with its data-plane counterpart, $\mathtt{XBSS_D}$. To enable a computationally efficient design, we also adopt the less-hashing technique~\cite{kirsch2006less}. Specifically, two base hash functions are selected to simulate all hash functions used in $\mathtt{STEM^2_C}$ and $\mathtt{STEM^2_D}$, thereby reducing the total number of hash computations for key lookups on $\mathtt{STEM^2_D}$ to two.

\subsection{Strategy for Choosing a Splitting Point}
\label{sec:splitting}

With the efficient binary separator $\mathtt{XBSS}$, we construct a tree-based structure, \sysname, to support MS-MQ. However, the choice of how multiple sets are partitioned at each tree node can lead to trees with substantially different shapes and depths, which in turn affects the lookup throughput, memory overhead, and update efficiency of \sysname. To jointly optimize these metrics, it is crucial to carefully select the splitting point used to partition the sets at each tree node.

We first analyze that splitting among a set is not optimal. $\mathtt{XBSS}$ is built to separate keys to two partitions, $L$ and $R$. As discussed in Section~\ref{sec:mem}, The memory efficiency of $\mathtt{XBSS}$ improves as the ratio $r = |R| / |L|$ increases, where $|L|$ and $|R|$ are the sizes of sets $L$ and $R$, respectively. However, when $r$ is small (especially when $|R| \approx |L|$), the memory cost of the $\mathtt{XBSS}$ ($M_o = (2.08\ln \frac{|R|}{|L|} + 4.65)\left|L\right|$) may become prohibitively high. An intuitive approach to mitigating this is to introduce a threshold $\theta$, and adjust the splitting point to make $r = \theta$, by allowing a set to be split across two partitions. Taking this strategy, we will first divide $L$ into two sets $L_1$ and $L_2$ to make $r=(|R|+|L_2|)/|L_1|=\theta$. Then for sets $R$ and $L_2$, we may take the same step until two sets $L_n$ and $R$ meet the ratio requirement. Thus, during this process, the original two sets are divided into multiple sets $L_1, L_2, ..., L_n, R$ and we need to build $n-1$ $\mathtt{XBSS}$ instances to finally separate $L$ and $R$. The total memory cost for this approach will be $M_i = 2.08\sum_{i=1}^{n-1}(\ln\frac{|L_{i+1}|+...+|L_n|+|R|}{|L_i|} + 4.65)\left|L_i\right| + (2.08\ln\frac{|R|}{|L_n|} + 4.65)\left|L_n\right|$. Given $|L_1|+|L_2|+|L_3|+...+|L_n| = |L|$, $M_i-M_o = 2.08\sum_{i=1}^{n-1}\ln\frac{|L_{i+1}|+...+|L_n|+|R|}{|L_i|}\left|L_i\right| + 2.08\ln\frac{|R|}{|L_n|}\left|L_n\right|-2.08\sum_{i=1}^{n}\ln \frac{|R|}{|L|}\left|L_i\right| > 0$. This shows that splitting a single set into multiple partitions to obtain a higher $r$ does not reduce memory cost; instead, it introduces more $\mathtt{XBSS}$ instances and consequently increases lookup latency. Therefore, when designing the splitting strategy, we do not place splitting points within a set.

%\textcolor{red}{On the other hand, splitting across the set will result in more $\mathtt{XBSS}$ and increase the query latency. In conclusion, splitting within a set is a subconsciously viable option but lacks practical feasibility.} 

We then propose two splitting strategies: fully greedy splitting ($\mathtt{FGS}$) strategy and balanced-then-greedy splitting (BGS) strategy.

\subsubsection{FGS strategy.}
%\noindent \textbf{FGS strategy.} 
As discussed in Section~\ref{sec:mem}, for a $\mathtt{XBSS}$, selecting a splitting point that maximizes the size ratio $r = \mathtt{n_l}/\mathtt{n_s}$ between the two groups $\mathtt{D_S}$ and $\mathtt{D_L}$ minimizes the memory overhead of $\mathtt{XBSS_D}$. Accordingly, applying a greedy strategy that minimizes the memory cost of each $\mathtt{XBSS}$ at tree node—referred to as fully greedy splitting ($\mathtt{FGS}$)—yields the minimum memory overhead of $\mathtt{STEM^2_D}$. Formally, consider $n$ sets ${S_0, S_1, \ldots, S_{n-1}}$ sorted in nondecreasing order of size, i.e., $|S_0| \le |S_1| \le \ldots \le |S_{n-1}|$, with total size $T = \sum_{i=0}^{n-1} |S_i|$. Under the $\mathtt{FGS}$ strategy, the $k$-th split ($0 \le k \le n-2$) partitions the current collection ${S_k, S_{k+1}, \ldots, S_{n-1}}$ into a left group ${S_k}$ and a right group ${S_{k+1}, \ldots, S_{n-1}}$. The total memory cost of the resulting tree is $ M_1 = \sum_{k=0}^{n-2}\!\bigl(2.08|S_k| \ln \tfrac{R_k}{|S_k|}+ 4.65|S_k|\bigr) $, where $R_k = \sum_{i=k+1}^{n-1} |S_i|$. 
\iffalse
Since $R_k \le T - |S_k|$, we have 
\begin{align*}
 M_1
&\le 2.08 \sum_{k=0}^{n-2} |S_k| \ln \frac{T - |S_k|}{|S_k|} + 4.65(T - |S_{n-1}|) \\
&\le 2.08 \sum_{k=0}^{n-1} |S_k| \ln \frac{T}{|S_k|} + 4.65T \\
&= 2.08 T \sum_{k=0}^{n-1} p_k \ln \frac{1}{p_k} + 4.65T,
\quad \text{where } p_k = \frac{|S_k|}{T}, \\ 
&\le 2.08T\ln n + 4.65T
\end{align*}
\fi
%thus $M_1 = \Theta(T)$.

\subsubsection{BGS strategy.} 
To reduce tree depth and improve lookup throughput as well as dynamic update efficiency, we first construct a balanced tree and then apply a greedy strategy at each node to minimize the local memory cost, following the approach described in Section~\ref{sec:tb}. This method, referred to as balanced-then-greedy splitting ($\mathtt{BGS}$), reduces the tree depth to $\lceil \log_2(n) \rceil$ for $n$ sets but increases the data plane memory overhead. Under the $\mathtt{BGS}$ strategy, the data-plane memory cost of $\mathtt{STEM^2_D}$ is $M_2 = \sum_{nodes}(2.08\ln (n_r/n_l) + 4.65)n_l$ where $n_r$ and $n_l$ denote the number of keys in the two subgroups at each node. %Since the depth of the tree is $\lceil \log_2(n) \rceil$ and approximately half of the sets are contained in $n_l$ at each level, we have $M_2 = \Theta(T \log n)>M_1$.

\subsubsection{Analysis.}
Because the overall memory cost of $\mathtt{STEM^2_D}$ is dominated by the linear merging component, the efficiency of each strategy depends on how often a set of size $|S_k|$ is accumulated into the left partition. For $M_1$, the linear cost is
$ C_{lin}(M_1)=4.65\sum_{k=0}^{n-2}|S_k|,$
which ensures that each set---except the largest one, $|S_{n-1}|$---is counted exactly once in the memory accumulation. In contrast, $\mathtt{BGS}$ organizes keys as a balanced hierarchical tree, where smaller sets are repeatedly assigned to the left subtree. Let $z(k)$ denote the number of times $|S_k|$ is placed in the left partition across the hierarchy. Then, the linear cost of $M_2$ can be expressed as $C_{lin}(M_2)=4.65\sum_{k=0}^{n-1} z(k)\,|S_k|.$
Since $z(k)\geq 1$ for all $k<n-1$, and can be as large as $\log_2 n$ for the smallest sets, these set sizes are repeatedly accumulated across multiple tree levels. Consequently, $M_1$ achieves a strictly smaller memory footprint than $M_2$, i.e.,
$C_{lin}(M_1)<C_{lin}(M_2)$.

However, from the perspectives of lookup throughput and dynamic update efficiency, applying the $\mathtt{FGS}$ strategy at each node produces a highly unbalanced tree with depth $n$. Consequently, querying or updating keys belonging to larger sets requires traversing more tree levels, which degrades both lookup throughput and update efficiency. For example, a query for a key belonging to $S_{n-1}$ must pass through all $n-1$ $\mathtt{XBSS}$ nodes, significantly increasing lookup latency and reducing throughput. Moreover, the $\mathtt{FGS}$ strategy increases the control-plane memory overhead of $\mathtt{XBSS_C}$, which in turn raises the overall memory cost of $\mathtt{STEM^2_C}$. The main reason is that a larger ratio $r=\mathtt{n_l}/\mathtt{n_s}$ results in a larger index table in the $\mathtt{CBF}$, because the indices of all keys in $\mathtt{D_L}$ must be maintained in that table. In contrast, the $\mathtt{BGS}$ strategy generates more balanced partitions at each node, resulting in a smaller size ratio $r=\mathtt{n_l}/\mathtt{n_s}$ when constructing $\mathtt{XBSS}$ nodes. This smaller ratio reduces the size of the index table in the control-plane $\mathtt{CBF}$ of $\mathtt{XBSS_C}$, and thus lowers the overall memory overhead of $\mathtt{STEM^2_C}$.

%Moreover, compared with $\mathtt{FGS}$, the $\mathtt{BGS}$ strategy produces more balanced partitions at each node, resulting in a smaller size ratio $r = \mathtt{n_l}/\mathtt{n_s}$ when constructing $\mathtt{XBSS}$ nodes. A smaller ratio $r$ leads to a smaller index table in the $\mathtt{CBF}$ of the control plane ($\mathtt{XBSS_C}$), and consequently reduces the memory overhead for $\mathtt{STEM_C}$.

In summary, users can select between the two splitting strategies according to their specific requirements, further enhancing the flexibility of \sysname deployments. Unless otherwise specified, we adopt the $\mathtt{BGS}$ strategy in the implementation of \sysname to achieve higher lookup throughput and more efficient updates.

\subsection{Operations of \sysname}
\label{sec:stem_operation}

%The proposed binary set separator $\mathtt{XBSS}$, tree-based framework, and the BGS splitting strategy help to build \sysname. 
%\wang{In this section, we present the operations supported by \sysname.}

\subsubsection{Operations of $\mathtt{STEM^2_C}$}
\label{sec:stemc_ope}
The control plane is responsible for constructing $\mathtt{STEM^2_C}$ and updating it in response to key updates.

%With the less hashing technique, $\mathtt{STEM^2_C}$ requires only two base hash functions to simulate the multiple hash functions used in $\mathtt{CBF}$ and Othello.
\noindent \textbf{Construction.}
Given multiple sets $\{S_1, S_2, \ldots, S_n\}$, we first select the two base hash functions and use them to construct a balanced tree, applying the splitting strategy at each node to determine the splitting point. This process partitions the sets into left and right groups, which are then used to build a binary set separator $\mathtt{XBSS}$ at each node. $\mathtt{STEM^2_C}$ invokes the $\mathtt{XBSS_C.build}$ function to construct the binary separator. Specifically, the hash functions used in each $\mathtt{XBSS_C}$ are derived from the two base functions by setting different indices $i$, as defined in Equation~\ref{equ:hashing}. Notably, if an Othello of $\mathtt{XBSS}$ fails to find simulated hash functions to build an acyclic graph after exceeding a predefined number of attempts, then $\mathtt{STEM^2_C}$ will reselect two base functions and retry the construction.

%Updates to set membership are common in MS-MQ applications. $\mathtt{STEM^2_C}$ supports the following update operations:

\noindent \textbf{Insertion and deletion.} Inserting or deleting a key $x$ from a set triggers updates to all $\mathtt{XBSS}$ instances along the path from the target leaf node to the root. At each node, the corresponding $\mathtt{XBSS}$ invokes $\mathtt{XBSS_C.insert(x)}$ to add the key or $\mathtt{XBSS_C.delete(x)}$ to remove it.

\noindent \textbf{Inter-set key migration.} A key migration refers to changing the set ID of a key. Suppose a key $x$ is migrated from set $i$ to set $j$; this operation is equivalent to inserting $x$ into set $j$ and deleting $x$ from set $i$. Thus, the migration is performed by executing the corresponding insertion and deletion operations.

%Updating keys one by one incurs high overhead, as a single update requires modifying multiple $\mathtt{XBSS_C}$ instances along the path from the leaf node to the root.
\noindent \textbf{Batch update.} A batch update mechanism is introduced to reduce amortized update latency. Each $\mathtt{XBSS_C}$ in \sysname caches the SetID in its two partitions. For each key update, \sysname first uses the cache information to identify the affected $\mathtt{XBSS_C}$ instances. Then, \sysname derives the corresponding operations—i.e., insertion, deletion, flipping—for each affected instance. For a key insertion or deletion, the corresponding update is the insertion or deletion operations on all affected $\mathtt{XBSS_C}$ instances. A key migration is decomposed into deleting the key from its original set and inserting it into the target set, yielding two ordered lists of separator-level operations. These two lists are then merged from bottom to top until the first common $\mathtt{XBSS_{C}}$ is encountered, at which point the insertion and deletion are combined into a flipping operation. For example, if a key $x$ is migrated from $S_1$ to $S_4$ in Figure~\ref{fig:sys}, \sysname generates the operations $\mathtt{{XBSS_{C3}}.delete(x)}$, $\mathtt{{XBSS_{C4}}.insert(x)}$, and $\mathtt{{XBSS_{C1}}.flip(x)}$. After all key updates have been analyzed, the generated operations are grouped by each $\mathtt{XBSS_{C}}$ and executed in batches.

%First, the batch of update requests is analyzed to identify the required insertions and deletions for each affected binary set separator. \textcolor{red}{The process of identifying which binary set separators are affected by an update operation is straightforward: we perform a query on the key being updated and record the indices of all binary separators traversed along the query path. Given \sysname’s high query efficiency, the overhead of this query is negligible relative to the update latency itself.} Then, for each separator, the corresponding operations are performed in a single step.

\subsubsection{Operations of $\mathtt{STEM^2_D}$}
%The data plane data structure $\mathtt{STEM^2_D}$ supports key lookups. 

\noindent \textbf{Lookup.}
To determine the set ID of a given key $x$, the query begins from the $\mathtt{XBSS_D}$ instance at the root node. The result of $\mathtt{XBSS_D.lookup}(x)$ indicates whether $x$ belongs to the left or right partition, thereby determining the next $\mathtt{XBSS_D}$ instance to query. This process continues recursively until a leaf node is reached, at which point the set ID associated with that leaf is returned as the final result.

%% file: 5_eval.tex
\begin{figure*}[tp!]
    \centering
    \subfigure[Lookup thro. vs. \# of keys]{
        \includegraphics[width=0.23\textwidth]{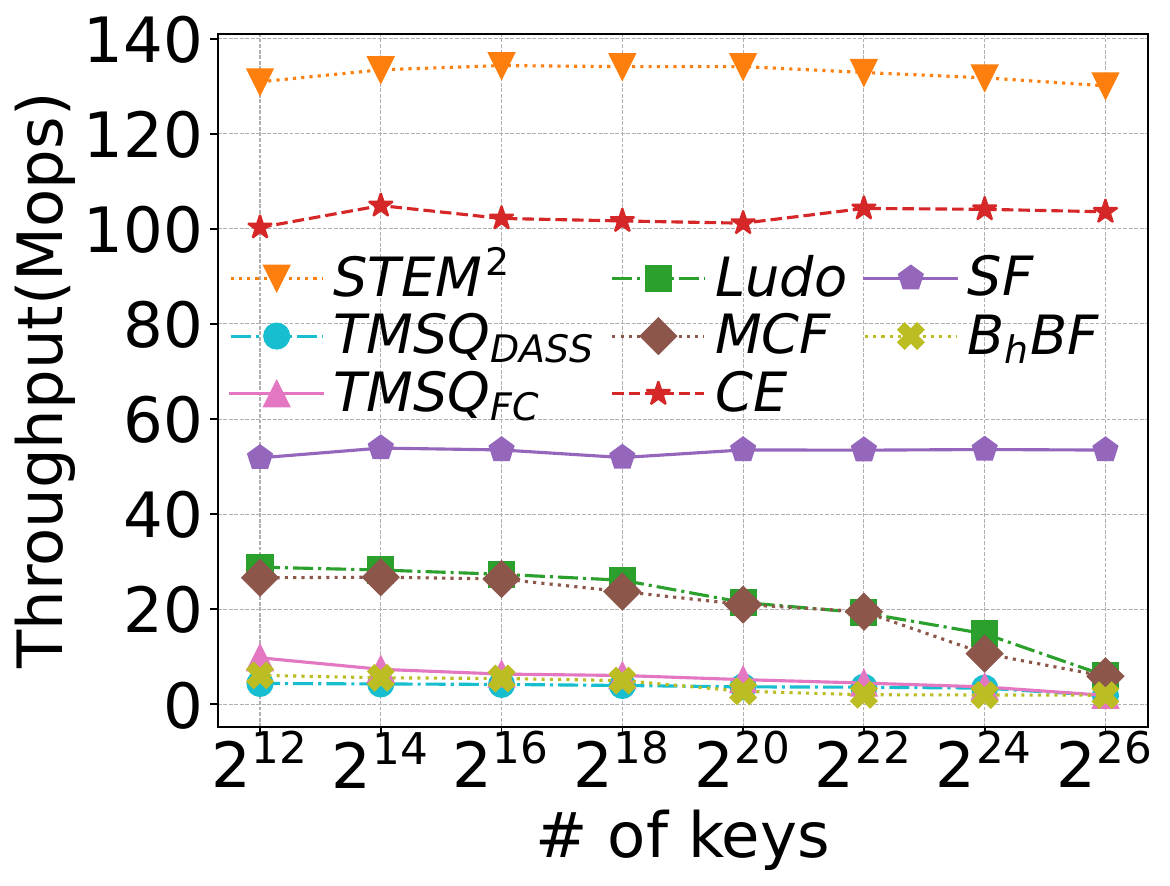}
        \label{fig:TKZ}
    }
    %\hfill
    \subfigure[Memory cost vs. \# of keys]{
        \includegraphics[width=0.23\textwidth]{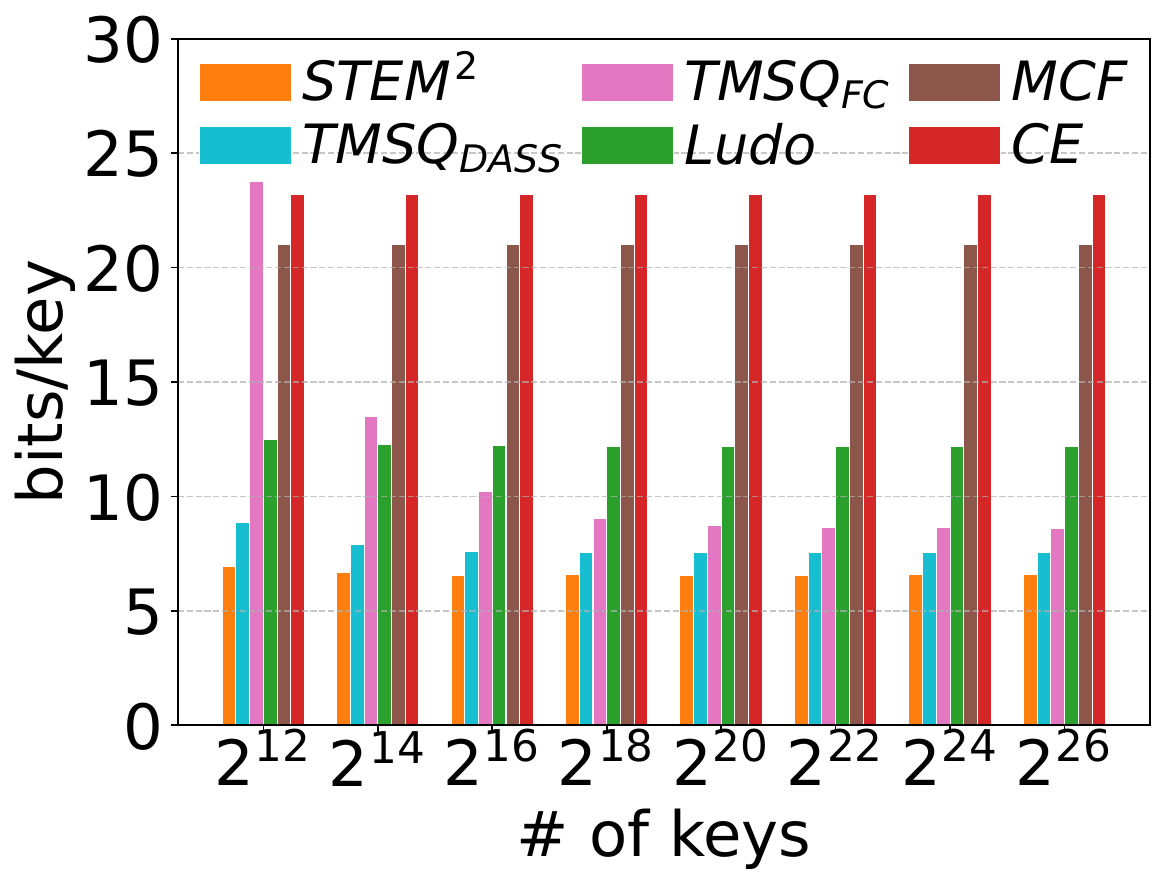}
        \label{fig:MKZ}
    }
    \subfigure[Lookup thro. vs. \# of sets]{
        \includegraphics[width=0.23\textwidth]{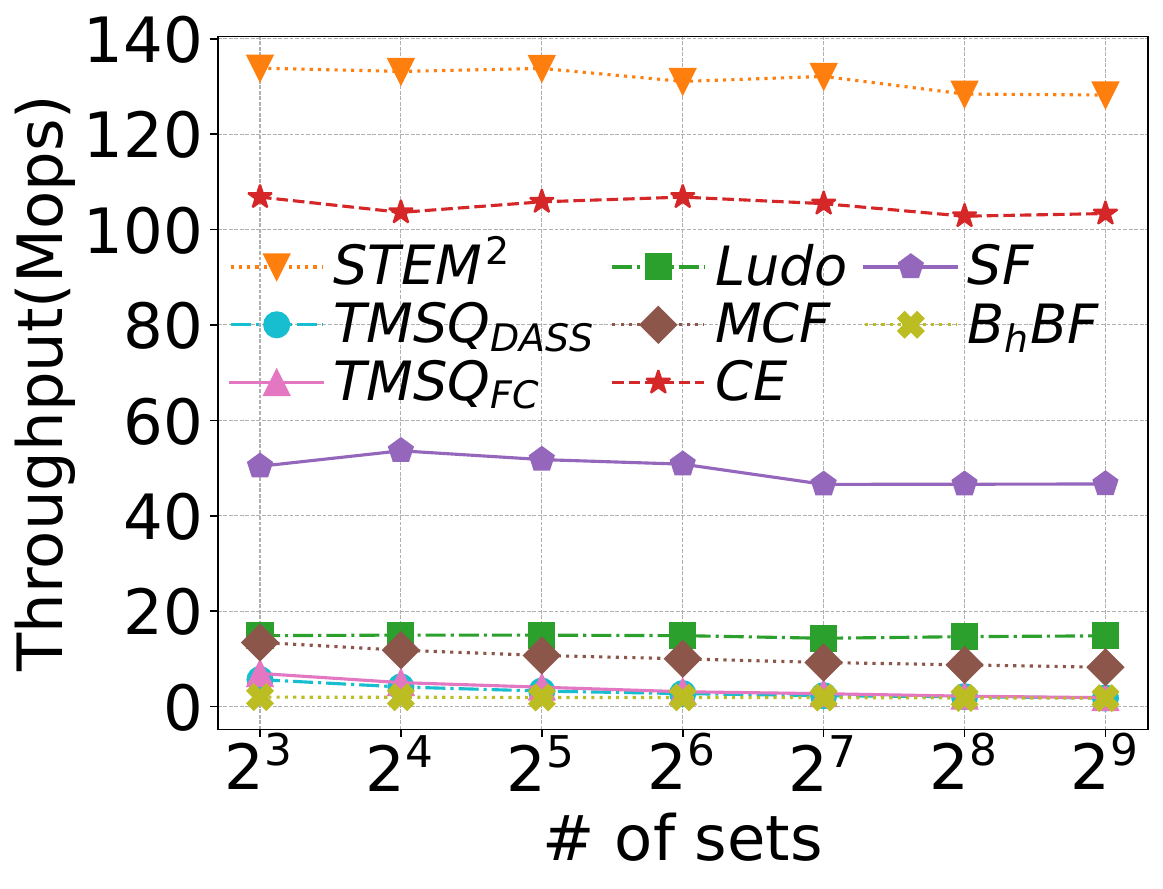}
        \label{fig:TSZ}
    }
    \subfigure[Memory cost vs. \# of Sets]{
        \includegraphics[width=0.23\textwidth]{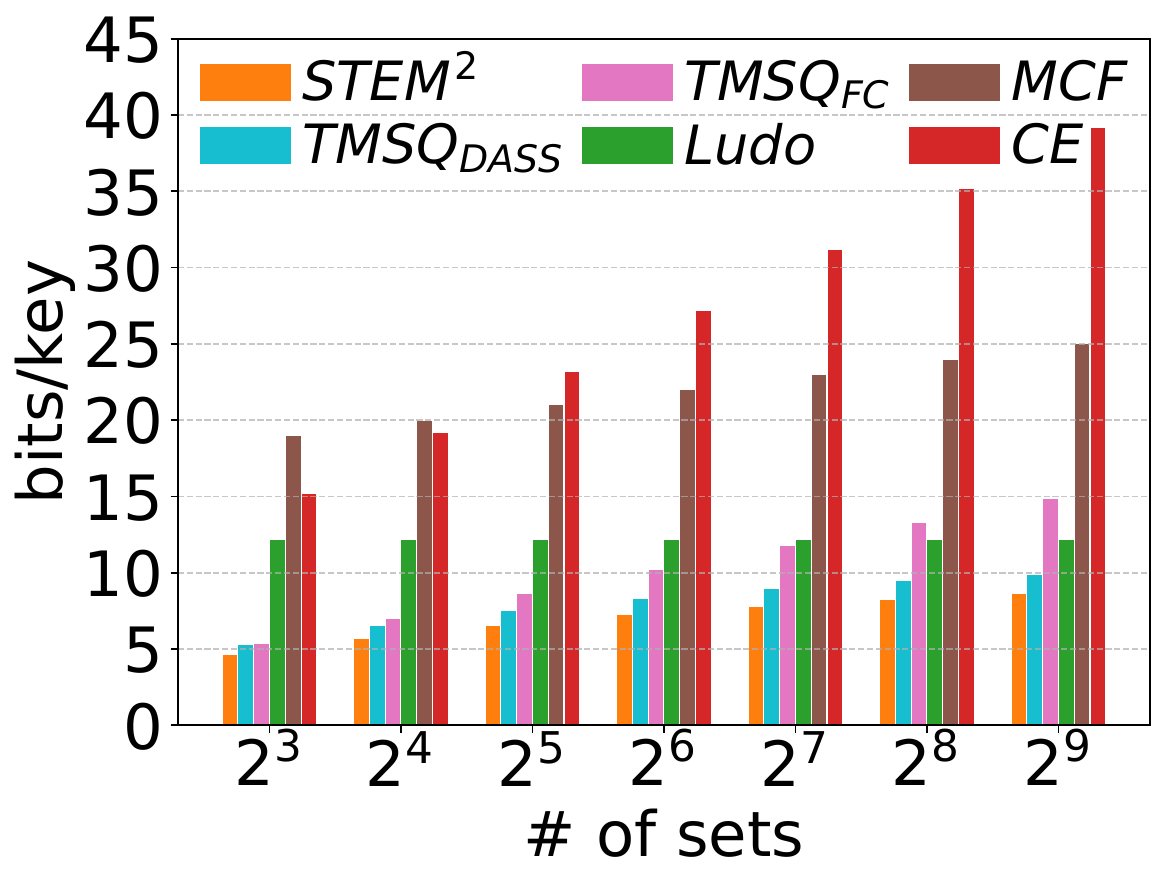}
        \label{fig:MSZ}
    }
    \vspace{-3.5ex}
    \caption{Lookup Throughput and Memory Cost Using Datasets under Zipfian Distribution.}

    \label{fig:EvaluationZipfian}
\end{figure*}

\section{Evaluation}
\label{sec:eval}

%The false positive rate of the (Counting) Bloom filter of $\mathtt{XBSS}$ is configured as $\epsilon = \frac{0.8925}{r}$ to get the minimized memory cost, where $r (r > 1)$ is the size ratio of the two partition groups, and each counter in a counting Bloom filter is allocated 4 bits.

\subsection{Implementation and Experiment Setup}
\label{sec:imple}
We implement a complete software prototype of \sysname in C++. The false positive rate $\epsilon$ of the (counting) Bloom filter in each $\mathtt{XBSS}$ is the only parameter that needs to be configured. We set $\epsilon = \frac{0.8925}{r}$ to minimize memory cost, where $r$ ($r > 1$) denotes the size ratio between the two partition groups. Each counter in the $\mathtt{CBF}$ is allocated 4 bits. Once $\epsilon$ is determined, the remaining parameters of the $\mathtt{CBF}$ and Othello within $\mathtt{XBSS}$ are automatically derived, as analyzed in Section~\ref{sec:mem}. We conduct two types of performance evaluation. 1) Synthetic-dataset evaluation, where the sizes of different sets follow Zipfian, normal, and uniform distributions. 2) Case study of \sysname on two applications, including packet forwarding and transaction tag query. All experiments are run on an Exxact Valence full-tower workstation with AMD Ryzen Threadripper PRO 5965WX, 3.8GHz, 128 MB L3 cache and Ubuntu 22.04.

\noindent\textbf{Metrics.} %We evaluate the BiMS by the following metrics: 1) Lookup throughput: number of keys be processed in query operation in one second. 2) Memory consumption: average bits consumed to store a key. 3. Update efficiency: average times (us) spent to insert/remove a keys. It is worth note that we do not evaluate the \textbf{accuracy} since BiMS achieves fully accuracy and outperforms the benchmarks with false positive in query.
We employ the following metrics to evaluate \sysname. \textbf{1) Lookup throughput}: quantified  by the number of lookups a data structure on the data plane can execute per second, reported in million operations per second (Mops). \textbf{2) Memory cost}: amortized number of bits consumed per key on the data plane. \textbf{3) Update efficiency}: amortized latency to insert/delete/migrate a key on the control plane. It is worth noting that we do not evaluate \textbf{lookup accuracy}, as \sysname guarantees 100\% correctness.

\noindent\textbf{Baseline methods}. %To compare our data structure with the state-of-the-art, we introduce several algorithms that have been proposed in recent years and are wildly recognized as the most efficient ones for multi-set query.
We compare \sysname with existing state-of-the-art data structures and algorithms for MS-MQ, including Ludo hashing (Ludo) \cite{ludo}, coloring embedder (CE) \cite{coloringembedder}, Shifting Filter (SF) \cite{shiftingfilter}, $B_h$ sequence-based Bloom filter ($\mathtt{B_hBF}$) \cite{bhbf}, and Marked Cuckoo filter (MCF) \cite{luo2021mcfsyn}. We use the publicly available C++ implementations of Ludo \cite{ludo_code} and CE \cite{CE_code}. For Shifting Filter, we rewrite its public Python implementation \cite{SF_code} in C++. We receive the implementation of $\mathtt{B_hBF}$ in Java from its authors and then rewrite it in C++. Finally, we implement MCF in C++ based on the design described in its paper \cite{luo2021mcfsyn}. Meanwhile, we also compare \sysname with $\mathtt{TMSQ_{FC}}$ and $\mathtt{TMSQ_{DASS}}$ introduced in Section~\ref{sec:tb}. Among those algorithms, only Ludo, $\mathtt{TMSQ_{FC}}$ and $\mathtt{TMSQ_{DASS}}$ can achieve 100\% lookup accuracy. To ensure a fair comparison and eliminate the influence of hash function variability on algorithm performance, we use Google FarmHash ~\cite{farmhash} as the hash function for all evaluated algorithms.

\noindent\textbf{Datasets}. We generate synthetic datasets with varying numbers of sets and total keys, where the set size distributions follow Zipfian, normal, and uniform distributions, respectively. Each key is 64 bits.

For two case studies, we employ two real world datasets, and Table \ref{fig:table1} summarizes the statistical information of the real datasets.

%including the number of keys and sets, and the set sizes.
%and generate plenty of synthetic datasets for experiments. The brief introduction of two real is as follows:

\noindent \textbf{Network traffic~\cite{garcia2020iot23}.} We utilize a network traffic dataset from Internet of Things (IoT) devices for packet forwarding, where the packet ID is the key and the port number serves as the set ID. %We use the unique timestamp of the connection event as the key to identify packet and the destination port number as the value. All entries with the same value is classified into the same set. Besides, we only keep sets with more than 15 elements.
%Finally, we obtain 303 sets with totally 5.7 million keys.

\noindent \textbf{Credit card transactions~\cite{Ryan2024credit}.} This dataset stores the transaction history of users that includes the amount for each transaction. We divide the transactions into different sets according to the transaction amount. For example, the transaction amount between 0 and 5 are assigned to set 0, amounts between 6 and 10 to set 1, and so on. Transactions with amounts greater than 2500 are grouped into one set. Notably, empty sets are discarded.

\begin{table}[tp!]
\centering
\caption{Statistical Information of Real Datasets.}
\vspace{-2ex}
\begin{tabular}{|c|c|c|c|c|}
\hline
Dataset & \#Sets & \#Keys & min set size & max set size \\
\hline
Network & 302 & 570397 & 16 & 271561 \\
\hline
Transactions& 503 & 1048575 & 1 & 141886 \\
\hline

\end{tabular}
\label{fig:table1}
\end{table}

\subsection{Data Plane Evaluation of \sysname}
\label{sec:dp_eval}

In this section, we evaluate the lookup throughput and memory cost of \sysname using synthetic datasets on the data plane.

\begin{figure*}[t!]
    \centering
    \subfigure[Lookup thro. vs. \# of keys]{
        \includegraphics[width=0.23\textwidth]{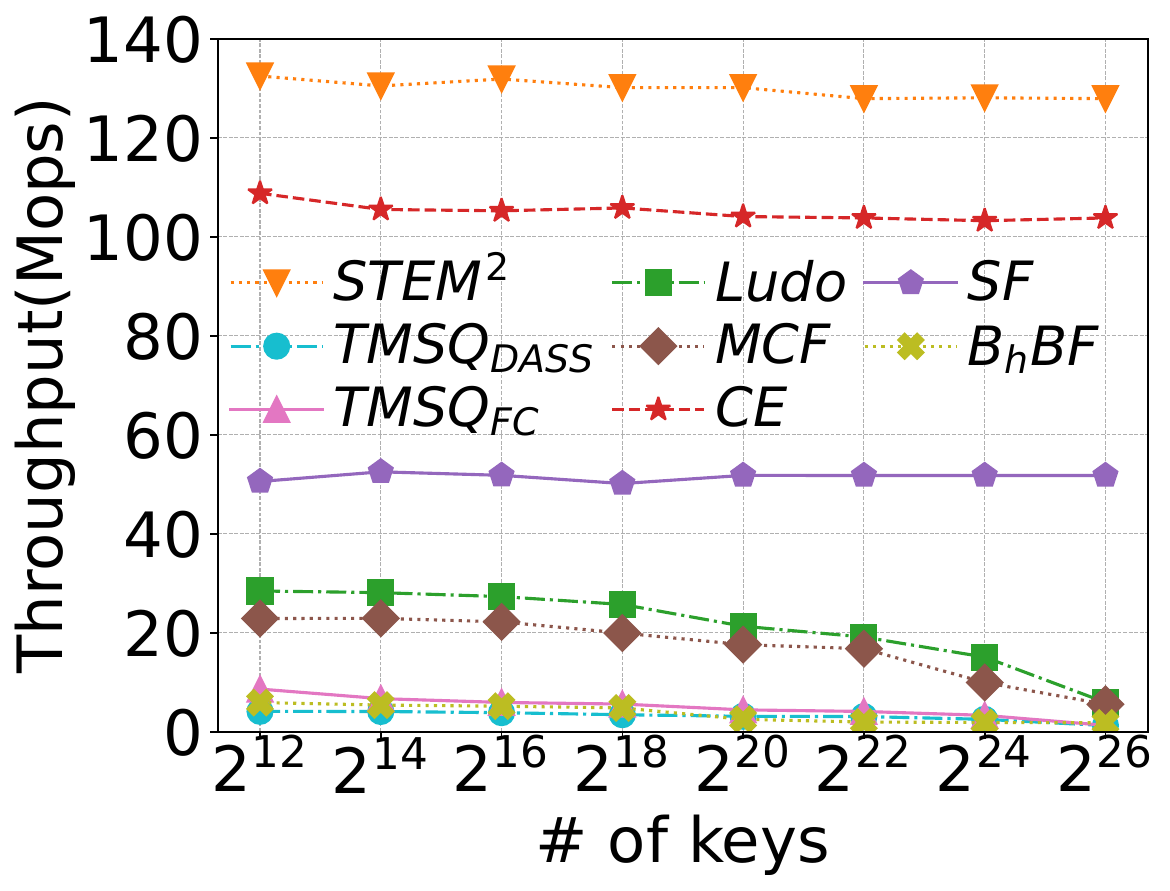}
        \label{fig:TKN}
    }
    %\hfill
    \subfigure[Memory cost vs. \# of keys]{
        \includegraphics[width=0.23\textwidth]{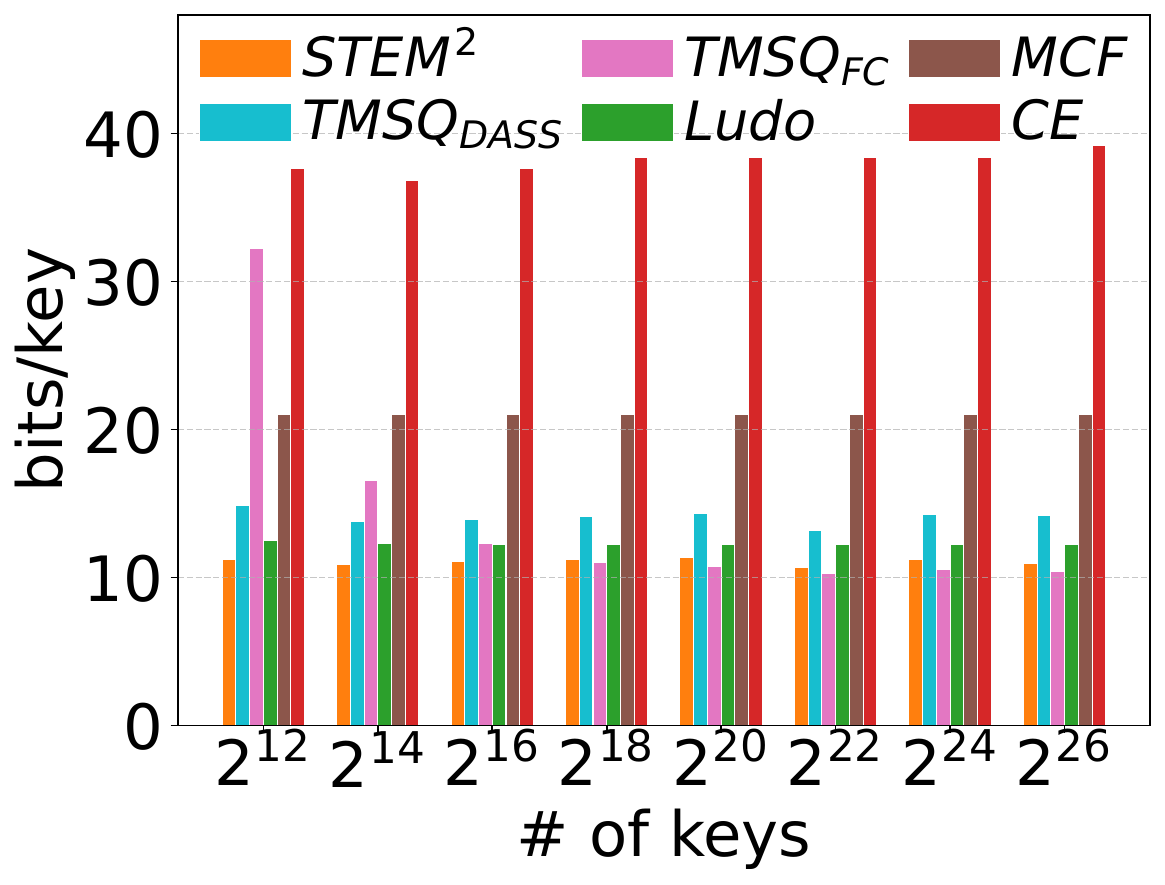}
        \label{fig:MKN}
    }
    \subfigure[Lookup thro. vs. \# of sets]{
        \includegraphics[width=0.23\textwidth]{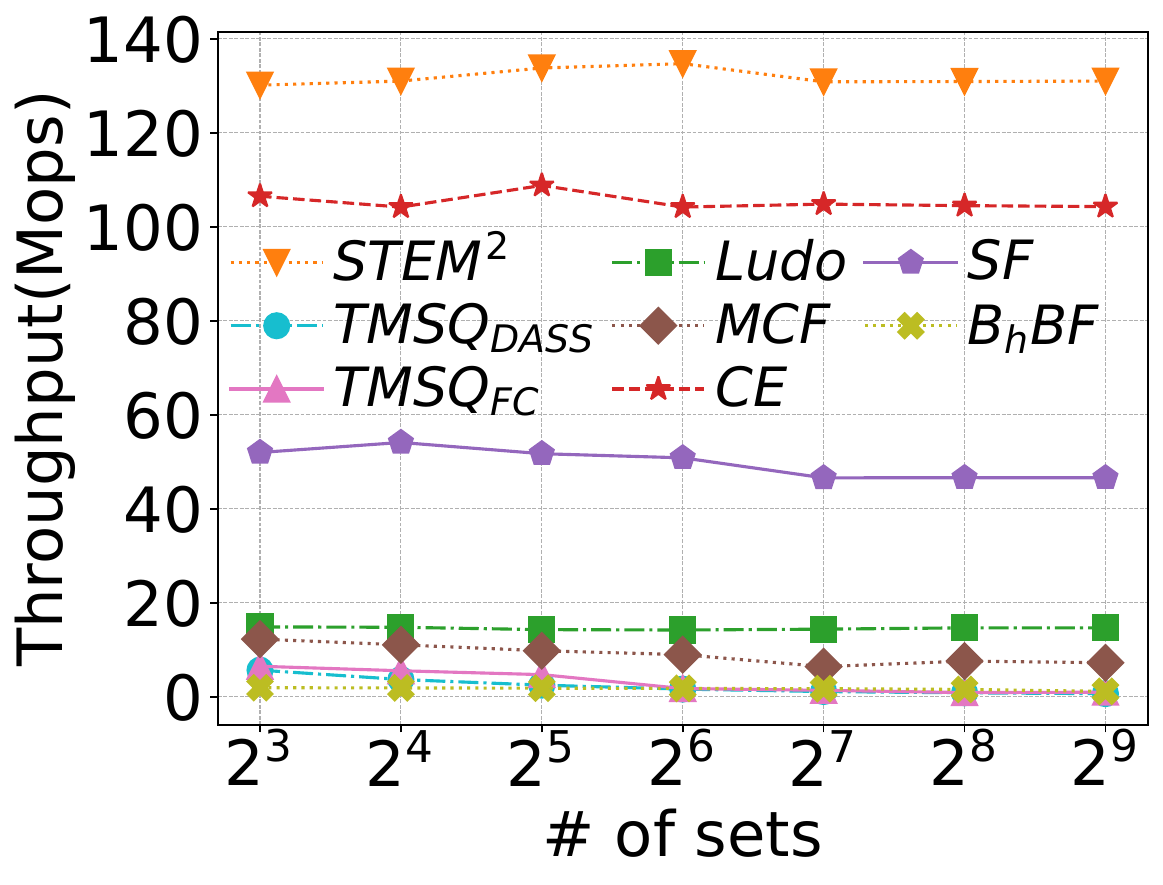}
        \label{fig:TSN}
    }
    \subfigure[Memory cost vs. \# of Sets]{
        \includegraphics[width=0.23\textwidth]{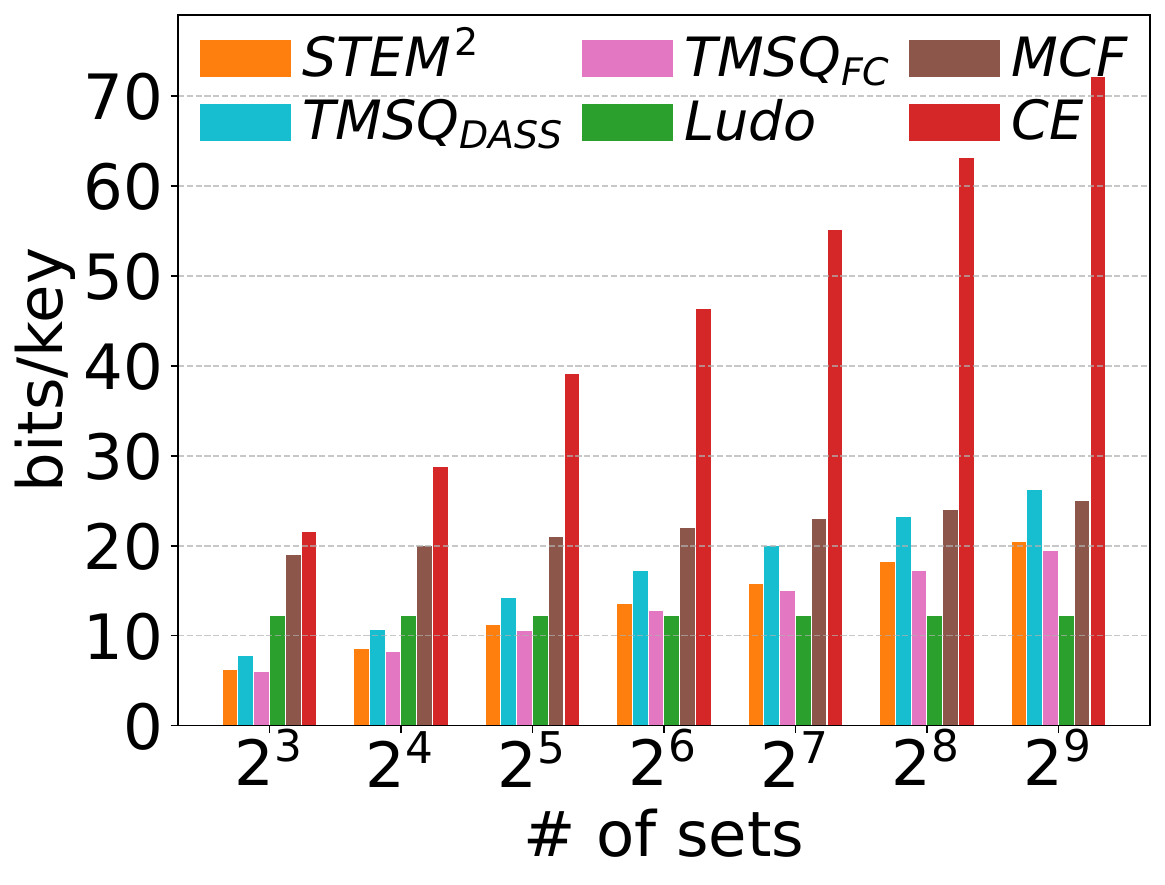}
        \label{fig:MSN}
    }
    \vspace{-3.5ex}
    \caption{Lookup Throughput and Memory Cost Using Datasets under Normal Distribution.}
    \label{fig:EvaluationNorm}
\end{figure*}

\subsubsection{Performance on Zipfian-distributed dataset} We generate synthetic datasets with varying number of
sets and total keys, where the set sizes for each dataset follow a Zipfian distribution. 
%\textcolor{red}{which is the most favorable dataset to \sysname according to our design}.
%Then we evaluate the lookup throughput and memory cost of \sysname, in comparison with other baseline methods.

\noindent \textbf{Lookup throughput and memory cost vs. number of keys}. Figure~\ref{fig:TKZ} and Figure~\ref{fig:MKZ} present the lookup throughput and memory cost as the total number of keys increases from $2^{12}$ to $2^{26}$, with the number of sets fixed at $2^5$. The results show that \textbf{\sysname achieves exceptionally high lookup throughput}, exceeding 130 Mops, which is more than 25.6\%--32.5\% higher than the best baseline CE. Moreover, \sysname achieves 4.54$\times$--21.63$\times$ higher lookup throughput compared to Ludo, which also guarantees fully accurate query. The lookup throughput of other methods ranges from 2 to 55 Mops. As number of keys grows, \sysname experiences a negligible degradation in throughput, due to an increased number of cache misses caused by larger data structures when handling more keys. Ludo and MCF show noticeable performance degradation as key size grows. 

Figure~\ref{fig:MKZ} shows the memory cost of \sysname on the data plane as the number of keys varies. The results show that \textbf{\sysname is the most space-efficient data structure under Zipfian distribution compared to other methods.} Specifically, \sysname consumes 6.56--6.95 bits per key with varying number of keys. $\mathtt{TMSQ_{DASS}}$, has the same tree-based framework as \sysname but uses different binary separators, requires 7.54--8.85 bits per key, demonstrating the memory efficiency of the tree-based framework. The memory cost of CE is relatively large compared with \sysname, which requires around 23.2 bits per key, although it achieves the best lookup throughput among baselines. We do not show memory cost of $\mathtt{B_hBF}$ and SF in the figures, as both require significantly more memory space compared to other methods. Specifically, $\mathtt{B_hBF}$ is reported to require an average of 80 bits per key based on the bit-field compression, which incurs additional costs in the form of memory alignment overhead and bitwise operation overhead. Allocating a full field such as uint32\_t for each counter significantly worsens memory cost (nearly 422 bits per key) but also significantly increases lookup throughput. Our evaluation for throughput and memory cost of $\mathtt{B_hBF}$ is based on the full field counter version. SF requires an average of 94.8 bits per key to achieve 95\% query accuracy.

\noindent \textbf{Lookup throughput and memory cost vs. number of sets}. Figure~\ref{fig:TSZ} and Figure~\ref{fig:MSZ} show the lookup throughput and memory cost as the number of sets increases from $2^3$ to $2^9$, with the number of keys fixed at $2^{24}$. We observe that the number of sets has a negligible effect on the lookup throughput, and \textbf{\sysname consistently achieves the highest performance}. Specifically, \sysname achieves 22.7\%–28.5\% higher lookup throughput than CE, and outperforms Ludo by 8.65$\times$--9.23$\times$. As the number of sets increases, a query must traverse more $\mathtt{XBSS}$ instances, however, the throughput shows only slight fluctuations. The key reason is that \sysname requires only two hash computations in a query process with the less-hashing technique, regardless of the number of $\mathtt{XBSS}$ instances in the tree. 

In terms of memory cost, \textbf{\sysname consumes the least memory} among all evaluated methods on the Zipfian-distributed dataset as shown in Figure~\ref{fig:MSZ}. As the number of sets increases, \sysname's memory usage rises moderately from 4.62 to 8.63 bits per key, primarily due to the introduction of additional $\mathtt{XBSS}$ instances for more sets. Among the baselines, $\mathtt{TMSQ_{DASS}}$, the most memory-efficient baseline, requires around 5.33--9.91 bits per key. CE, which achieves the highest lookup throughput among baselines, incurs significantly higher memory cost, ranging from 15.2--39.2 bits per key.

\begin{figure*}[t!]
    \centering
    \subfigure[Lookup thro. vs. \# of keys]{
        \includegraphics[width=0.23\textwidth]{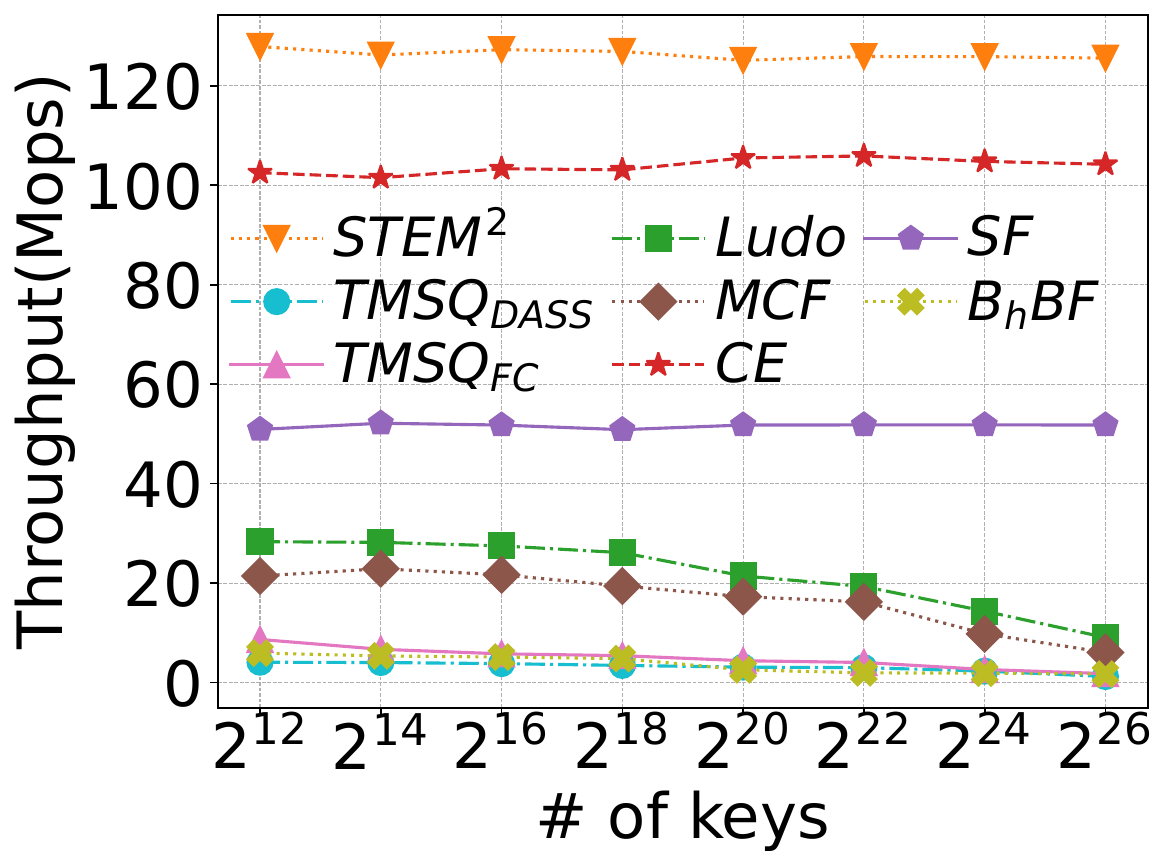}
        \label{fig:TKU}
    }
    %\hfill
    \subfigure[Memory cost vs. \# of keys]{
        \includegraphics[width=0.23\textwidth]{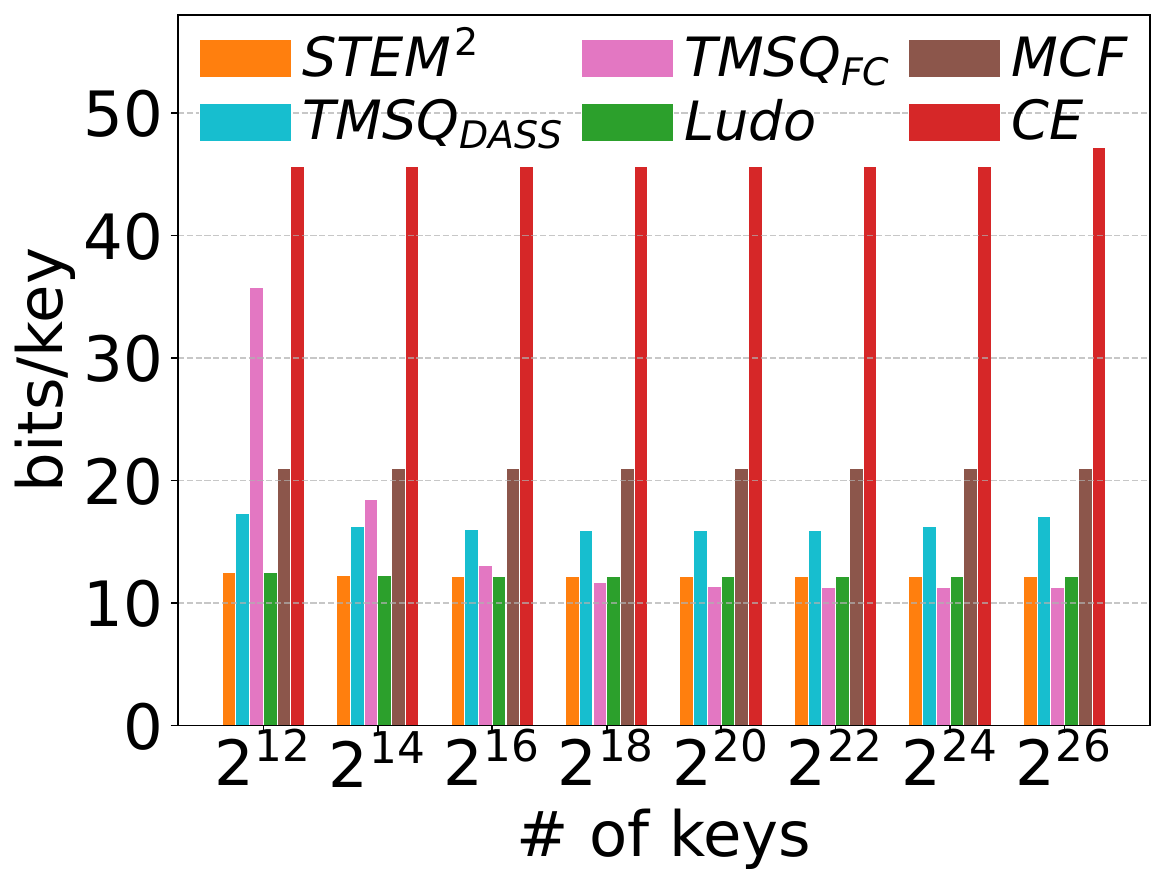}
        \label{fig:MKU}
    }
    \subfigure[Lookup thro. vs. \# of sets]{
        \includegraphics[width=0.23\textwidth]{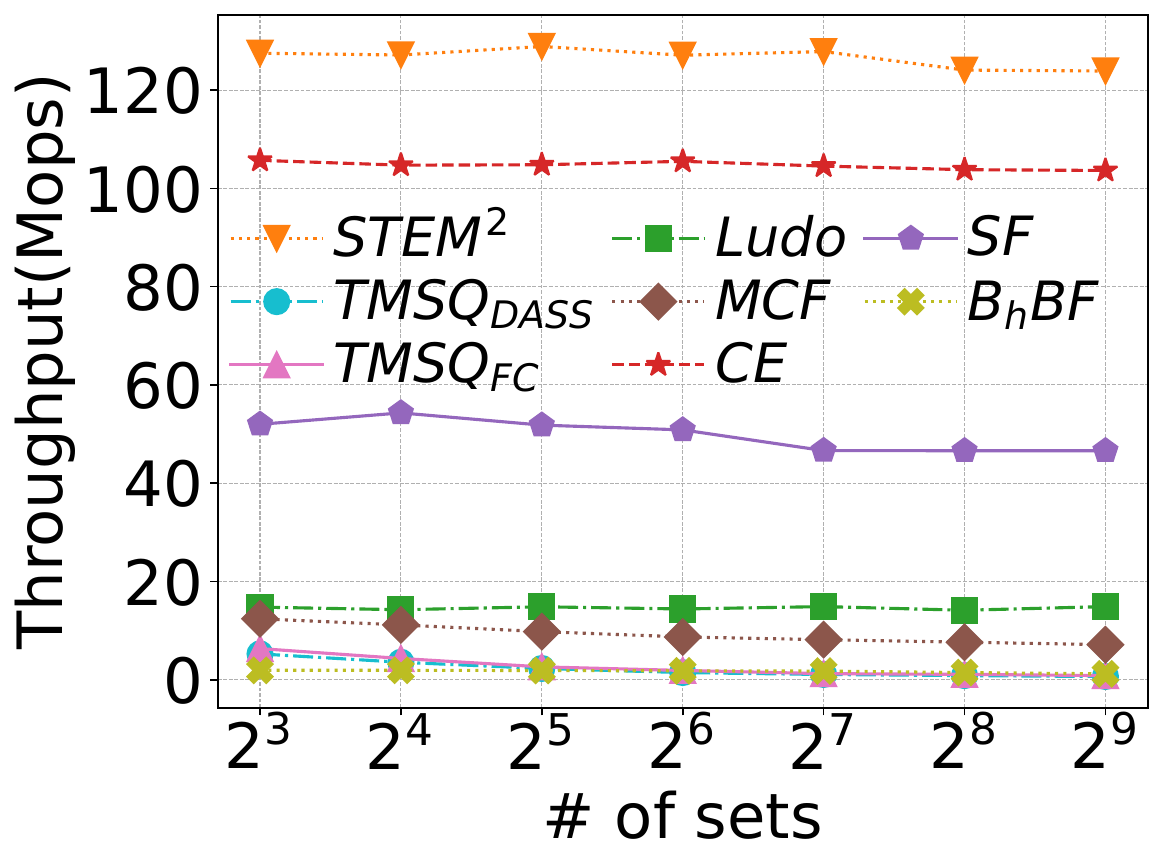}
        \label{fig:TSU}
    }
    \subfigure[Memory cost vs. \# of sets]{
        \includegraphics[width=0.23\textwidth]{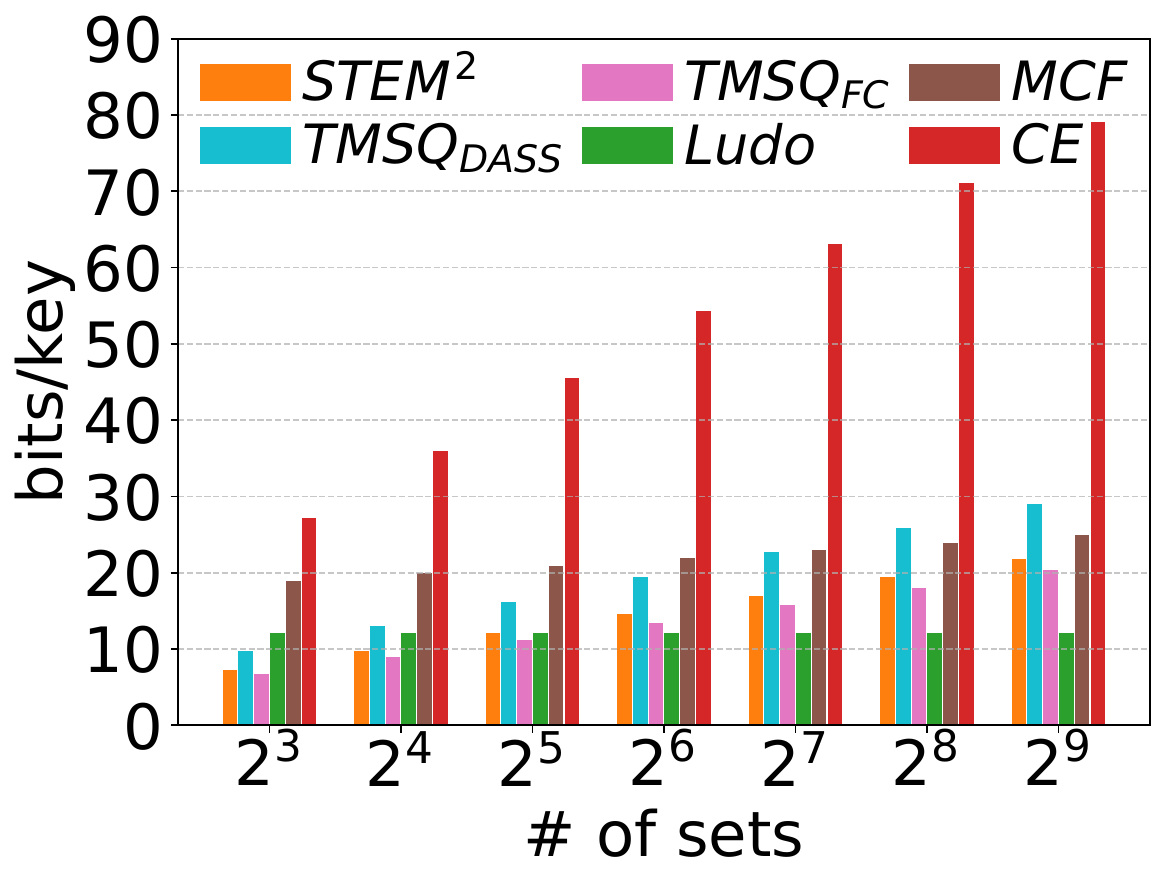}
        \label{fig:MSU}
    }
    \vspace{-3.5ex}
    \caption{Lookup Throughput and Memory Cost Using Datasets under Uniform Distribution.}
    \label{fig:EvaluationUniform}
\end{figure*}

\begin{figure*}[t!]
\centering

% ---------- Left Figure ----------
\begin{minipage}[t]{0.48\textwidth}
    \centering
    \subfigure[Amortized insert. latency vs. \# of keys]{
        \includegraphics[width=0.45\textwidth]{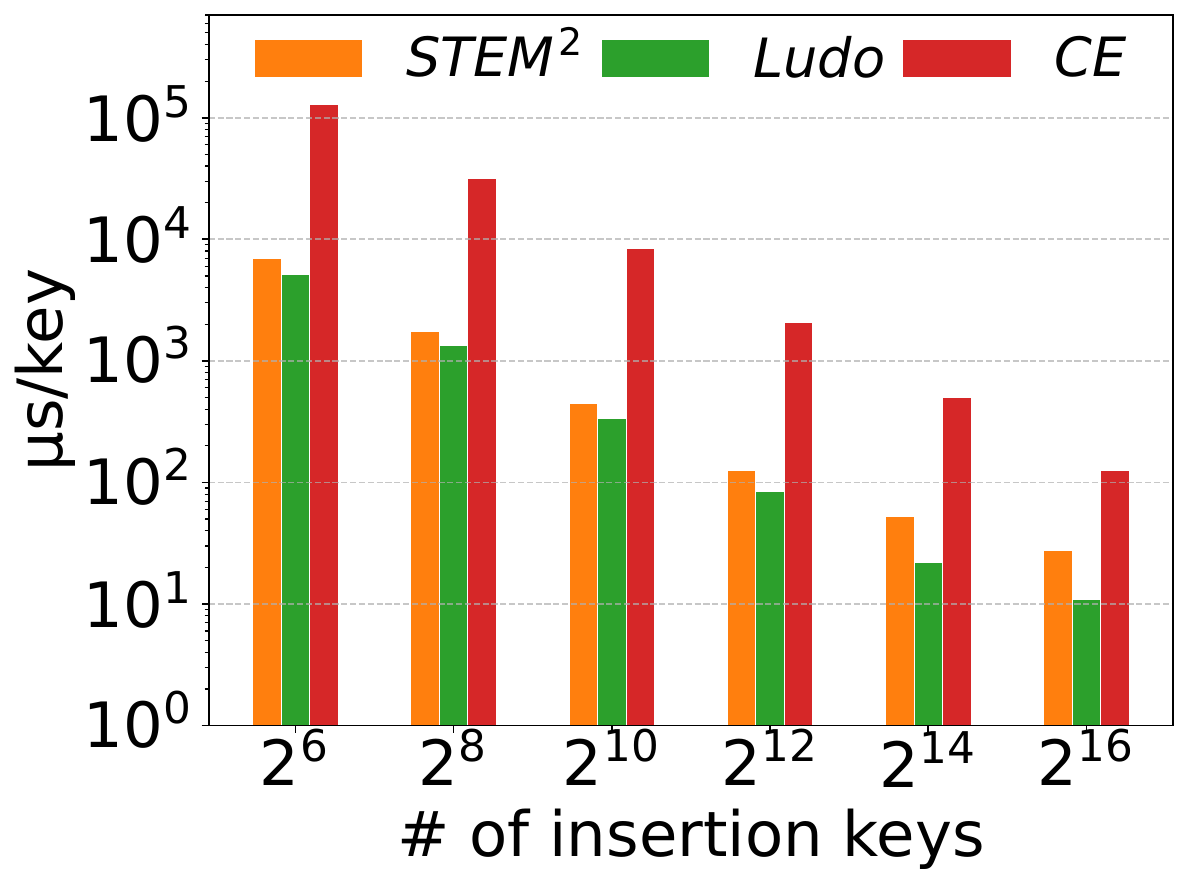}
        \label{fig:Insertvskeys}
    }
    \hfill
    \subfigure[Amortized insert. latency vs. \# of sets]{
        \includegraphics[width=0.45\textwidth]{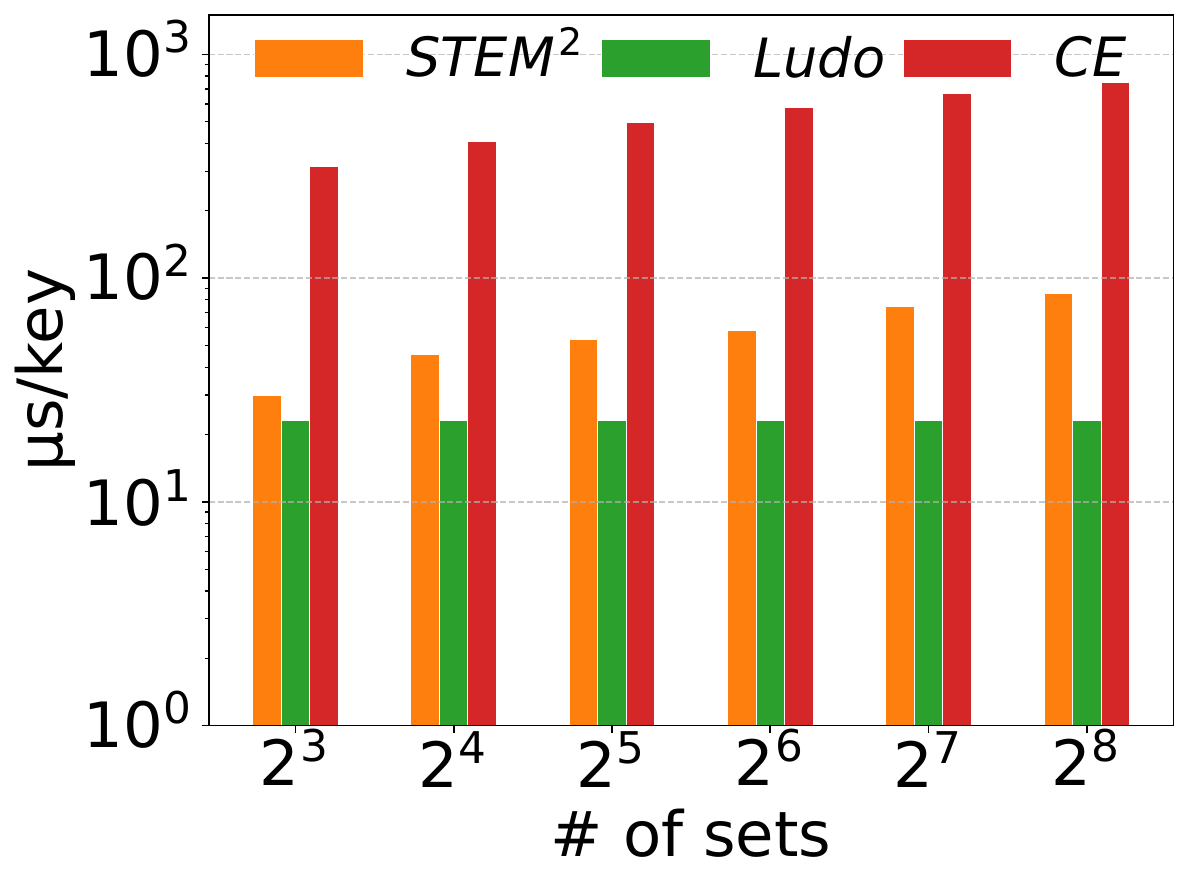}
        \label{fig:Insertvssets}
    }
    \vspace{-3.5ex}
    \caption{Amortized Insertion Latency Per Key.}
    \label{fig:insertefficiency} 
\end{minipage}
\hfill
% ---------- Right Figure ----------
\begin{minipage}[t]{0.48\textwidth}
    \centering
    \subfigure[Migration latency vs. \# of keys]{
        \includegraphics[width=0.45\textwidth]{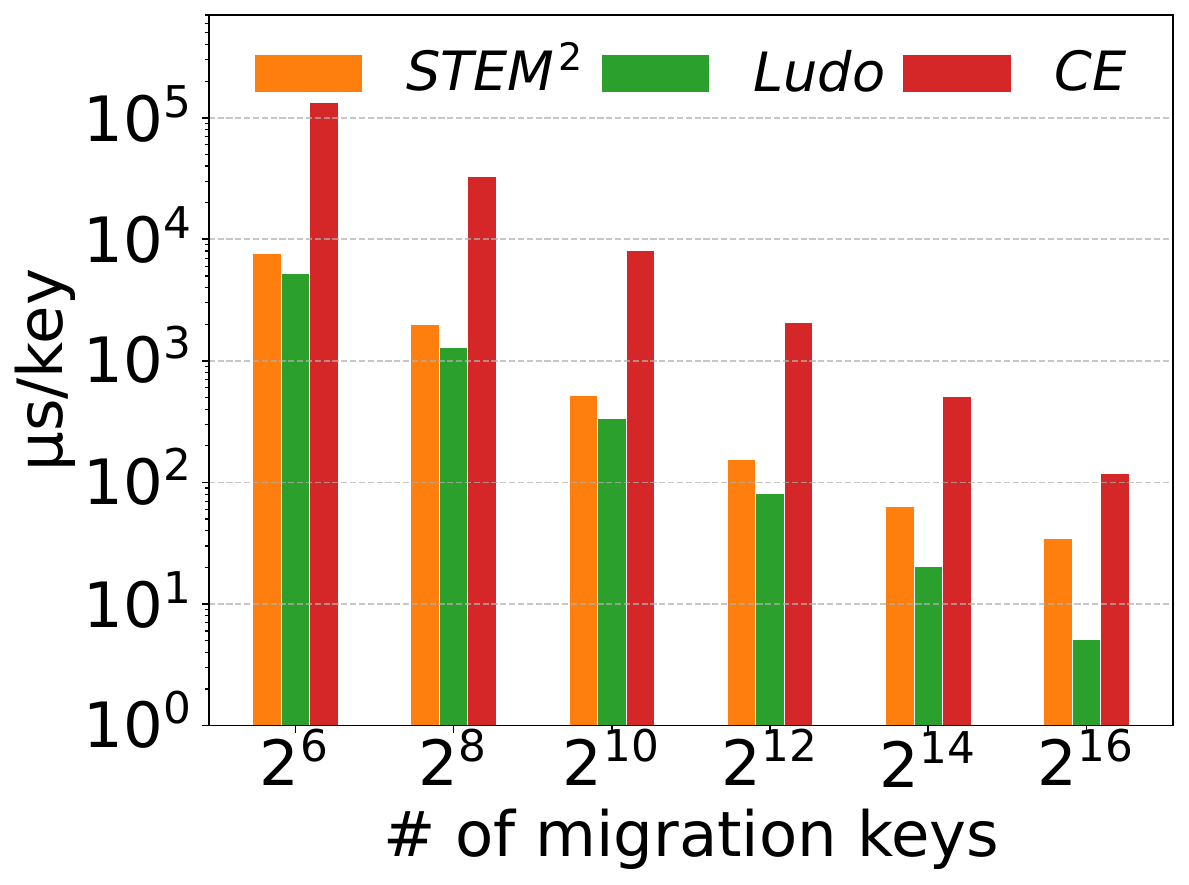}
        \label{fig:movevskeys}
    }
    \hfill
    \subfigure[Migration latency vs. \# of sets]{
        \includegraphics[width=0.45\textwidth]{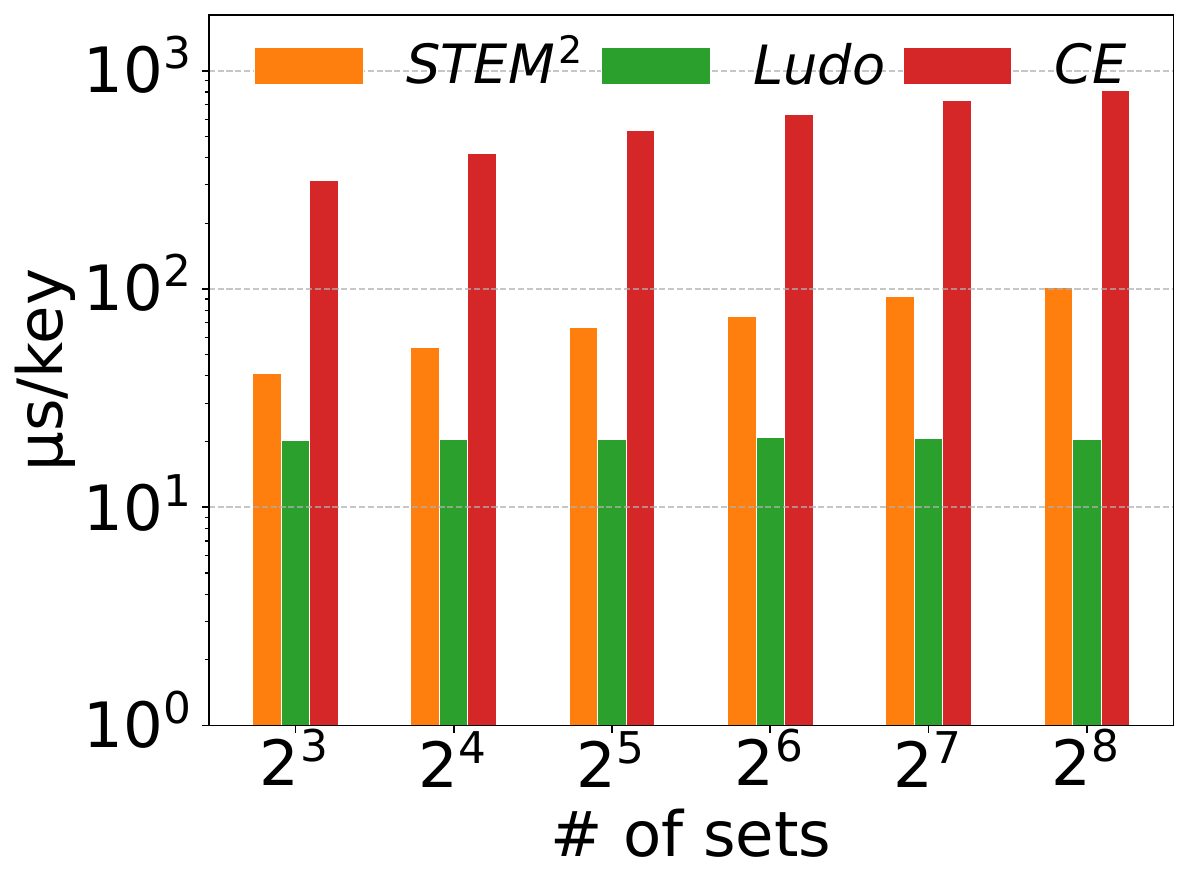}
        \label{fig:Movevssets}
    }
    \vspace{-3.5ex}
    \caption{Amortized Migration Latency Per Key.}
    \label{fig:moveefficiency}
\end{minipage}

\end{figure*}

\subsubsection{Performance on normally distributed dataset}
In some real-world scenarios, the set sizes exhibit a normal-like distribution. To evaluate \sysname under such conditions, we construct synthetic datasets with different numbers of sets and total keys, where the set sizes for each dataset follow a normal distribution.

\noindent \textbf{Lookup throughput and memory cost vs. number of keys}. Figure~\ref{fig:TKN} and Figure~\ref{fig:MKN} show the lookup throughput and memory cost of different methods as the total number of keys grows from $2^{12}$ to $2^{26}$, while keeping the number of sets fixed at $2^{5}$. \textbf{\sysname achieves the highest lookup throughput}, which is 21.8\%--25.3\% higher than the best baseline method CE. All methods exhibit lookup throughput trends that are consistent with those observed under Zipfian-distributed datasets as the number of keys increases.

For memory cost, \textbf{\sysname remains the most efficient solution}, with memory usage ranging from 10.65 to 11.34 bits per key. Among baselines, the most efficient design, Ludo, requires an average of 12.2 bits per key on the data plane. Although the normally distributed dataset is less skewed than the Zipfian-distributed one, \sysname presents only negligible performance degradation, demonstrating the adaptability and robustness of our design.

%The memory overhead of \sysname and the benchmarks are evaluated in the normal distributed dataset, with the number of sets maintained as $2^{5}$ and the key size varies from $2^{12}$ to $2^{26}$. \textbf{\sysname is still the most efficient solution in memory overhead}, which falls into the range of 10.65 - 11.34 bits per key. As the best benchmark, Ludo's data plane needs an average of 12.2 - 12.5 bits to store a key. Although the normal distributed dataset is not as skewed as the zipfian distributed dataset, the performance of \sysname has only negligible degradation, which proves the adaptability of our solution.

\noindent \textbf{Lookup throughput and memory cost vs. number of sets}. Figure~\ref{fig:TSN} and Figure~\ref{fig:MSN} show the lookup throughput and memory cost as the number of sets grows from $2^3$ to $2^9$, while keeping the total number of keys at $2^{24}$. \textbf{\sysname achieves the highest lookup throughput} among all methods and the increasing number of sets has negligible impact on its throughput.

\sysname consumes 6.22--20.46 bits memory per key as the number of sets grows, mainly because more $\mathtt{XBSS}$ instances are required. Since the normal distribution is less skewed than the Zipfian distribution, \sysname experiences some degradation in memory efficiency under this setting. In contrast, Ludo exhibits the lowest and most stable memory usage, requiring 12.2 bits per key regardless of the number of sets. $\mathtt{TMSQ_{FC}}$, as the second-best benchmark in this scenario, requires 6.06--19.47 bits per key. Although \sysname does not achieve the lowest memory cost when handling a large number of sets, its memory usage remains comparable to the most efficient methods (Ludo and $\mathtt{TMSQ_{FC}}$) and still outperforms other baselines. Notably, the \textbf{superior lookup throughput of \sysname} makes it a compelling and practical solution.
 
%With the number of sets increasing, \sysname's memory consumption increased from 6.22 to 20.46 bits per key, this is because more binary separators are introduced in this process. Since the normal distributed dataset is not as biased as the zipfian distribution, there is a degradation in memory efficiency. In contrast, Ludo requires the least memory space in this scenario. It requires 12.2 bits per key, regardless of how the number of sets changes. Besides, $TMSQ_{FC}$ as the second best benchmark, requires 6.06 - 19.47 bits per key. Although \sysname cannot gain an absolute advantage in terms of memory overhead when dealing with large number of sets, its memory overhead is very close to the most efficient solution in such scenario, such as Ludo and $TMSQ_{FC}$ and outperform other benchmarks. In addition, we cannot ignore the absolute advantage of \sysname in terms of query throughput, so \sysname remains a very valuable algorithm.

\subsubsection{Performance on uniformly distributed dataset}

The lookup throughput and memory cost on skewed datasets with Zipfian and normal distributions demonstrate the exceptional advantages of \sysname. One key reason is that the memory usage of our designed separator, $\mathtt{XBSS}$, decreases as the size ratio between the two partitioned subgroups increases, as analyzed in Section~\ref{sec:mem}. To further evaluate the applicability of \sysname in more general scenarios, we construct datasets with varying numbers of sets and total keys, where the set sizes for each dataset follow a uniform distribution, and examine \sysname's performance on these datasets.

%The lookup throughput and memory cost on skewed datasets with Zipfian and normal distributions demonstrate the exceptional advantages of \sysname. One key reason is that the memory usage of our designed binary set separator, XBSS, decreases as the size ratio between the two partitioned subgroups increases, as analyzed in Section~\ref{sec:mem}. To further evaluate the applicability of \sysname in \sout{more general scenarios} \textcolor{red}{least favorable workloads}, we construct datasets with varying numbers of sets and total keys, where
%the set sizes for each dataset follow a uniform distribution, and examine \sysname's performance on these datasets.

%\textbf{Uniform distribution.} According to our design principle and analysis, the more biased the dataset is, the better \sysname performs. Therefore, to test the worst performance of our work, we run \sysname and benchmarks on the uniform distribution.

\noindent \textbf{Lookup throughput and memory cost vs. number of keys}. For the experimental setting, the number of sets is set as $2^5$ and the number of keys varies from $2^{12}$ to $2^{26}$. Figure~\ref{fig:TKU} and Figure~\ref{fig:MKU} show the lookup throughput and memory cost respectively. The results show that \textbf{\sysname achieves the highest lookup throughput} compared with other methods. \textbf{\sysname and Ludo incur the lowest memory cost, with \sysname requiring only 12.18--12.46 bits per key and Ludo requires 12.2 bits per key.} Notably, although $\mathtt{TMSQ_{FC}}$ consumes 35.75 bits per key when the dataset size is $2^{12}$, its memory efficiency improves significantly for larger datasets (e.g. $2^{18}$ to $2^{26}$), requiring only an average of 11.26 bits per key. The reason $\mathtt{TMSQ_{FC}}$ performs poorly in terms of memory cost on small datasets is the width of each bloom filter in the filter cascade is set with a fixed lower bound to prevent the number of layers from growing too rapidly. Consequently, even if the current Bloom filter produces only one false positive key, another Bloom filter layer must be added. In addition, the total number of keys is too small to amortize the relatively large memory cost.

%The 2 metrics of \sysname and the benchmarks are evaluated in the uniform distributed dataset, with \textbf{The outstanding throughput performance of the \sysname remains within our expectations.} For memory overhead, \sysname (12.18-12.46 bits per key) \textbf{requires the least memory space to store the dataset and outperforms all other algorithms}. Ludo, as the best benchmark, requires an average of 12.2-12.5 bits to store a key, which is very close to \sysname. It should be noted that although $TMSQ_{FC}$ requires 35.75 bits per keys to store a key when total dataset size is $2^{12}$, when dataset size is large (such as $2^{24}$ and $2^{16}$), only an average of 11.26 bits is required to store a key.

%The 2 metrics of \sysname and the benchmarks are evaluated in the uniform distributed dataset, with the number of sets $n$ is set as $2^5$ and the key size varies from $2^{12}$ to $2^26$. \textbf{The outstanding throughput performance of the \sysname remains within our expectations.} For memory overhead, \sysname (12.18-12.46 bits per key) \textbf{requires the least memory space to store the dataset and outperforms all other algorithms}. Ludo, as the best benchmark, requires an average of 12.2-12.5 bits to store a key, which is very close to \sysname. It should be noted that although $TMSQ_{FC}$ requires 35.75 bits per keys to store a key when total dataset size is $2^{12}$, when dataset size is large (such as $2^{24}$ and $2^{16}$), only an average of 11.26 bits is required to store a key.

\noindent \textbf{Lookup throughput and Memory cost vs. number of sets}. We set the number of keys as $2^{24}$ and vary the number of sets from $2^{3}$ to $2^9$. As shown in Figure~\ref{fig:TSU} and Figure~\ref{fig:MSU}, \sysname achieves over 120 Mops lookup throughput, \textbf{outperforming all other methods}. As the number of sets grows, the memory cost of \sysname increases from 7.31 to 21.92 bits per key. In contrast, Ludo requires an average of 12.2 bits to store a key, regardless of how the number of sets changes. $\mathtt{TMSQ_{FC}}$ as the second-best benchmark, requires an average of 6.06--19.47 bits to store a key, which is very close to \sysname. 

\noindent\textbf{Summary}. \sysname achieves the highest lookup throughput across all datasets. It also incurs the lowest memory cost on skewed datasets under both Zipfian and normal distributions. Even on the uniformly distributed datasets—its worst case in terms of memory usage—\sysname exhibits only slightly higher memory overhead than Ludo and remains comparable to $\mathtt{TMSQ_{FC}}$, while still outperforming all other baselines.

\subsection{Control Plane Update Evaluation of \sysname}
\label{sec:eval_update}

\begin{figure*}[t!]
\centering

% ---------- Left Figure ----------
\begin{minipage}[t]{0.48\textwidth}
    \centering
    \subfigure[Throughput vs. \# of keys]{
        \includegraphics[width=0.45\textwidth]{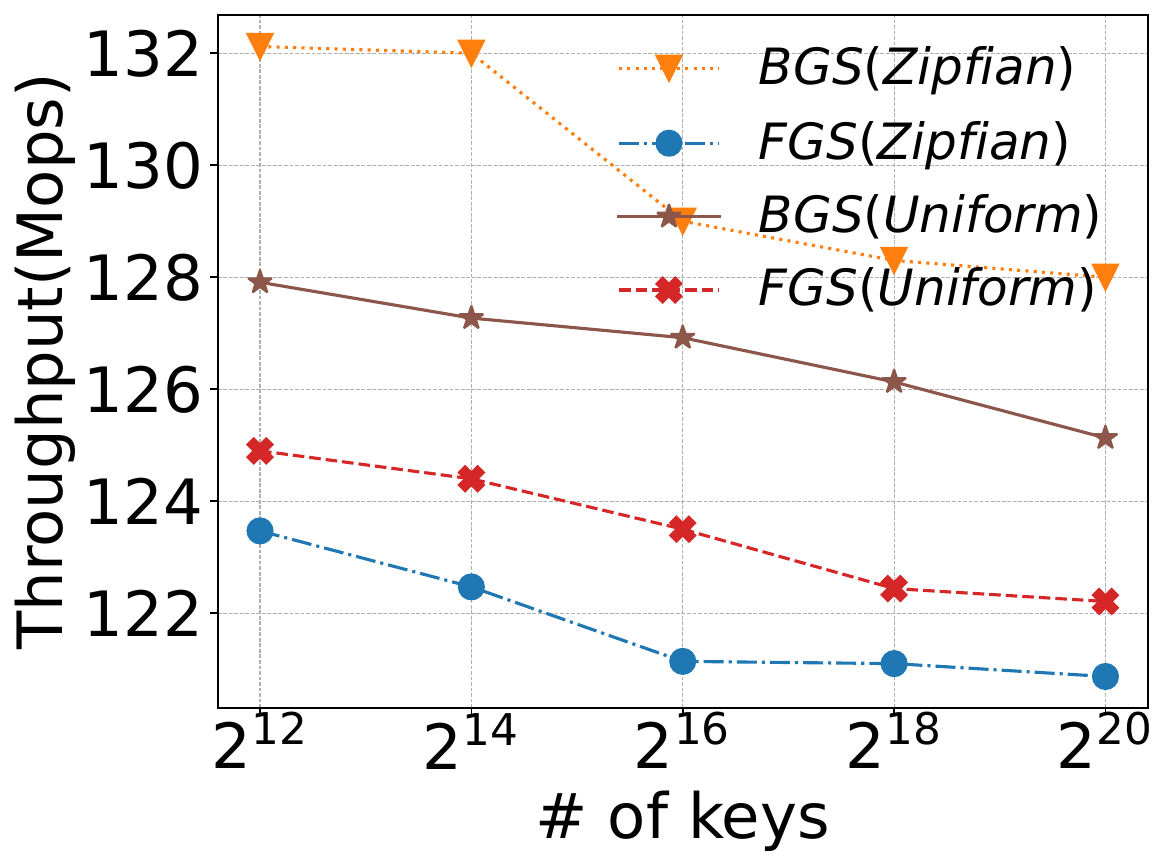}
        \label{fig:greedythroughputvskeys}
    }
    \hfill
    \subfigure[Throughput vs. \# of sets]{
        \includegraphics[width=0.45\textwidth]{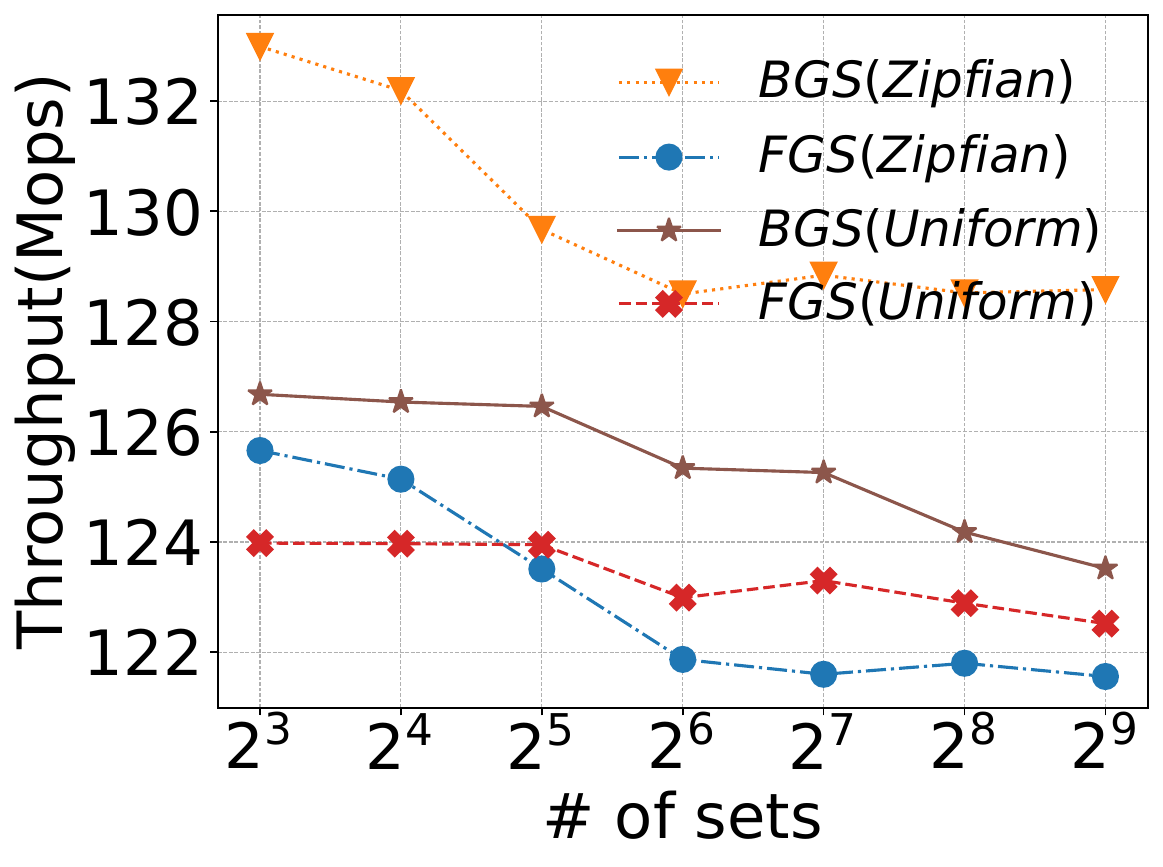}
        \label{fig:greedythroughputvssets}
    }
    \vspace{-3.5ex}
    \caption{Thro. Comparison Using Two Splitting Strategies.}
    \label{fig:greedythroughput}  
\end{minipage}
\hfill
% ---------- Right Figure ----------
\begin{minipage}[t]{0.48\textwidth}
    \centering
    \subfigure[DP mem. cost vs. \# of keys]{
        \includegraphics[width=0.45\textwidth]{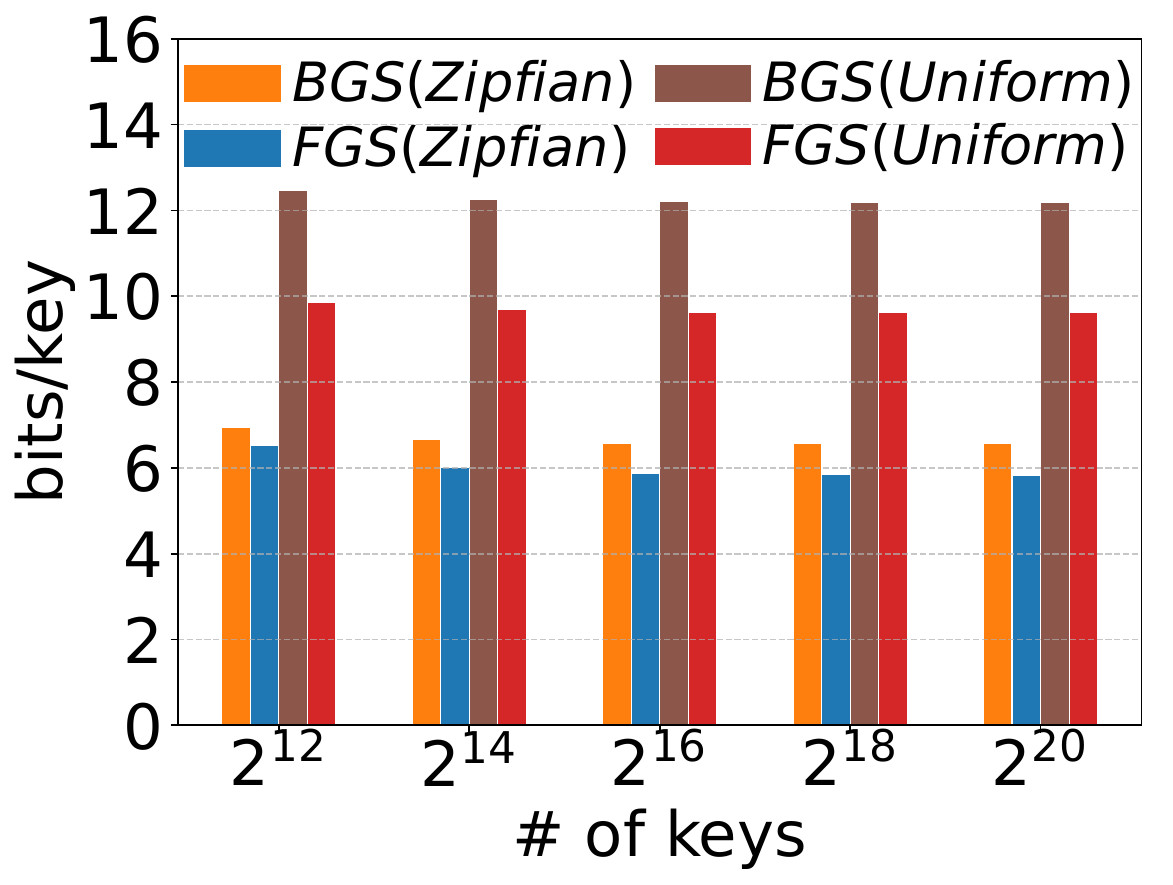}
        \label{fig:greedyDataMemvskeys}
    }
    \hfill
    \subfigure[CP mem. cost vs. \# of sets]{
        \includegraphics[width=0.45\textwidth]{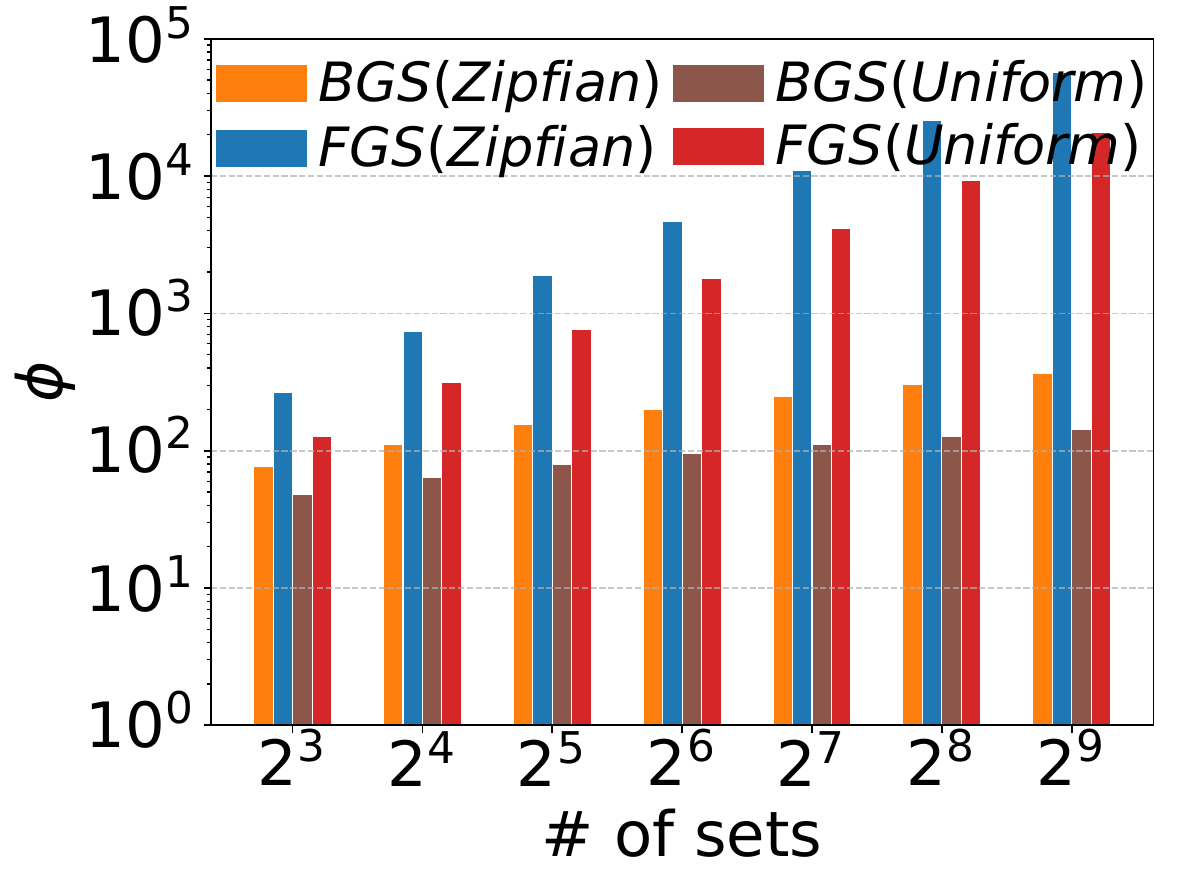}
        \label{fig:greedyConMemvssets}
    }
    \vspace{-3.5ex}
    \caption{Mem. Cost Using Two Splitting Strategies.}
    \label{fig:greedyDataPlane}
\end{minipage}

\end{figure*}

\begin{figure*}[t!]
\centering

% % ---------- Left Figure ----------
% \begin{minipage}[t]{0.48\textwidth}
%     \centering
%     \subfigure[CP mem. cost vs. \# of keys]{
%         \includegraphics[width=0.475\textwidth]{figures/Memory_control_greedy1.pdf}
%         \label{fig:greedyConMemvskeys}
%     }
%     \hfill
%     \subfigure[CP mem. cost vs. \# of sets]{
%         \includegraphics[width=0.475\textwidth]{figures/Memory_control_greedy2.pdf}
%         \label{fig:greedyConMemvssets}
%     }
%     \vspace{-3.5ex}
%     \caption{CP Mem. Cost Using Two Splitting Strategies.}
%     \label{fig:greedyConMem}
% \end{minipage}
% \hfill
% ---------- Right Figure ----------
\begin{minipage}[t]{0.48\textwidth}
    \centering
    \subfigure[Insertion latency vs. \# of keys]{
        \includegraphics[width=0.45\textwidth]{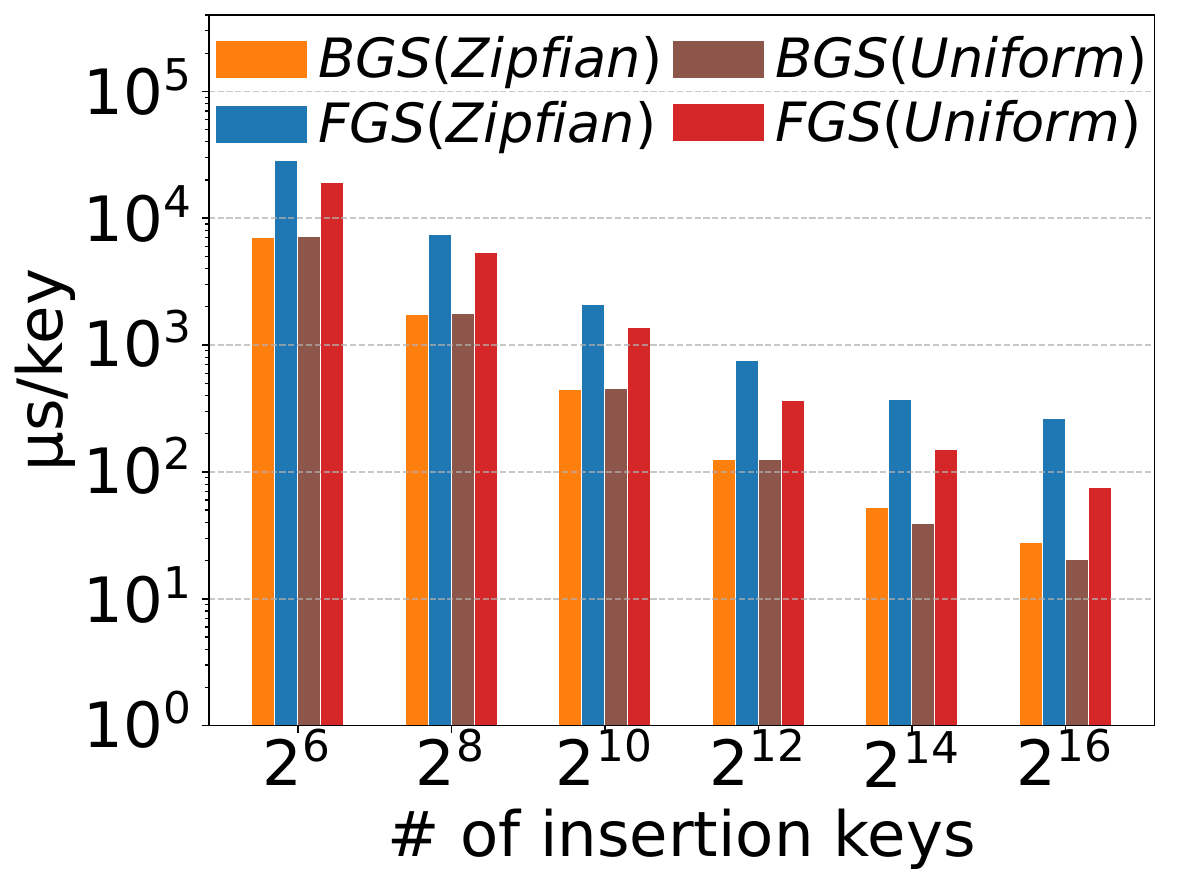}
        \label{fig:InsertGreedyvskeys}
    }
    \hfill
    \subfigure[Insertion latency vs. \# of sets]{
        \includegraphics[width=0.45\textwidth]{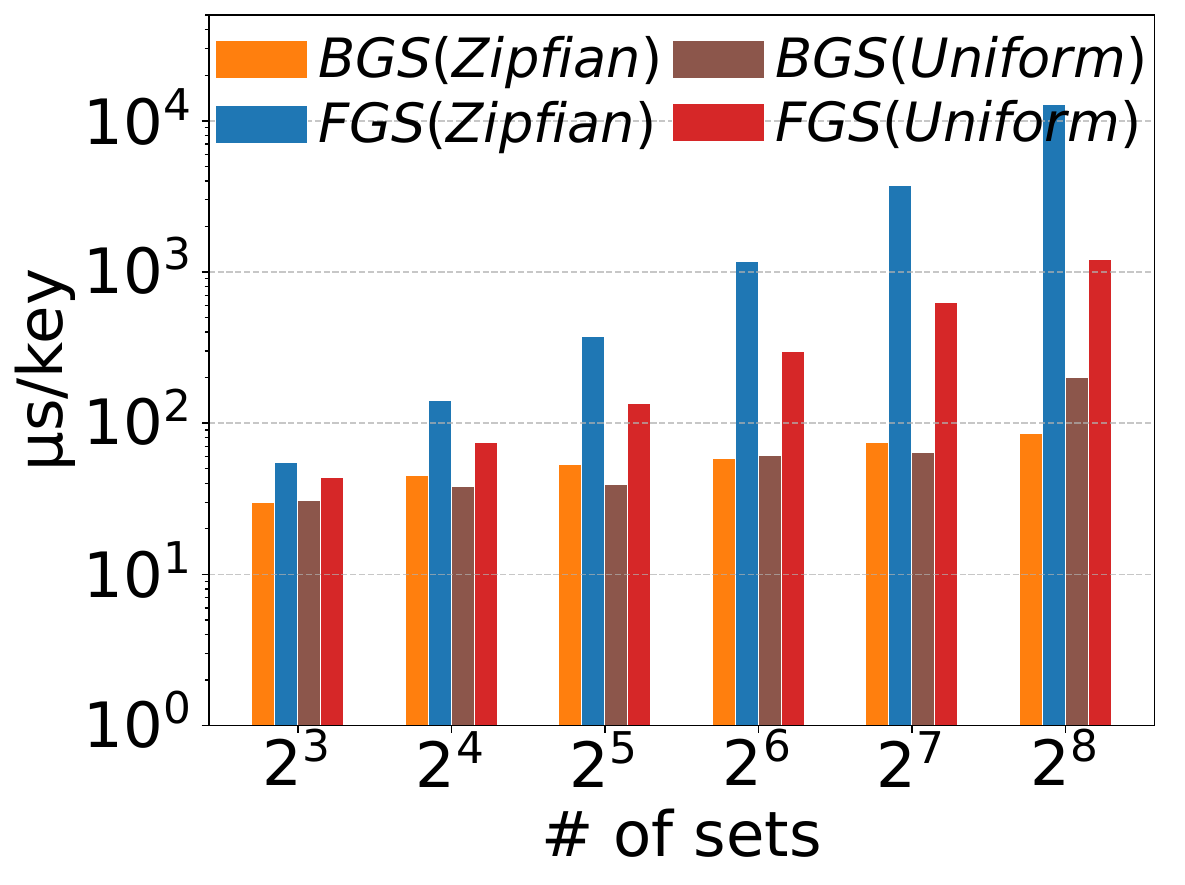}
        \label{fig:InsertGreedyvssets}
    }
    \vspace{-3.5ex}
    \caption{Insertion Latency of Two Splitting Strategies.}
    \label{fig:insertgreedy}
\end{minipage}
\hfill
\begin{minipage}[t]{0.48\textwidth}
    \centering
    \subfigure[Migration latency vs. \# of keys]{
        \includegraphics[width=0.45\textwidth]{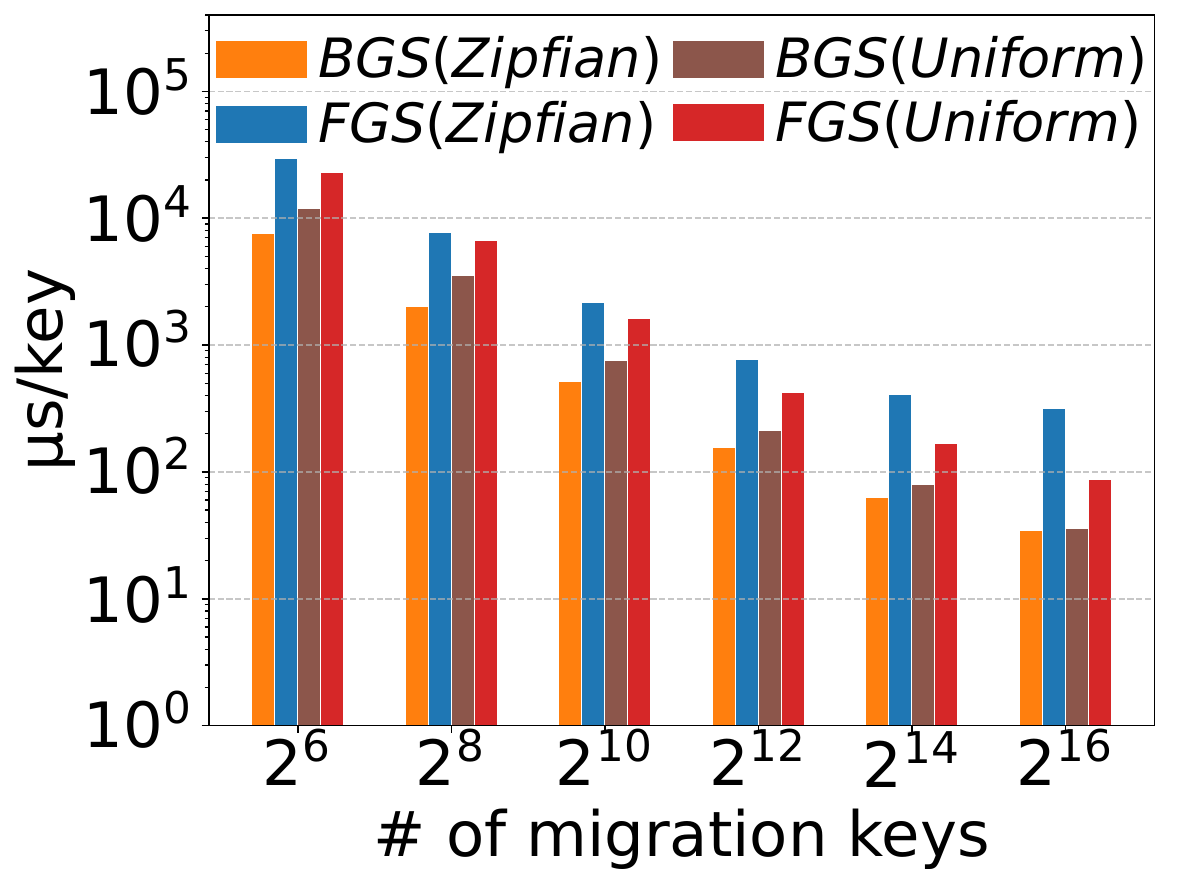}
        \label{fig:movevskeysgreedy}
    }
    \hfill
    \subfigure[Migration latency vs. \# of sets]{
        \includegraphics[width=0.45\textwidth]{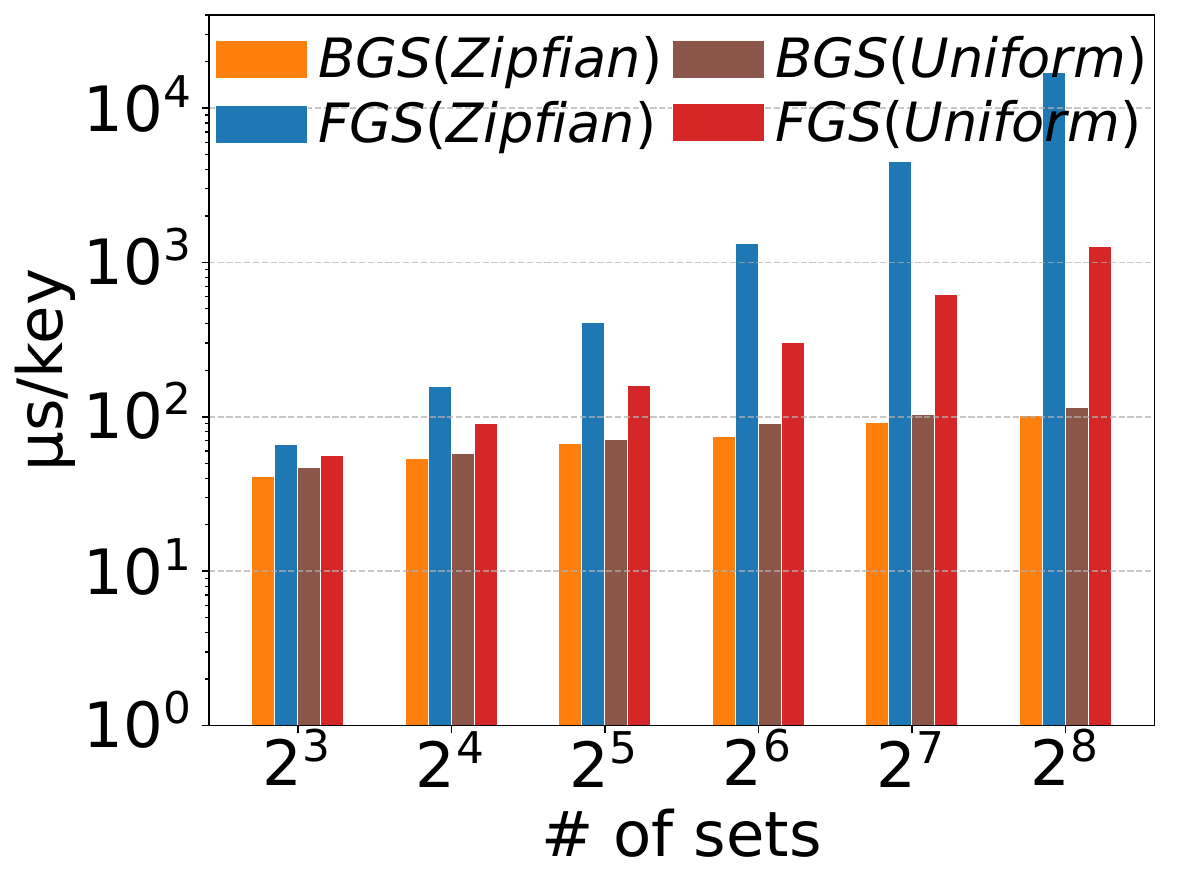}
        \label{fig:movevssetsgreedy}
    }
    \vspace{-3.5ex}
    \caption{Migration Latency of Two Splitting Strategies.}
    \label{fig:movegreedy}
\end{minipage}
\end{figure*}

%\wang{4. experiments: update, also add messages needs to be transmitted to update the structure: delta update}
The control plane of \sysname is responsible for constructing the data structure and dynamically updating it if there are membership changes. Among these baselines, only Ludo, $\mathtt{TMSQ_{FC}}$ and $\mathtt{TMSQ_{DASS}}$ can provide 100\% query accuracy. Since $\mathtt{TMSQ_{FC}}$ and $\mathtt{TMSQ_{DASS}}$ have the same structural framework as \sysname, and \sysname outperforms them in prior evaluations, we select Ludo as the baseline for control plane update evaluation. In addition, to compare \sysname with the approximate data structures, we also include CE, which achieves the highest lookup throughput among evaluated approximate data structures. All evaluations in this subsection are conducted on datasets with Zipfian and uniform distribution.

%STEM: [7003, 1744, 446, 126, 52.2, 27.6]
%Ludo: [5164, 1332, 333, 85, 22, 11]
%CE: [128388, 31945, 8389, 2076, 501, 125.52]
\noindent \textbf{Insertion latency vs. \# of insertion keys}. We evaluate the amortized insertion latency as the number of insertion keys increases. In the experimental setup, the original structure stores $2^{20}$ keys across $2^5$ sets. We then insert more keys, ranging from $2^6$ to $2^{16}$ in total, each insertion key is randomly assigned to one of the sets. As shown in Figure~\ref{fig:Insertvskeys}, Ludo is the fastest one in insertion among the three methods. \sysname incurs 1.31$\times$--2.51$\times$ the amortized insertion latency compared to Ludo. CE is the slowest in insertion, causing 4.55$\times$--18.81$\times$ the latency compared to \sysname, as it does not support dynamic updates and requires a full rebuild upon each new insertion.
%Specifically, Ludo achieves 1.31×–2.51× higher insertion throughput than \sysname, while \sysname achieves 4.55×–18.81× higher insertion throughput than CE. when $2^{16}$ keys are inserted, the amortized per-key migration latency for \sysname, Ludo and CE to move a key is 27.6 $\mu s$, 11 $\mu s$ and 125.52 $\mu s$ respectively. 
For all methods, when the number of new keys increases, the amortized latency decreases. When $2^{14}$ keys are inserted, the amortized per-key insertion latency for \sysname, Ludo and CE is 52.2 $\mu s$, 22 $\mu s$ and 501 $\mu s$, while the latencies are reduced to 27.6 $\mu s$, 11 $\mu s$ and 125.52 $\mu s$ respectively for inserting $2^{16}$ keys. Additionally, both Ludo and \sysname rely on Othello structures, which may require rebuilding during insertions. Due to \sysname’s adoption of hash function reuse across multiple XBSS, the probability of triggering an Othello rebuild is higher than that in Ludo. This design trade-off explains why Ludo achieves better update efficiency overall.

%\sysname and two benchmarks store $2^{20}$ keys from $2^5$ sets. We insert additional $K$ new keys with random set ID ($K$ varies from $2^{6}$ to $2^{16}$) to \sysname and benchmarks to test the insertion efficiency. 
%When the number of new keys grows, the amortized overhead decreases. Besides, the insertion performance of \sysname is close to that of Ludo in almost all cases, although \sysname's performance is slightly lower. In contrast, CE takes several times longer than Ludo and \sysname to complete a key. Specially, when $2^{16}$ new keys need to be inserted, the amortized time for \sysname, Ludo and CE to insert a key is 27.6 $\mu s$, 11 $\mu s$ and 125.52 $\mu s$. When $2^{14}$ new keys need to be inserted, the amortized time for \sysname, Ludo and CE to insert a key is 52.2 $\mu s$, 22 $\mu s$ and 501 $\mu s$. According to the original paper, CE does not support dynamic update, thus a new rebuild is needed every time there is new key. In addition, it should be note that the Othello part of the Ludo and \sysname may require rebuild in key insertion. Even worse, since \sysname introduces hash reuse, the rebuild probability of Othello in \sysname is greater than that in Ludo, which explains the fact that the update efficiency of Ludo is better.

\noindent \textbf{Insertion latency vs. number of sets.} In this experimental setup, we fix the number of insertion keys to be $2^{14}$. The original data structure stores a total of $2^{20}$ keys, with the number of sets varying from $2^3$ to $2^8$. Figure~\ref{fig:Insertvssets} shows the amortized insertion latency per key. When the number of sets grows, the amortized insertion time of \sysname grows from 30 $\mu s$ to 85.46 $\mu s$, because more $\mathtt{XBSS}$s need to be traversed during the insertion. In contrast, the amortized insertion time of Ludo remains stable at 23 $\mu s$, while that of CE increases from 314 $\mu s$ to 746.21 $\mu s$. In summary, the batch insertion of \sysname is highly efficient, with an average per-key insertion latency remaining in the microsecond range.

%In this evaluation, we insert $2^{14}$ keys into \sysname and two benchmarks, with fixed key size ($2^{20}$) and varying number of sets (from $2^3 to 2^8$). When the number of sets grows, the amortized insertion time of \sysname grows from 30 $\mu s$ to 85.46 $\mu s$, because more binary separators need to be traversed for a query process. In contrast, the amortized insertion time of Ludo maintains as 23 $\mu s$; amortized insertion time of CE grows from 314 $\mu s$ to 746.21 $\mu s$. Overall, batch insertion of \sysname has been shown to be efficient since the average time overhead for a single insertion is in the microsecond range.
\begin{figure}[t!]
\centering
\begin{minipage}[t]{0.48\textwidth}
    \centering
    \subfigure[Throughput]{
        \includegraphics[width=0.45\textwidth]{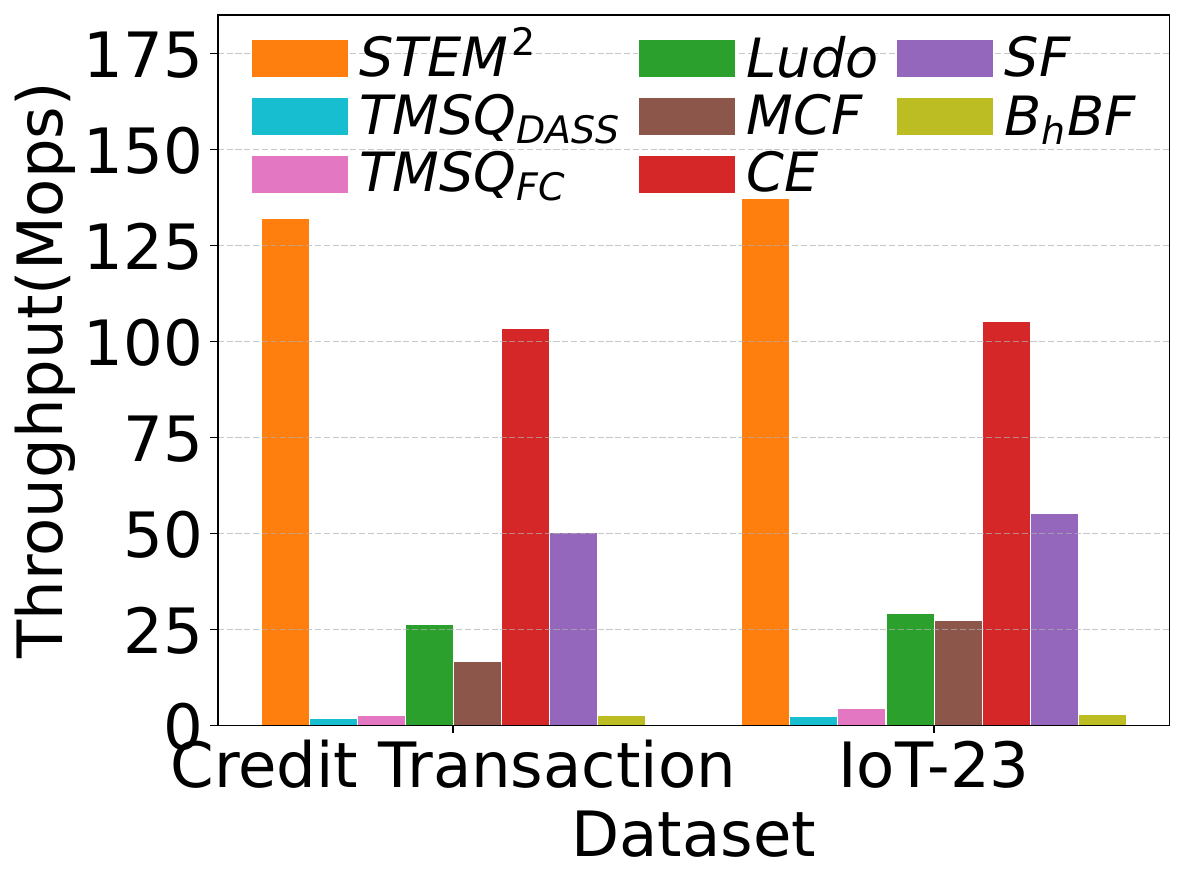}
        \label{fig:rel_thro}
    }
    \hfill
    \subfigure[Memory]{
        \includegraphics[width=0.45\textwidth]{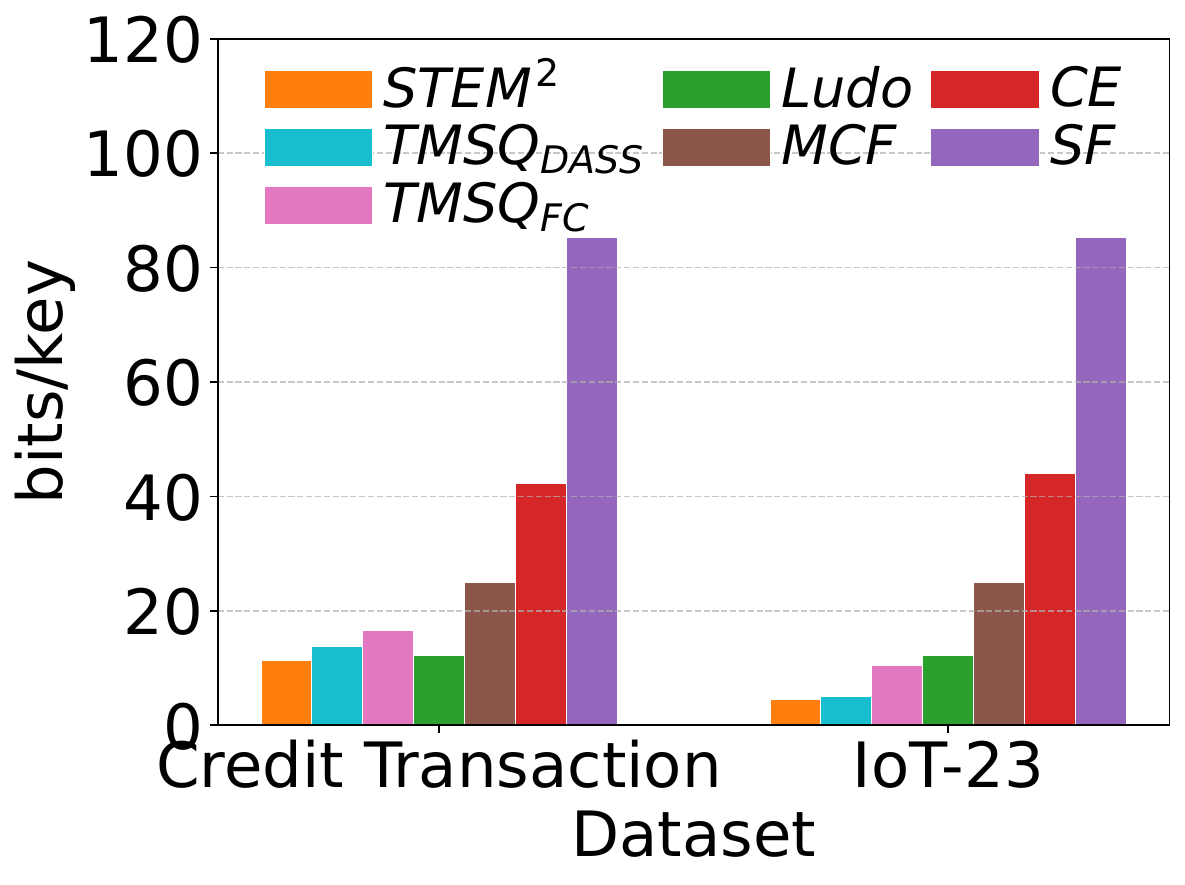}
        \label{fig:rel_mem}
    }
    \vspace{-3.5ex}
    \caption{Lookup Thro. and Memory Cost on Case Studies.}
    \label{fig:real}
\end{minipage}

\end{figure}

%which involves both insertion and deletion operations.
We further evaluate key migration latency, which involves both insertion and deletion operations.

% STEM: [7643, 1997, 518.19, 154.43, 63, 34.5]
% Ludo: [5230, 1284, 335.5, 81.43, 20.48, 5.15]
% CE: [132510, 33040, 8074, 2051, 508, 118]
    
\noindent\textbf{Key migration latency vs. number of keys.} We evaluate the amortized migration latency that varies with the different numbers of migration keys. In the experimental setup, the original structure stores $2^{20}$ keys across $2^5$ sets. We then randomly select a number of keys ranging from $2^6$ to $2^{16}$, with each selected key randomly migrated to another set. As shown in Figure~\ref{fig:movevskeys}, Ludo is the fastest one in migration, followed by \sysname, and CE incurs the highest migration latency. Specifically, \sysname causes 1.46$\times$--6.7$\times$ the migration latency of Ludo, while CE causes 3.42$\times$--17.34$\times$ the migration latency of \sysname. When $2^{16}$ keys are migrated, the amortized per-key migration latency for \sysname, Ludo and CE is 34.5 $\mu s$, 5.15 $\mu s$ and 118 $\mu s$ respectively. 

%When $2^{14}$ keys are migrated, the corresponding latencies are 63 $\mu s$, 20.48 $\mu s$ and 508 $\mu s$.

%In this evaluation, \sysname and two benchmarks store $2^{20}$ keys from $2^5$ sets. We select $K$ keys randomly (K varies from $2^4$ to $2^{14}$) and move them from their original sets to another different random sets. As the number of keys to be moved increases, the amortized movement time gradually decreases. Besides, the moving performance of \sysname is close to that of Ludo in almost all cases, although \sysname's performance is slightly lower. In contrast, CE takes several times longer than Ludo and \sysname to complete a key move. Specially, when $2^{16}$ new keys need to be move from one set to other set, the amortized time for \sysname, Ludo and CE to move a key is 34.5 $\mu s$, 5.15 $\mu s$ and 118 $\mu s$. When $2^{14}$ keys need to be moved, the amortized time for \sysname, Ludo and CE to insert a key is 63 $\mu s$, 20.48 $\mu s$ and 508 $\mu s$.

% STEM:[41.13, 54.08, 66.9, 74.9, 92.52, 102.1]
% Ludo:[20.33, 20.53, 20.4, 20.9, 20.7, 20.5]
% CE:[314, 418, 534, 632, 732, 811]
\noindent\textbf{Key migration latency vs. number of sets.} In this experimental setup, we fix the number of migration keys to be $2^{14}$. The original data structure stores a total of $2^{20}$ keys, with the number of sets varying from $2^3$ to $2^8$. As shown in Figure~\ref{fig:Movevssets}, as the number of sets increases, the amortized key migration latency of \sysname increases from 41.13 $\mu s$ to 102.1 $\mu s$, because more $\mathtt{XBSS}$ instances need to be traversed during the migration process. In contrast, the amortized migration latency of Ludo remains nearly constant at around 21 $\mu s$, while that of CE increases significantly from 314 $\mu s$ to 811 $\mu s$. Comparing Figure~\ref{fig:Insertvssets} and Figure~\ref{fig:Movevssets}, we can find that the key migration process in \sysname involves both deleting the key from its original set and inserting it into the new set, which results in a longer delay compared to the pure insertion operation.

\noindent\noindent \textbf{Discussion.} Although \sysname has slightly higher update latency than Ludo, particularly when the number of sets is large, Ludo provides much lower query throughput. Thus, for query-intensive or latency-sensitive applications with infrequent updates, \sysname achieves a better trade-off.
%Although the update latency of \sysname is higher than that of Ludo—particularly when the number of sets becomes large—Ludo suffers from significant limitations in query throughput. Therefore, for query-intensive or latency-sensitive applications where dynamic updates occur infrequently, \sysname provides a more preferable trade-off.}

%when the number of sets increases, the amortized key migration time of \sysname increases from 41.13 $\mu s$ to 102.1 $\mu s$, because more binary separators need to be traversed for a query process. In contrast, the amortized insertion time of Ludo maintains as nearly 21 $\mu s$; amortized insertion time of CE grows from 314 $\mu s$ to 811 $\mu s$.

\subsection{Performance of Splitting Strategies}
\label{sec:split_eval}
We evaluate two splitting strategies, $\mathtt{FGS}$ and $\mathtt{BGS}$, for binary-tree construction on datasets with Zipfian and Uniform distributions.
%\textcolor{red}{ which are the most favorable and least favorable workloads for \sysname.}

%We analyzed that fully greedy splitting can minimize the data plane memory overhead, but increase the depth of the tree and result in an extremely unbalanced tree.

%As we mentioned above, there are two splitting strategies when we build a binary tree for \sysname, the first strategy is greedy strategy to minimize the data plane memory overhead, but will increase the depth of the binary tree and result in an extremely unbalanced tree. The second strategy is to divide multiple sets equally at each separation, which we call balanced splitting strategy. As we analysis above, greedy splitting will decrease the building, searching and update efficiency because a normal operation of \sysname will go through more binary set separators (non-leaf nodes of the tree). We compare the two strategies as follows and all evaluations are completed in zipfian distributed dataset.

\noindent\textbf{Lookup throughput.} We evaluate lookup throughput as the number of keys and sets varies. Figure~\ref{fig:greedythroughputvskeys} shows the results when the number of keys increases from $2^{12}$ to $2^{20}$ with the number of sets fixed at $2^5$, and Figure~\ref{fig:greedythroughputvssets} shows the results when the number of sets increases from $2^{3}$ to $2^{9}$ with the number of keys fixed at $2^{20}$. The results show $\mathtt{BGS}$ achieves 5.0\%--7.8\% higher lookup throughput than $\mathtt{FGS}$ on Zipfian-distributed datasets, and 1.1\%--3.0\% higher throughput on uniformly distributed datasets. Under $\mathtt{FGS}$, a query with $2^9$ sets traverses up to 512 $\mathtt{XBSS}$ instances, whereas $\mathtt{BGS}$ traverses only 9. Since less-hashing technique keeps the per-$\mathtt{XBSS}$ lookup overhead low, the throughput gap remains modest.

\noindent\textbf{Data plane (DP) memory cost.} We use the same experimental settings to evaluate data-plane memory cost under the two splitting strategies. Figure~\ref{fig:greedyDataMemvskeys} shows the results when the number of sets is fixed at $2^{5}$ and the number of keys increases from $2^{12}$ to $2^{20}$. On Zipfian-distributed datasets, the memory cost decreases from 6.95 to 6.56 bits/key under $\mathtt{BGS}$ and from 6.51 to 5.83 bits/key under $\mathtt{FGS}$, giving $\mathtt{FGS}$ a 6.3\%--11.1\% advantage. On uniformly distributed datasets, the memory cost decreases from 12.46 to 12.18 bits/key under $\mathtt{BGS}$ and from 9.86 to 9.63 bits/key under $\mathtt{FGS}$, corresponding to a larger gap. In summary, $\mathtt{FGS}$ consistently achieves lower DP memory cost than $\mathtt{BGS}$, and its advantage is modest on Zipfian datasets but substantially larger on uniformly distributed datasets. This is because the more aggressive greedy partitioning in $\mathtt{FGS}$ is especially beneficial when the set sizes are less skewed.

\noindent\textbf{Control plane (CP) memory cost.} As described in Section~\ref{sec:xbss_c}, \sysname introduces an index table for each counting Bloom filter to avoid re-testing all keys during update operations. This index table is maintained as part of the control plane. We adopt the same experimental settings as that in evaluating the lookup throughput and data plane memory cost. To quantify the control-plane memory overhead, we introduce a new metric, denoted by $\phi$, defined as the ratio between the number of keys stored in the index table and the total number of keys. This metric captures the relative memory burden imposed by the index table, which incurs substantially higher memory cost than the core data structure of \sysname.

As shown in Figure~\ref{fig:greedyConMemvssets}, when the number of keys is fixed at $2^{20}$ and the number of sets increases from $2^3$ to $2^9$: the index-table memory cost of $\mathtt{FGS}$ is 3.48$\times$--157.18$\times$ that of $\mathtt{BGS}$ on Zipfian-distributed datasets and 2.65$\times$--144.8$\times$ on uniformly distributed datasets. This gap arises because $\mathtt{FGS}$ increases the ratio $r=|D_L|/|D_S|$ for each $\mathtt{XBSS}$ in \sysname, which results in a larger $\mathtt{CBF}$ index table. Besides, under the same strategy and dataset settings, uniformly distributed datasets incur lower CP memory overhead than Zipfian datasets.

\noindent\textbf{New key insertion latency.} Figure~\ref{fig:insertgreedy} compares the insertion latency of the two splitting strategies. Because $\mathtt{FGS}$ introduces more levels in the binary tree, its average per-key insertion latency is substantially higher than that of the $\mathtt{BGS}$ strategy used in \sysname. Specifically, when inserting $2^6$--$2^{16}$ new keys into \sysname with $2^5$ sets and an initial size of $2^{20}$ keys, $\mathtt{FGS}$ incurs 4.06$\times$--9.60$\times$ higher insertion latency than $\mathtt{BGS}$ on Zipfian-distributed datasets, and 2.66$\times$--3.84$\times$ higher latency on uniformly distributed datasets.

%Figure~\ref{fig:insertgreedy} shows the insertion latency of two different splitting strategies. Since the FGS strategy introduces more binary tree levels, the average insertion time per key is significantly higher than that of the BGS strategy used in \sysname. Specifically, when inserting $2^6$--$2^{16}$ new keys into the \sysname that originally stores $2^{20}$ keys for $2^5$ sets (Zipfian distributed), the key insertion latency of the FGS strategy is 4.06$\times$--9.60$\times$ that of the BGS strategy. \textcolor{red}{Besides, when datasets are Uniformly distributed, the key insertion latency of the FGS strategy is 2.66$\times$--3.84$\times$ that of the BGS strategy.}

When \sysname initially stores $2^{20}$ keys and the number of sets increases from $2^{3}$ to $2^{8}$, inserting $2^{14}$ new keys into the Zipfian dataset causes $\mathtt{FGS}$ to incur 1.82$\times$--149.59$\times$ the insertion latency of $\mathtt{BGS}$, indicating a substantial gap. On uniformly distributed datasets, the corresponding gap is smaller but still significant, with $\mathtt{FGS}$ incurring 1.43$\times$--9.79$\times$ higher insertion latency than $\mathtt{BGS}$.

%When the original \sysname stores $2^{20}$ keys and the number of sets varies from $2^{3}$ to $2^{8}$ (datasets are Zipfian distributed), the latency of the FGS strategy when inserting $2^{14}$ new keys is 1.82$\times$ -- 149.59$\times$ that of the BGS strategy, revealing a substantial performance gap. \textcolor{red}{Besides, when datasets are Uniformly distributed, the key insertion latency of the FGS strategy is 1.43$\times$--9.79$\times$ that of the BGS strategy.}

\noindent\textbf{Key migration latency.} We use the same experimental setup as in the key insertion latency evaluation. As shown in Figure~\ref{fig:movegreedy}, $\mathtt{FGS}$ consistently incurs higher migration latency than $\mathtt{BGS}$. When migrating $2^6$--$2^{16}$ random keys in \sysname with $2^5$ sets and an initial size of $2^{20}$ keys, the amortized migration latency of $\mathtt{FGS}$ is 3.84$\times$--9.24$\times$ that of $\mathtt{BGS}$ on Zipfian-distributed datasets, and 1.87$\times$--2.41$\times$ on uniformly distributed datasets. When the original structure stores $2^{20}$ keys and the number of sets increases from $2^{3}$ to $2^{8}$, the gap becomes even larger. On Zipfian-distributed datasets, the migration latency of $\mathtt{FGS}$ is 1.61$\times$--165.41$\times$ that of $\mathtt{BGS}$. On uniformly distributed datasets, $\mathtt{FGS}$ still incurs 1.2$\times$--10.98$\times$ higher migration latency. This difference arises because $\mathtt{FGS}$ produces a deeper and more unbalanced tree, causing each migration to update many more $\mathtt{XBSS}$ instances, and the penalty becomes especially pronounced as the number of sets increases.

%We use the same experimental setup as in the key insertion latency evaluation. As shown in Figure~\ref{fig:movegreedy}, $\mathtt{FGS}$ incurs substantially higher latency than $\mathtt{BGS}$. Specifically, when migrating $2^6$--$2^{16}$ random keys in \sysname with $2^5$ sets and an initial size of $2^{20}$ keys, the amortized migration latency of $\mathtt{FGS}$ is 3.84$\times$--9.24$\times$ that of $\mathtt{BGS}$ on Zipfian-distributed datasets, \wang{and 1.87$\times$--2.41$\times$ on uniformly distributed datasets.}

%When the original structure stores $2^{20}$ keys and the number of sets increases from $2^{3}$ to $2^{8}$, the key migration latency of $\mathtt{FGS}$ is 1.61$\times$--165.41$\times$ that of $\mathtt{BGS}$ on Zipfian-distributed datasets. \wang{This large gap arises mainly because $\mathtt{FGS}$ requires updating many more $\mathtt{XBSS}$ instances. On uniformly distributed datasets, the gap is smaller but still notable, with $\mathtt{FGS}$ incurring 1.2$\times$--10.98$\times$ the migration latency of $\mathtt{BGS}$.}

\noindent \textbf{Discussion}. Overall, $\mathtt{FGS}$ and $\mathtt{BGS}$ expose different trade-offs. $\mathtt{FGS}$ achieves lower DP memory cost, with the advantage being especially noticeable on uniformly distributed datasets, but it incurs substantially higher CP memory overhead and update latency due to its deeper and more unbalanced tree structure. In contrast, $\mathtt{BGS}$ provides higher lookup throughput and much better update efficiency, while incurring only a modest increase in data-plane memory cost.

\subsection{Evaluation on Real Dataset}
We evaluate the performance of \sysname in two case studies using the aforementioned network traffic and credit card transaction datasets. Figure~\ref{fig:real} shows the lookup throughput and memory cost. The results show that \textbf{\sysname achieves the highest lookup throughput} among all methods, exceeding 130 Mops. For memory cost, \sysname requires an average of 11.41 and 4.49 bits per key, respectively, achieving the lowest memory cost among all methods.
%For memory cost, \sysname requires an average of 11.41 bits and 4.49 bits \textcolor{red}{\sout{keys}} to store a key, respectively, \textbf{achieving the lowest memory cost} among all methods.

%% file: 6_conclusion.tex
\section{Conclusion}
\label{sec:conclu}
%\subsection{Conclusion}

This paper presents a holistic design for exact multi-set membership queries, centered on the proposed \sysname data structure. Through a decoupled architecture, \sysname separates maintenance and query processing into two planes: a maintenance-oriented control plane for structure construction and incremental updates, and a query-optimized data plane for high-throughput lookups. This design is enabled by our novel Exact Binary Set Separator ($\mathtt{XBSS}$) and the less-hashing technique that requires only two hash computations, jointly ensuring 100\% query accuracy without sacrificing performance. Extensive experiments show that \sysname outperforms state-of-the-art hashing- and filter-based solutions in both throughput and memory efficiency, providing a robust and scalable foundation for high-performance networking and database applications.